\documentclass[twocolumn,floatfix,times,trackchanges]{aastex63}

\def\sun{\hbox{$\odot$}}
\def\earth{\hbox{$\oplus$}}
\def\lesssim{\mathrel{\hbox{\rlap{\hbox{%
 \lower4pt\hbox{$\sim$}}}\hbox{$<$}}}}
\def\gtrsim{\mathrel{\hbox{\rlap{\hbox{%
 \lower4pt\hbox{$\sim$}}}\hbox{$>$}}}}

\def\arcmin{\hbox{$^\prime$}}
\def\arcs{\hbox{$^{\prime\prime}$}}

\def\micron{\hbox{$\mu$m}}

\newcommand{\Ntot}{$N_\mathrm{tot}$}

\newcommand{\persqcm}{\mbox{~cm$^{-2}$}}

\newcommand{\kmpers}{\mbox{~km~s$^{-1}$}}

\newcommand{\Eu}{$E_\mathrm{u}$}
\newcommand{\gu}{$g_\mathrm{u}$}

\newcommand{\vlr}[3]{${#1}^{+#2}_{-#3}$}

\usepackage{CJKutf8}
\usepackage{graphicx}
\usepackage{longtable} \LTcapwidth=\textwidth
\usepackage{multirow}
\usepackage{subfigure}
\usepackage{ulem}
\usepackage{soul}
\usepackage{lipsum, babel}
\usepackage{enumitem}
\usepackage{amsmath}
\usepackage{euscript}
\usepackage{verbatim}

\setstcolor{red}

\accepted{\today}
\submitjournal{ApJS}

\shorttitle{QWOSCO. Overview, Isotopologues, Isomers, and Complex Organics}
\shortauthors{Hsu et al.}
\graphicspath{{./}{figures/}}

\begin{document}

\begin{CJK*}{UTF8}{bsmi}
\title{
Q/W-band Observations toward Starless Cores in Orion (QWOSCO) I. Overview, Isotopologues, Isomers, and Complex Organics\\
}

\author[0000-0002-1369-1563]{Shih-Ying Hsu}
\email{seansyhsu@gmail.com}
\affiliation{Institute of Astronomy and Astrophysics, Academia Sinica, No.1, Sec. 4, Roosevelt Rd, Taipei 106319, Taiwan (R.O.C.)}

\author[0000-0001-8315-4248]{Xunchuan Liu}
\email{liuxunchuan001@gmail.com}
\affiliation{Shanghai Astronomical Observatory, Chinese Academy of Sciences, Shanghai 200030, PR China}

\author[0000-0012-3245-1234]{Sheng-Yuan Liu}
\email{syliu@asiaa.sinica.edu.tw}
\affiliation{Institute of Astronomy and Astrophysics, Academia Sinica, No.1, Sec. 4, Roosevelt Rd, Taipei 106319, Taiwan (R.O.C.)}

\author[0000-0002-5286-2564]{Tie Liu}
\affiliation{Shanghai Astronomical Observatory, Chinese Academy of Sciences, Shanghai 200030, PR China}

\author[0000-0001-9304-7884]{Naomi Hirano}
\affiliation{Institute of Astronomy and Astrophysics, Academia Sinica, No.1, Sec. 4, Roosevelt Rd, Taipei 106319, Taiwan (R.O.C.)}

\author[0000-0002-5809-4834]{Mika Juvela}
\affiliation{Department of Physics, P.O.Box 64, FI-00014, University of Helsinki, Finland}

\author[0000-0003-2412-7092]{Kee-Tae Kim}
\affiliation{Korea Astronomy and Space Science Institute, 776 Daedeokdae-ro, Yuseong-gu, Daejeon 34055, Republic of Korea}
\affiliation{University of Science and Technology, Korea (UST), 217 Gajeong-ro, Yuseong-gu, Daejeon 34113, Republic of Korea}

\author[0000-0002-3024-5864]{Chin-Fei Lee}
\affiliation{Institute of Astronomy and Astrophysics, Academia Sinica, No.1, Sec. 4, Roosevelt Rd, Taipei 106319, Taiwan (R.O.C.)}

\author[0000-0003-1275-5251]{Shanghuo Li}
\affiliation{School of Astronomy and Space Science, Nanjing University, 163 Xianlin Avenue, Nanjing 210023, People's Republic of China}
\affiliation{Key Laboratory of Modern Astronomy and Astrophysics (Nanjing University), Ministry of Education, Nanjing 210023, People's Republic of China}

\author[0000-0002-6868-4483]{Sheng-Jun Lin}
\affiliation{Institute of Astronomy and Astrophysics, Academia Sinica, No.1, Sec. 4, Roosevelt Rd, Taipei 106319, Taiwan (R.O.C.)}

\author{Sheng-Li Qin}
\affiliation{School of Physics and Astronomy, Yunnan University, Kunming 650091, People's Republic of China}

\author[0000-0002-4393-3463]{Dipen Sahu}
\affiliation{Physical Research laboratory, Navrangpura, Ahmedabad, Gujarat 380009, India}

\author[0000-0002-8149-8546]{Ken'ichi Tatematsu}
\affil{Nobeyama Radio Observatory, National Astronomical Observatory of Japan, National Institutes of Natural Sciences, Nobeyama, Minamimaki, Minamisaku, Nagano 384-1305, Japan}
\affiliation{Astronomical Science Program, SOKENDAI (The Graduate University for Advanced Studies), 2-21-1 Osawa, Mitaka, Tokyo 181-8588, Japan}

\author[0000-0001-5950-1932]{Fengwei Xu}
\affiliation{Max Planck Institute for Astronomy, Heidelberg, Germany}

\author[0000-0002-5310-4212]{L. Viktor T\'oth}
\affiliation{Institute of Physics and Astronomy, E\"otv\"os Lor\`and University, P\'azm\'any P\'eter s\'et\'any 1/A, H-1117 Budapest, Hungary}
\affiliation{Faculty of Science and Technology, University of Debrecen, H-4032 Debrecen, Hungary}

\begin{abstract} 
Molecular inventories in starless cores are powerful tools for probing the physical and chemical structures at the earliest stages of star formation. 
Wide-band spectral scans are invaluable for obtaining a comprehensive view of the chemical composition.
In this paper, we present the first results from the project Q/W-band Observations toward Starless Cores in Orion (QWOSCO), which uses the Yebes 40-m telescope to survey 23 starless cores in the Orion cloud at the Q (31.0--50.5 GHz) and W (71.1--91.4 GHz) bands with a total bandwidth of 40 GHz. 
We detect approximately 40 molecular species and derive their column densities, with each species exhibiting a characteristic spread of roughly one order of magnitude.
The derived isomer and isotopologue column density ratios, including {\textit{A}}/{\textit{E}}, {\textit{ortho}}/{\textit{para}}, {\textit{cyclic}}/{\textit{linear}}, HNC/HCN, $^{12}$C/$^{13}$C, $^{14}$N/$^{15}$N, $^{16}$O/$^{18}$O, $^{32}$S/$^{34}$S, and D/H, are consistent with expectations for starless environments.
Our results together with the literature suggest that the complex organic molecules (COMs) CH$_3$OH and CH$_3$CHO are both likely ubiquitous in starless cores. 
The column density ratio of CH$_3$CHO with respect to CH$_3$OH in starless cores are comparable or lower by a factor of around 25 than those in hot corinos at the protostellar stages if the CH$_3$OH column density is directly derived or rescaled from that of $^{13}$CH$_3$OH, respectively. 
Accordingly, we discuss the possible roles of methanol opacity and chemical mechanisms across the starless and protostellar stages.
\end{abstract}


\section{Introduction}
\label{sec:Intro}

\begin{deluxetable*}{llllcrl}
\label{tab:coord}
\caption{
Information of the targets.
}
\tablehead{
\colhead{Name} & \colhead{Short Name} & \colhead{$\alpha$} & \colhead{$\delta$} & \colhead{Cloud} & \colhead{$v_\mathrm{LSR}$}  & \colhead{JCMT Name} \\
\colhead{}     & \colhead{}           & \colhead{(J2000)} & \colhead{(J2000)} & \colhead{} & \colhead{(km s$^{-1}$)}  & \colhead{}
}
\startdata
G198.69-09.12N1 & G198.69N1 & 05:52:29.61 & +08:15:37.0 & $\lambda$ Orionis & 11.10 & G198.69-09.12North1 \\
G198.69-09.12N2 & G198.69N2 & 05:52:25.30 & +08:15:09.0 & $\lambda$ Orionis & 10.70 & G198.69-09.12North2 \\
\hline
G203.21-11.20E1 & G203.21E1 & 05:53:51.00 & +03:23:07.3 & Orion B           & 10.30 & G203.21-11.20East1 \\
G203.21-11.20E2 & G203.21E2 & 05:53:47.48 & +03:23:11.3 & Orion B           & 10.20 & G203.21-11.20East2 \\
G205.46-14.56M3 & G205.46M3 & 05:46:05.98 & -00:09:32.3 & Orion B           & 10.00 & G205.46-14.56North1$^{\dagger}$ \\
G206.21-16.17N  & G206.21N  & 05:41:39.54 & -01:35:52.2 & Orion B           &  9.80 & G206.21-16.17North \\
G206.21-16.17S  & G206.21S  & 05:41:36.37 & -01:37:43.6 & Orion B           &  9.80 & G206.21-16.17South \\
\hline
G207.36-19.82N4 & G207.36N4 & 05:30:44.55 & -04:10:27.4 & Orion A           & 11.20 & G207.36-19.82North4 \\
G208.68-19.20N2 & G208.68N2 & 05:35:20.47 & -05:00:50.4 & Orion A           & 11.10 & G208.68-19.20North2 \\
G209.29-19.65N1 & G209.29N1 & 05:35:00.38 & -05:39:59.7 & Orion A           &  8.50 & G209.29-19.65North1 \\
G209.29-19.65S1 & G209.29S1 & 05:34:55.99 & -05:46:04.0 & Orion A           &  8.70 & G209.29-19.65South1 \\
G209.29-19.65S2 & G209.29S2 & 05:34:53.81 & -05:46:17.6 & Orion A           &  7.60 & G209.29-19.65South2 \\
G209.55-19.68N2 & G209.55N2 & 05:35:07.50 & -05:56:42.4 & Orion A           &  8.20 & G209.55-19.68North2 \\
G209.77-19.40E3 & G209.77E3 & 05:36:35.90 & -06:02:42.2 & Orion A           &  8.20 & G209.77-19.40East3 \\
G209.79-19.80W  & G209.79W  & 05:35:10.70 & -06:13:59.3 & Orion A           &  5.80 & G209.79-19.80West \\
G209.94-19.52N  & G209.94N  & 05:36:11.55 & -06:10:44.7 & Orion A           &  8.20 & G209.94-19.52North \\
G209.94-19.52S1 & G209.94S1 & 05:36:24.96 & -06:14:04.7 & Orion A           &  8.00 & G209.94-19.52South1 \\
G210.37-19.53N  & G210.37N  & 05:36:55.03 & -06:34:33.2 & Orion A           &  6.40 & G210.37-19.53North \\
G210.82-19.47N2 & G210.82N2 & 05:38:00.00 & -06:57:15.5 & Orion A           &  5.20 & G210.82-19.47North2 \\
G211.16-19.33N4 & G211.16N4 & 05:38:55.68 & -07:11:25.9 & Orion A           &  4.50 & G211.16-19.33North4 \\
G211.16-19.33N5 & G211.16N5 & 05:38:46.00 & -07:10:41.9 & Orion A           &  4.30 & G211.16-19.33North5 \\
G211.72-19.25S1 & G211.72S1 & 05:40:21.21 & -07:36:08.8 & Orion A           &  4.30 & \nodata \\
G212.10-19.15N1 & G212.10N1 & 05:41:21.34 & -07:52:26.9 & Orion A           &  4.30 & G212.10-19.15North1
\enddata
\tablecomments{
The targeted coordinates are based on the continuum peak in at 1.3~mm observations reported by \citet{2020Dutta_ALMASOP}. 
The ``JCMT Name'' is the name of the source used in \citet{2018Yi_PGCC_SCUBA2_II}. 
The dagger ($^{\dagger}$) denotes those JCMT names different from the ALMA names. 
}
\end{deluxetable*}

Low- and intermediate-mass starless cores are the potential precursors of solar-like stars. 
A starless core is a dense condensation of gas and dust within a molecular cloud that lacks any embedded protostar. 
Among these, the cores that are expected to overcome turbulent, thermal, and magnetic support—thus destined to collapse and form protostars—are referred to as prestellar cores \citep[e.g.,][]{2007DiFrancesco_review}.
Understanding the physical and chemical evolution of starless cores is fundamental to constraining the initial conditions of star formation. 

Molecular line observations are powerful tools for probing the physical and chemical structures of starless cores. 
For example, N$_2$H$^+$ traces dense, CO-depleted gas because its main destroyer, CO, freezes efficiently onto dust grains. 
Molecular D/H ratios can serve as chemical clocks, since deuteration is greatly enhanced at the low temperatures ($\sim$10 K) typical of prestellar cores \citep[e.g., ][]{2015Kong_deuteration,2025Lin_G205.46-14.56M3_H2Dp}. 
The column density ratio between CCS and NH$_3$ was suggested to serve as an indicator of core evolution in low-mass star-forming regions \citep[e.g., ][]{1992Suzuki_chem_survey}. 
Similarly, \citet{2014Tatematsu_OrionA} suggested that the column density ratio between CCS and N$_2$H$^+$ can also serve as an indicator of core evolution. 
Different molecules trace distinct physical and chemical environments; as a result, multi-species surveys are essential for constructing a comprehensive picture of core evolution.

Starless cores also represent the envelopes from which protostars will eventually form.
Therefore, constraining their chemical composition is crucial for understanding the chemistry of subsequent protostellar and planetary systems. 
Carbon plays a particularly central role, as it is a key element in organic chemistry and is thus linked to the origin of life.
In the early diffuse-cloud stage, carbon exists primarily as C$^+$ in the gas phase, since interstellar UV radiation penetrates deeply into the cloud. 
These C$^+$ ions participate in a network of gas-phase reactions, producing abundant carbon-chain molecules \citep[CCMs; e.g.,][]{2013Sakai_review_carbon-chain}. 
As extinction increases, carbon becomes locked in CO, suppressing carbon-chain chemistry.
CO that freezes onto icy dust mantles can subsequently form CH$_3$OH, a precursor of complex organic molecules (COMs) defined by saturated organic molecules having at least six atoms \citep{2009Herbst_COM_review}.  
Through various desorption processes, COMs formed on dust grains can be released back into the gas phase. 
Consequently, starless cores are rich in both unsaturated (CCM) and saturated (COM) species, making them excellent laboratories for studying organic chemistry in the interstellar medium (ISM). 

Observations have revealed that CCM–rich and COM-rich sources may represent distinct chemical evolutionary stages \citep{2013Sakai_review_carbon-chain}. 
In particular, the so-called warm carbon-chain chemistry (WCCC) sources exhibit enhanced CCM near young protostars, while hot corino chemistry (HCC) sources are dominated by COMs released from grain mantles \citep[e.g.,][]{2009Herbst_COM_review,2020Hsu_ALMASOP,2022Hsu_ALMASOP}. 
Investigating the molecular content of starless cores, the potential precursors of both types, thus provides key insights into how these divergent chemistry originate.

To obtain a comprehensive view of chemical composition, wide-band spectral scans are invaluable.
A milestone was achieved by \citet{2004Kaifu_TMC1_chem}, who conducted a spectral line survey between 8.8 and 50.0~GHz toward the TMC-1 cyanopolyyne peak (TMC-1 CP) using the 45-m radio telescope at the Nobeyama Radio Observatory. 
This survey detected 38 molecular species, including 11 new ones, and provided their column densities. 
Such wide-band observations have offered a holistic view of the chemical inventory of TMC-1 and have served as benchmarks for subsequent surveys such as GOTHAM \citep{2020McGuire_TMC-1_GOTHAM} and QUIJOTE \citep{2022Cernicharo_TMC-1_QUIJOTE}.
While these studies have provided detailed chemical inventories, they mostly focus on individual and well-known starless cores, such as TMC-1, insight into the population-wide diversity of starless cores remains to be explored.
Systematic surveys targeting multiple cores within a single molecular cloud remain scarce. 
A coherent sample observed under uniform observation conditions is essential for disentangling intrinsic chemical diversity from environmental effects and for establishing statistically robust trends in molecular abundances and evolutionary states.

To this end, our project, Q/W-band Observations toward Starless Cores in Orion (QWOSCO), conducted observations of a sample of 23 starless cores in the Orion molecular cloud using the Yebes 40~m telescope.
This paper presents the first results from these observations.
Section~\ref{sec:methods} describes the methodology, including sample selection, observing programs, spectral cleaning, and the procedures used to derive molecular column densities.
Section~\ref{sec:results} presents the results and discussion, focusing on isotopologue ratios, isomer ratios, and COMs.
Finally, Section~\ref{sec:Conclusions} summarizes the main findings of this work.



\begin{figure*}[htb!]
\centering
\includegraphics[width=\linewidth]{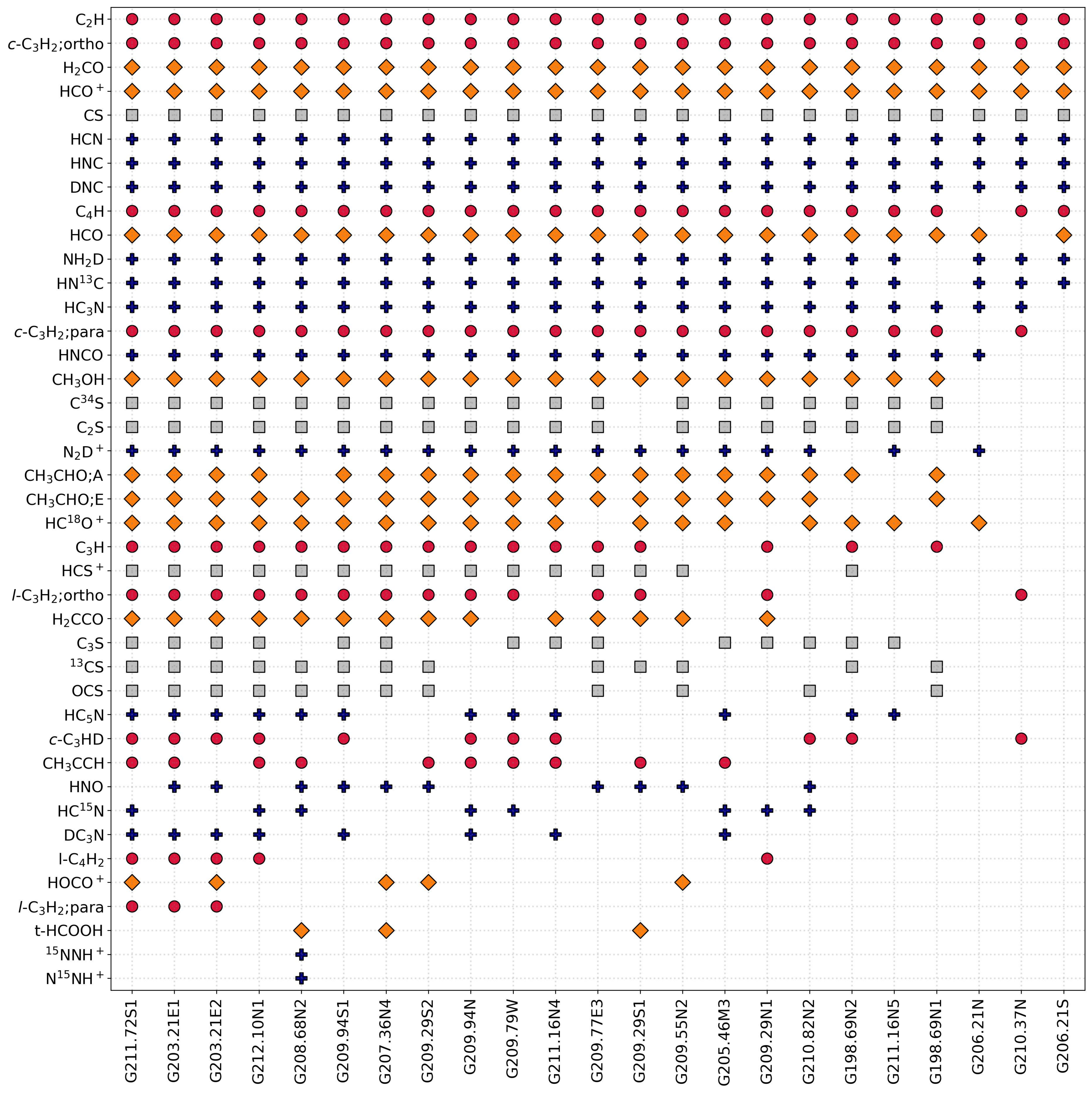}
\caption{\label{fig:scatter_detection} 
Detection statistics from this study.
The sources on the x-axis are ordered by the number of molecular species detected in each source, while the molecular species on the y-axis are ordered by the number of sources in which they are detected. 
The colors and styles of the markers indicate chemical families: silver square for S-bearing molecules, navy blue plus for N-bearing molecules, orange diamond for O-bearing molecules, and crimson red circle for hydrocarbons.
}
\end{figure*}

\section{Methods}
\label{sec:methods}

\subsection{Sample Selection}
\label{sec:methods:sample}

In our QWOSCO project, we investigated 23 starless cores drawn from the ALMA Survey of Orion PGCCs (ALMASOP) project \citep{2020Dutta_ALMASOP}. 
The Planck Galactic Cold Clump (PGCC) catalogue provides an all-sky inventory of cold (10-20 K), dense clumps characterized by molecular hydrogen column densities of $N(H_2) > 10^{20}$ cm$^{-2}$ at an angular resolution of 5\arcmin\ \citep{2016Planck_PGCC}. 
Based on the observations with the James Clerk Maxwell Telescope (JCMT) using its Submillimetre Common User Bolometer Array-2 (SCUBA-2) instrument, 119 dense cores were identified by their 850~\micron\ (dust continuum) emission within 96 PGCCs located in the Orion A, Orion B, and $\lambda$ Orionis clouds \citep{2018Yi_PGCC_SCUBA2_II}. 
From these 119 cores, the ALMASOP project selected 72 compact and high-density cores and ultimately cataloged 23 starless cores and 56 protostellar cores \citep{2020Dutta_ALMASOP}. 
Our sample consists of these 23 starless ALMASOP cores, which provide a good starting point for the study of starless cores in the Orion cloud.
Table~\ref{tab:coord} shows the information of the targets. 
For more related literature, please see Appendix~\ref{appx:sample} and Table~\ref{tab:lit}.  

\subsection{Observations}
\label{sec:methods:obs}

This study utilized data obtained from the Yebes 40-m telescope under Project ID 22A010 (PI: Xunchuan Liu) in the Q band (31.00-50.50~GHz) and 22B025 (PI: Xunchuan Liu) in the W band (71.10-91.40~GHz). 
Both programs observed all 23 starless cores with the dual linear polarization receiver in frequency-switching mode. 
In the Q band, the integration time per source was 180 minutes, and the frequency throw was 10.52 MHz. 
In the W band, the integration time per source was 150 minutes, and the frequency throw was 5.67 MHz. 
The spectral resolutions in both frequency bands were 38~kHz. 
The spectral coverage of both bands consists of eight consecutive sub-bands overlapped by 250~MHz (between sub-band 1 and 2) or 150~MHz (the others).
Data reduction was performed using the \texttt{CLASS} program of the \texttt{GILDAS}\footnote{https://www.iram.fr/IRAMFR/GILDAS/} package \citep{2005GILDAS}. 
With a 40-m aperture, the angular resolutions are approximately 40\arcs and 20\arcs\ for Q and W bands, respectively. 
The noise level is in general around 5~mK. 

\subsection{Spectrum Cleaning}
\label{sec:methods:spec}

To obtain the spectrum of each molecular transition, we performed baseline subtraction on the received spectra using the Python package \texttt{pybaselines}.
The baseline extraction was carried out in several steps. 
First, we applied the Statistics-sensitive Non-linear Iterative Peak-clipping (SNIP) method from \texttt{pybaselines} to estimate the primary baseline of each sub-band spectrum. 
The baseline-subtracted sub-band spectra were then concatenated to construct the full-band spectrum. 
For overlapping channels between adjacent sub-bands, we calculated their average values. 
The outer 250 channels (corresponding to $\sim$9.5~MHz) at both ends of each sub-band were excluded, as baseline estimates are often unreliable near the spectral edges. 
Finally, we applied the Noise Median method from \texttt{pybaselines} to the full-band spectrum to estimate and subtract a secondary baseline.

To better estimate the line intensities, we applied two rounds of baseline subtraction by first inspecting the eight sub-band spectra and flagging bad channels and strong lines (both positive and negative) and applying the baseline subtraction.
After obtaining a preliminarily cleaned full-band spectrum, we repeated the same line feature identification and baseline extraction.


Our observations were carried out in frequency-switching mode.
In this mode, an emission line at frequency $f_\mathrm{c}$ would appear as a positive and a negative feature at $f_\mathrm{c}+\Delta f / 2$ and $f_\mathrm{c}-\Delta f / 2$, respectively.
We averaged two versions of the spectrum: one shifted by $-\Delta f / 2$, and the other shifted by $+\Delta f / 2$ with the flipped amplitude.
In the averaged (folded) spectrum, each emission line appears as three features: a central line at $f_\mathrm{c}$ and two symmetric negative side  features at $f_\mathrm{c}\pm\Delta f$, each with approximately half the amplitude of the central line.

\subsection{Molecular Transition Identification}
\label{sec:methods:id}

We scanned the averaged spectrum to search for emission lines, identified by their characteristic frequency-switching signature of three features (negative-positive-negative). 
An emission line was further considered detected if its peak intensity exceeded a signal-to-noise ratio (SNR) of seven or its three channels exceeded an SNR of three. 
The localized noise level was derived from the standard deviation of the flux within a window of $\pm50$ \kmpers\ (corresponding to $\sim$6.7~MHz in the Q band and $\sim$13.3~MHz in the W band). 
Channels near the three features of each candidate line were excluded from the noise estimation.

Possible molecular candidates were examined using the databases of Cologne Database of Molecule Spectroscopy \citep[CDMS][]{2005CDMS} and the Jet Propulsion Laboratory Millimeter and Submillimeter Spectral Line catalogue \citep[JPL][]{1988JPL}. 
A molecule was considered identified if it satisfied one of the following criteria: (i) multiple transitions were detected simultaneously, or (ii) the transition has previously been reported in other starless cores. 
In addition, CH$_3$OH was regarded as detected based on the transition with the strongest line strength, which has the lowest upper-state energy (2.3 K).
Table~\ref{tab:trans} lists the parameters, and Figure set~\ref{fig:appx:spec_all_trans_0} shows the spectra of all identified features.

\subsection{Column Density Evaluation}
\label{sec:methods:colDens}

We applied rotation diagram analysis \citep{1999Goldsmith_popdiagram} to estimate the column densities of detected molecular species.
The rotation diagram is described as a function between $\ln(N_\mathrm{u}/g_\mathrm{u})$ and $E_\mathrm{u}$:
\begin{equation}
\label{eq:rot}
    \ln(\frac{N_\mathrm{u}}{g_\mathrm{u}}) = -\frac{E_\mathrm{u}}{k_\mathrm{B}T_\mathrm{rot}} + \ln(\frac{N_\mathrm{tot}}{Z}), 
\end{equation}
where $N_\mathrm{u}$ is the upper state column density, $g_\mathrm{u}$ is the upper state degeneracy, $E_\mathrm{u}$ is the upper state energy level, $k_\mathrm{B}$ is the Boltzmann constant, $T_\mathrm{rot}$ is the rotational temperature, $Z$ is the partition function at $T_\mathrm{rot}$, and $N_\mathrm{tot}$ is the total column density. 
Under the assumptions of local thermodynamic equilibrium (LTE) and optically thin emission, the rotation diagram will show a straight line. 
The slope of this line determines the rotational temperature, while the intercept is related to the total column density.
The upper state column density $N_\mathrm{u}$ can be derived from:
\begin{equation}
\label{eq:Nu}
    N_\mathrm{u} = \frac{8\pi k_\mathrm{B}}{hc^3} \frac{\nu^2}{A_\mathrm{ij}} W = \frac{8\pi k_\mathrm{B}}{hc^3} \frac{\nu^2}{A_\mathrm{ij}} \int T_\mathrm{B} \,\mathrm{d}v, 
\end{equation}
where $h$ and $c$ are the Planck constant and the speed of light, respectively, $\nu$ is the transition frequency, $A_\mathrm{ij}$ is the Einstein A coefficient, $W$ is the integrated intensity, $T_\mathrm{B}$ is the brightness temperature, and $v$ is the velocity. 
The velocity interval for each line was defined to include all channels with SNR greater than 3, with an additional two channels extended at both ends to ensure full coverage of the emission.
Table~\ref{tab:W} lists the values of $W$ derived in this study.
We note that for HCN, $W$ is derived from its high frequency hyperfine group (with a low line strength) with a scaling factor of nine in order to avoid the potential problem of optically thick lines. 

Given that most molecules have a limited number of detected transitions and a narrow range of upper-state energies, the slope of the rotation diagram is subject to significant uncertainty. 
To mitigate this, we fitted the diagrams with three linear relations in which the slope was fixed to correspond to rotational temperatures of 5, 7.5, and 10~K, while the intercept (related to the column density $N$) was allowed to vary. 
This is similar to the ``constant excitation temperature (CTEX)'' approximation \citep[e.g., ][]{2015Mangum_colDens,2020Scibelli_COM_Taurus}. 
Given the rotational temperature, for molecules with only one detected transition, the total column density (\Ntot) was directly derived using the rotational diagram equations (i.e., Eqs.~\ref{eq:rot} and \ref{eq:Nu}).
For molecules with more than one detected transition, \Ntot\ was obtained by minimizing the $\chi^2$ value of the rotational diagram data points.
In Table~\ref{tab:Ntot}, we show the derived column densities of each molecular species in each core.



\begin{figure*}[htb!]
\centering
\includegraphics[width=\linewidth]{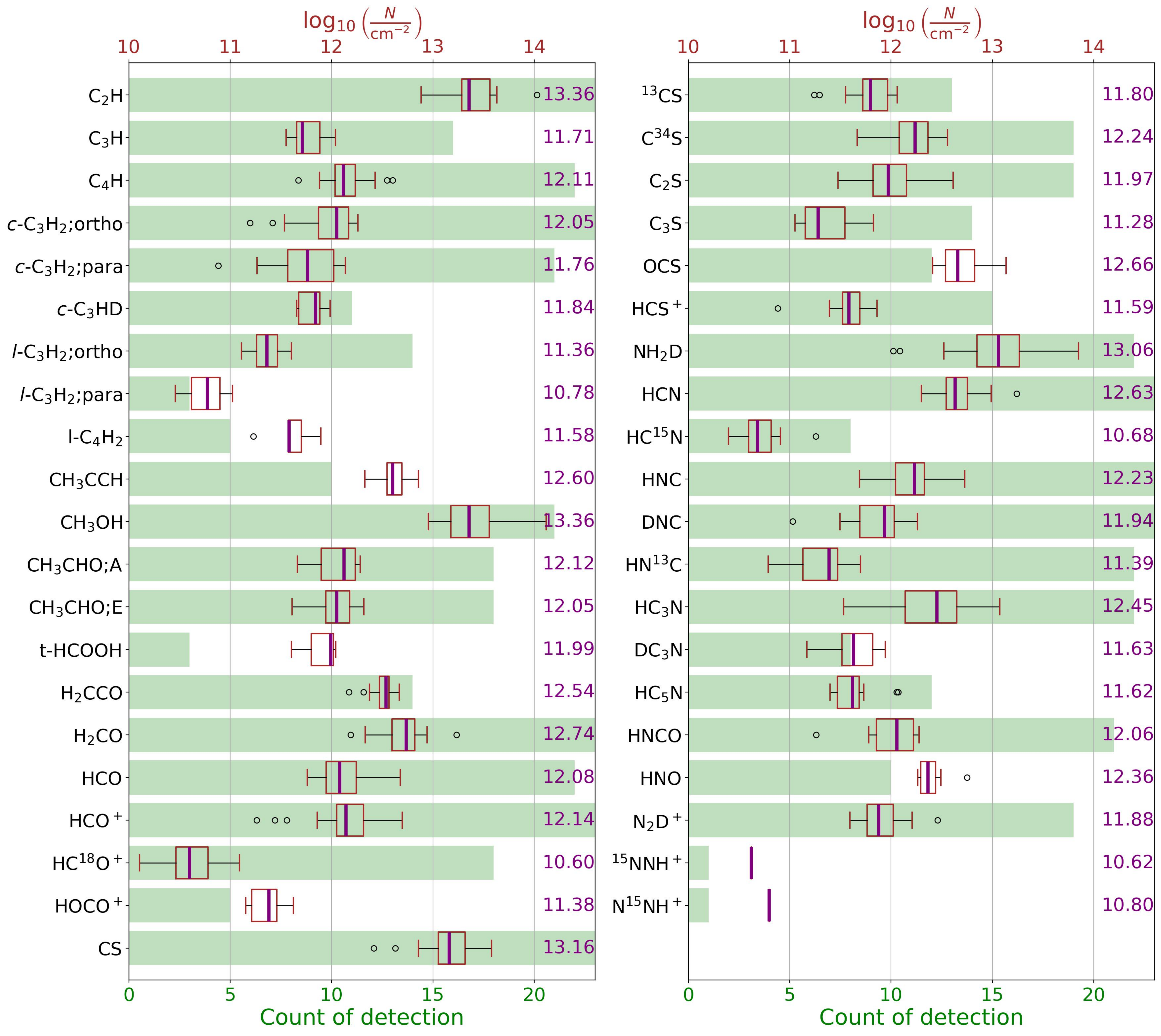}
\caption{\label{fig:bar_box_Ntot} The detection statistics and distribution of the molecular column densities. 
In the boxplots, the box spans the interquartile range (IQR), from the first quartile (Q1, 25\%) to the third quartile (Q3, 75\%).
The line within each box indicates the median (Q2, 50\%) which is also labeled at the right end of each row.
The whiskers extend from the quartiles to 1.5 times the IQR (i.e., values within Q1-1.5$\times$IQR and Q3+1.5$\times$IQR), while data points outside this range are treated as outliers and shown in circles. 
}
\end{figure*}

\section{Results and Discussions}
\label{sec:results}

\subsection{Overview}

In the following, we present the overall statistics of molecular detections and the column density ratios among selected species.
A detailed correlation analysis between molecular species will be deferred to a future paper.
Unless otherwise noted, the quoted column densities were estimated by adopting a rotational temperature of 7.5~K. 
The variations of column densities adopting rotational temperatures of 5 and 10 K mostly decrease and increase, respectively, within a factor of $\sim3$. 
Significant exceptions include HC$_5$N and NH$_2$D, which have opposite variation trend. 
In Tables~\ref{tab:isomer} and \ref{tab:isotope}, we list the column density ratios between isotopologues and isomers. 
The brackets denote the column density of each molecule (e.g., [HCN] represents the column density of HCN).
For each ratio we also list the mean values calculated from all data points and from the data after excluding outliers. 
The equations for calculating the average and standard errors are:
\begin{equation}
     \mu=\frac{1}{n}\sum_{i=1}^{n}x_i
\end{equation}
for the average ($\mu$) and 
\begin{equation}
    \sigma =\frac{1}{n}\sqrt{\sum_{i=1}^{n}\sigma_i^2}.
\end{equation}
for the left and right standard errors ($\sigma$). 

Figure~\ref{fig:scatter_detection} shows the molecular detections toward each target. 
The sources are ordered by the number of molecular species detected, while the molecular species are ordered by the number of sources in which they are detected. 
The triangular distribution of the markers suggests that most non-detections are likely due to limited sensitivity rather than true chemical absence.
Nevertheless, chemical segregation between sources can still be tentatively discerned.
For example, G209.29N1 and G209.29S1 seem to have mutually exclusive detection although they are residing within the same clump with a separation of around 2\arcmin. 
Detailed analysis of the potential chemical segregation will be studied in another paper.  

Figure~\ref{fig:bar_box_Ntot} presents the distributions of the derived column densities (boxplots) together with the detection counts (bar charts) for each molecular species.
In the boxplots, each box spans the interquartile range (IQR; Q1-Q3), with the line marking the median (Q2).
Whiskers extend to values within 1.5$\times$IQR, and points beyond this range are shown as outliers (circles) defined by values beyond 1.5 times the IQR from the quartiles (i.e., $<$ Q1 $-1.5\times$IQR or $>$ Q3 $+1.5\times$IQR). 
As shown in Figure~\ref{fig:bar_box_Ntot}, the IQR range of each molecule is less than half order of magnitude wide.
This suggests that these molecules, if detected, have their own consistent column density values in starless cores. 


\begin{deluxetable*}{lcccccccc}
\setlength{\tabcolsep}{3pt}
\tablecaption{\label{tab:isomer} Summary of the isomer column density ratios in this study. 
$\mu$ and $\mu'$ denote the mean values calculated from all data points and from the data after excluding outliers, respectively.
Outliers are defined as data points lying beyond 1.5 times the IQR from the first or third quartile.
}
\tablehead{\colhead{source} & \colhead{$\mathrm{\frac{[\it{c}\text{-}\text{C}_3\text{H}_2;ortho]}{[\it{c}\text{-}\text{C}_3\text{H}_2;para]}}$} & \colhead{$\mathrm{\frac{[\it{l}\text{-}\text{C}_3\text{H}_2;ortho]}{[\it{l}\text{-}\text{C}_3\text{H}_2;para]}}$} & \colhead{$\mathrm{\frac{[\it{c}\text{-}\text{C}_3\text{H}_2]}{[\it{l}\text{-}\text{C}_3\text{H}_2]}}$} & \colhead{$\mathrm{\frac{[\it{c}\text{-}\text{C}_3\text{H}_2;ortho]}{[\it{l}\text{-}\text{C}_3\text{H}_2;ortho]}}$} & \colhead{$\mathrm{\frac{[\it{c}\text{-}\text{C}_3\text{H}_2;para]}{[\it{l}\text{-}\text{C}_3\text{H}_2;para]}}$} & \colhead{$\mathrm{\frac{[CH_3CHO;A]}{[CH_3CHO;E]}}$} & \colhead{$\mathrm{\frac{[HCN]}{[HNC]}}$} & \colhead{$\mathrm{\frac{450\times[HC^{15}N]}{77\times[HN^{13}C]}}$}}
\startdata
G198.69N1 & 2.1$^{+0.9}_{-2.1}$ & \nodata & \nodata & \nodata & \nodata & 0.8$^{+0.1}_{-0.1}$ & 2.1$^{+2.2}_{-1.8}$ & \nodata \\
G198.69N2 & 1.2$^{+0.4}_{-0.6}$ & \nodata & \nodata & \nodata & \nodata & \nodata & 2.3$^{+2.2}_{-1.9}$ & \nodata \\
G203.21E1 & 1.2$^{+0.3}_{-0.3}$ & 6.9$^{+2.7}_{-4.4}$ & 10.6$^{+2.3}_{-2.7}$ & 6.7$^{+2.0}_{-2.0}$ & 37.9$^{+13.5}_{-23.6}$ & 1.1$^{+0.1}_{-0.1}$ & 3.2$^{+2.7}_{-2.6}$ & \nodata \\
G203.21E2 & 1.3$^{+0.5}_{-0.5}$ & 3.9$^{+1.6}_{-2.2}$ & 7.0$^{+1.9}_{-2.1}$ & 5.0$^{+1.9}_{-1.8}$ & 14.9$^{+6.0}_{-8.4}$ & 1.1$^{+0.2}_{-0.2}$ & 3.2$^{+2.7}_{-2.6}$ & \nodata \\
G205.46M3 & 1.6$^{+0.6}_{-0.7}$ & \nodata & \nodata & \nodata & \nodata & 0.7$^{+0.5}_{-0.4}$ & 2.1$^{+2.1}_{-1.8}$ & 1.4$^{+1.1}_{-1.1}$ \\
G206.21N & \nodata & \nodata & \nodata & \nodata & \nodata & \nodata & 2.1$^{+2.1}_{-1.8}$ & \nodata \\
G206.21S & \nodata & \nodata & \nodata & \nodata & \nodata & \nodata & 2.3$^{+2.5}_{-2.0}$ & \nodata \\
G207.36N4 & 1.8$^{+1.5}_{-1.1}$ & \nodata & 6.9$^{+4.4}_{-3.4}$ & 4.4$^{+3.7}_{-2.7}$ & \nodata & 1.0$^{+0.2}_{-0.3}$ & 3.0$^{+2.8}_{-2.5}$ & \nodata \\
G208.68N2 & 3.1$^{+0.5}_{-0.5}$ & \nodata & 6.8$^{+1.9}_{-2.7}$ & 5.2$^{+1.5}_{-2.1}$ & \nodata & \nodata & 3.2$^{+3.0}_{-2.7}$ & 2.1$^{+1.6}_{-1.6}$ \\
G209.29N1 & 2.6$^{+1.4}_{-2.4}$ & \nodata & 4.1$^{+3.5}_{-1.9}$ & 2.9$^{+2.4}_{-1.8}$ & \nodata & 0.7$^{+0.4}_{-0.4}$ & 2.9$^{+2.7}_{-2.4}$ & 1.7$^{+1.6}_{-1.4}$ \\
G209.29S1 & 3.8$^{+1.9}_{-3.6}$ & \nodata & 4.7$^{+3.8}_{-1.4}$ & 3.7$^{+3.0}_{-1.7}$ & \nodata & 1.0$^{+0.1}_{-0.2}$ & 3.1$^{+2.7}_{-2.5}$ & \nodata \\
G209.29S2 & 2.4$^{+1.9}_{-1.4}$ & \nodata & 5.4$^{+3.6}_{-2.8}$ & 3.8$^{+3.2}_{-2.3}$ & \nodata & 1.2$^{+0.2}_{-0.2}$ & 2.6$^{+2.4}_{-2.2}$ & \nodata \\
G209.55N2 & 1.8$^{+1.0}_{-1.6}$ & \nodata & \nodata & \nodata & \nodata & 1.0$^{+0.6}_{-0.6}$ & 2.4$^{+2.4}_{-2.0}$ & \nodata \\
G209.77E3 & 2.5$^{+0.3}_{-0.2}$ & \nodata & 10.5$^{+1.0}_{-0.9}$ & 7.5$^{+0.9}_{-0.8}$ & \nodata & 1.0$^{+0.0}_{-0.0}$ & 2.9$^{+2.7}_{-2.4}$ & \nodata \\
G209.79W & 1.9$^{+0.5}_{-0.4}$ & \nodata & 12.7$^{+4.1}_{-5.3}$ & 8.3$^{+3.0}_{-3.7}$ & \nodata & 1.6$^{+0.6}_{-1.0}$ & 1.6$^{+1.7}_{-1.4}$ & 0.8$^{+0.8}_{-0.7}$ \\
G209.94N & 1.4$^{+0.4}_{-0.6}$ & \nodata & 7.3$^{+1.6}_{-1.2}$ & 4.3$^{+0.9}_{-0.8}$ & \nodata & 1.1$^{+0.6}_{-0.4}$ & 3.2$^{+3.0}_{-2.6}$ & 0.5$^{+0.5}_{-0.4}$ \\
G209.94S1 & 1.6$^{+0.7}_{-0.5}$ & \nodata & 8.5$^{+2.4}_{-1.8}$ & 5.2$^{+2.3}_{-1.7}$ & \nodata & 1.0$^{+0.2}_{-0.1}$ & 3.7$^{+3.1}_{-2.9}$ & \nodata \\
G210.37N & 2.1$^{+0.3}_{-1.7}$ & \nodata & 3.1$^{+1.9}_{-1.3}$ & 2.1$^{+0.6}_{-0.8}$ & \nodata & \nodata & 2.5$^{+2.7}_{-2.2}$ & \nodata \\
G210.82N2 & 2.7$^{+0.2}_{-0.2}$ & \nodata & \nodata & \nodata & \nodata & 1.3$^{+3.5}_{-1.3}$ & 2.1$^{+2.2}_{-1.8}$ & 0.9$^{+0.9}_{-0.8}$ \\
G211.16N4 & 1.7$^{+0.4}_{-0.4}$ & \nodata & \nodata & \nodata & \nodata & 0.8$^{+0.5}_{-0.4}$ & 2.9$^{+2.6}_{-2.4}$ & \nodata \\
G211.16N5 & 2.4$^{+0.4}_{-0.5}$ & \nodata & \nodata & \nodata & \nodata & \nodata & 2.7$^{+2.5}_{-2.2}$ & \nodata \\
G211.72S1 & 1.0$^{+0.2}_{-0.2}$ & 1.2$^{+0.4}_{-0.6}$ & 9.6$^{+2.0}_{-2.7}$ & 8.7$^{+1.4}_{-1.2}$ & 10.7$^{+3.9}_{-5.9}$ & 1.8$^{+1.0}_{-0.6}$ & 4.1$^{+3.5}_{-3.3}$ & 1.6$^{+1.4}_{-1.3}$ \\
G212.10N1 & 1.1$^{+0.3}_{-0.5}$ & \nodata & 12.3$^{+3.2}_{-2.2}$ & 6.4$^{+0.6}_{-0.6}$ & \nodata & 1.6$^{+0.6}_{-0.6}$ & 3.2$^{+3.1}_{-2.7}$ & 1.0$^{+0.9}_{-0.8}$ \\
$\mu$ & 2.0$^{+0.2}_{-0.3}$ & 4.0$^{+1.0}_{-1.6}$ & 7.8$^{+0.8}_{-0.7}$ & 5.3$^{+0.6}_{-0.5}$ & 21.2$^{+5.1}_{-8.6}$ & 1.1$^{+0.2}_{-0.1}$ & 2.8$^{+0.5}_{-0.5}$ & 1.2$^{+0.4}_{-0.4}$ \\
$\mu^{'}$ & 2.0$^{+0.2}_{-0.3}$ & 4.0$^{+1.0}_{-1.6}$ & 7.8$^{+0.8}_{-0.7}$ & 5.3$^{+0.6}_{-0.5}$ & 21.2$^{+5.1}_{-8.6}$ & 1.0$^{+0.2}_{-0.1}$ & 2.8$^{+0.5}_{-0.5}$ & 1.2$^{+0.4}_{-0.4}$
\enddata
\end{deluxetable*}

\subsection{Isomers}
\label{sec:isomer}

Our observations cover several sets of isomers, including the structural isomer HCN/HNC, the {\textit{cyclic}}/{\textit{linear}} structural isomer, the {\textit{ortho}}/{\textit{para}} nuclear spin isomer, and the {\textit{A}}/{\textit{E}} nuclear spin isomers. 
In this section, we present and discuss these ratios, which are shown in Table~\ref{tab:isomer}.

\subsubsection{{\textit{A}}- and {\textit{E}}-type Spin Isomers}
\label{sec:isomer:A-E}

{\textit{A}}- and {\textit{E}}-type isomers are nuclear spin isomers of molecules with a methyl group ($-$CH$_3$), where the three equivalent hydrogen nuclei (nuclear spin = 1/2) couple to form distinct spin states. 
The A-type isomer, with a total nuclear spin of $I = 3/2$ (symmetric under C$_3$ rotation), has a statistical weight of 4. 
The E-type isomer, with a total nuclear spin of $I = 1/2$ (antisymmetric under C$_3$ rotation), due to degeneracy, has a total statistical weight of 4.
As a result, the {\textit{A}}-to-{\textit{E}} ratio is expected to be a unity, supported by for example, the ratio of CH$_3$OH column densities of $1.00\pm0.15$ in L1498 reported by \citet[][]{2017Dapra_L1498_methanolAE}. 

In our study, both methanol (CH$_3$OH) and acetaldehyde (CH$_3$CHO) have {\textit{A}}- and {\textit{E}}-type spin isomers. 
For CH$_3$OH, our observations include only one transition of the {\textit{A}}-type isomer, with an upper-state energy of $E_\mathrm{u} = 2.3$~K.
The transitions of the {\textit{E}}-type isomer, in contrast, have significantly higher upper-state energies (15.4, 28.8, and 40.4~K).
Therefore, we assume an {\textit{A}}-to-{\textit{E}} ratio of unity in our column density calculations.

For CH$_3$CHO, we have five and four transitions for the {\textit{E}}- and {\textit{A}}-type spin isomers, respectively. 
The upper energy ranges of the two spin isomers are comparable ($\sim2$--10~K). 
These enable us to have good estimates on the column densities of each spin type.
The average of the {\textit{A}}-to-{\textit{E}} ratio is \vlr{1.1}{0.2}{0.1}, suggesting that the {\textit{A}}- and {\textit{E}}-type spin isomers have the similar abundance. 

\subsubsection{{\textit{ortho}}- and {\textit{para}}-type Spin Isomers}
\label{sec:isomer:o-p}

Two molecular species in this study, cyclopropenylidene ($c$-C$_3$H$_2$) and propadienylidene ($l$-C$_3$H$_2$ or H$_2$CCC), possess two equivalent hydrogen nuclei (nuclear spin = 1/2), which combine to form {\textit{ortho}} (total nuclear spin = 1) and {\textit{para}} (total nuclear spin = 0) species. 
The {\textit{para}} state has a lower zero-point energy than {\textit{ortho}} state while the {\textit{ortho}}-to-{\textit{para}} ratio is presumably to be equal to 3 under thermodynamical equilibrium due the their spin statistical weights.
In starless core environments ($\sim$10~K), the conversion between the {\textit{ortho}} and {\textit{para}} states of the same molecular species is inefficient in the gas phase, leading the forms {\textit{ortho}} and {\textit{para}} to be treated as distinct species.
\citet{2001Takakuwa_OPR_C3H2} observed $c$-C$_3$H$_2$ lines at 3~mm toward the starless core TMC-1C and reported an {\textit{ortho}}-to-{\textit{para}} ratio (OPR) of \vlr{2.4}{0.1}{0.1}. 
They suggested that the relatively low OPR of $c$-C$_3$H$_2$ results from a low OPR of H$_2$ on the basis that $c$-C$_3$H$_3^+$, a precursor of $c$-C$_3$H$_2$, is formed via the reaction C$_3$H$^+$ + H$_2$. 
The low OPR of H$_2$, in turn, arises because the conversion from {\textit{ortho}}-H$_2$ to {\textit{para}}-H$_2$ through reactions involving H$^+$ or H$_3^+$ is not fully thermalized at temperatures below 20~K \citep[e.g., ][]{2006Flower_OPR_H2}.

We detected both {\textit{ortho}} and {\textit{para}} isomers of $c$-C$_3$H$_2$ simultaneously in 21 sources. 
Their upper energy ranges are comparable (4.1, 13.7, and 15.8~K for {\textit{ortho}}; 6.4, 8.7, and 16.1~K for {\textit{para}}). 
The average OPR for $c$-C$_3$H$_2$ is \vlr{2.0}{0.7}{0.7}, below the thermal equilibrium value of 3 and consistent with literature values in starless cores \citep{2001Takakuwa_OPR_C3H2}.

For $l$-C$_3$H$_2$, our data include two {\textit{ortho}} transitions with similar upper energy levels ($\sim$2.0~K) and one {\textit{para}} transition ($\sim$3.0~K).
The {\textit{para}} isomer is detected in only three sources, whereas the {\textit{ortho}} isomer is detected in 61\% of the sample (14/23).
The average OPR of \vlr{4.0}{1.0}{1.6} appears higher than the thermal equilibrium value of 3 but with large uncertainties. 
This ratio is only weakly dependent on the assumed temperature, as the upper energy levels of the three transitions are comparable.


\subsubsection{{\textit{cyclic}}- and {\textit{linear}}-type Structural Isomers}
\label{sec:isomer:c-l}

Cyclopropenylidene ($c$-C$_3$H$_2$) and propadienylidene
($l$-C$_3$H$_2$ or H$_2$CCC) are {\textit{cyclic}} and {\textit{linear}} isomers of C$_3$H$_2$, respectively. 
The {\textit{cyclic}}-to-{\textit{linear}} ratios of C$_3$H$_2$, defined by the ratio of their column densities, are diverse within starless cores. 
In TMC-1C and L1544, the ratios are $67 \pm 7$ and $32 \pm 4$, respectively \citep{2016Spezzano_L1544_TMC1C_C3H2}. 
In Serp S1a, L1521F, Lupus-1A, TMC1, and L1495B, the ratios are 28, 38, 40, 67, and 111, respectively \citep{2017Loison_C3H_C3H2}. 

\citet{2016Sipila_L1544_C3H2} demonstrated that the dissociative recombination reaction C$_3$H$_3^+$ + e$^-$ $\rightarrow$ C$_3$H$_2$ + H proceeds with a higher rate coefficient for the {\textit{cyclic}} isomer than for the {\textit{linear}} one, assuming that the {\textit{cyclic}} and {\textit{linear}} ions exclusively form their corresponding neutrals. 
They also showed that the {\textit{cyclic}}-to-{\textit{linear}} ratio of C$_3$H$_2$ evolves over time, peaking at $\sim$10$^5$ yr. 
In contrast, \citet{2017Loison_C3H_C3H2} emphasized that $c$-C$_3$H$_2$ can form from both $c$- and $l$-C$_3$H$_3^+$.
They further demonstrated that the {\textit{cyclic}}-to-{\textit{linear}} ratio of C$_3$H$_2$ varies not only with the core evolutionary time but also the core density \citep[see Figure 5 of ][]{2017Loison_C3H_C3H2}. 
The variations of the {\textit{cyclic}}-to-{\textit{linear}} ratio generally starts at values of 10--20 around $10^{4}$ yr, increases to a peak between $10^{4}$ and $10^{5}$ yr, and then decreases to 30-40 by $10^{7}$ yr.
Cores with lower densities ($\mathrm{H_2}$ number density $\sim 10^{4}$ cm$^{-3}$) show smaller ratios at early times but reach higher peak values than those with higher densities ($\mathrm{H_2}$ number density $\sim 10^{5}$ cm$^{-3}$).

The {\textit{cyclic}}-to-{\textit{linear}} ratios of C$_3$H$_2$, combining with the column densities of both {\textit{ortho}} and {\textit{para}} isomers, have an average value of \vlr{7.8}{0.8}{0.7}, lower than the values reported in the literature. 
As discussed in Sect.~\ref{sec:isomer:o-p}, the column densities of $l$-C$_3$H$_2$, particularly for the {\textit{para}} isomer, carry large uncertainties.
When derived separately, the {\textit{cyclic}}-to-{\textit{linear}} ratios for the {\textit{ortho}} and {\textit{para}} isomers are \vlr{5.3}{0.6}{0.5} and \vlr{21.2}{5.1}{8.6}, respectively.
Nevertheless, the average ratio derived from the {\textit{ortho}} isomers is still significantly lower than values reported in the literature. 
As shown in Table~\ref{tab:isomer}, all sources exhibit {\textit{cyclic}}-to-{\textit{linear}} ratios below 10, indicating that this is a general trend across our sample. 
According to the chemical model of \citet{2017Loison_C3H_C3H2}, such low {\textit{cyclic}}-to-{\textit{linear}} ratios occur only at an early evolutionary stage ($\sim$10$^4$ yr) and low H$_2$ number density ($\sim$10$^4$~cm$^{-3}$).
Alternatively, the systematically low ratios may reflect environmental effects in the Orion clouds, characterized by a stronger turbulence and a more intense UV radiation field \citep[e.g.,][]{2022Ha_SFR_turbulence,2022Xia_SFR_UV}. 

\subsubsection{HCN and HNC Structural Isomers}
\label{sec:isomer:HCN}

Hydrogen cyanide (HCN) and hydrogen isocyanide (HNC) are structural isomers that are both commonly detected in starless cores. 
The two species are often observed to have comparable abundances in cold environments ($\sim$10 K).
For instance, \citet{2025Tasa-Chaveli_GEMS} reported [H$^{13}$CN]/[HN$^{13}$C] = $1.24\pm0.44$ and [HC$^{15}$N]/[H$^{15}$NC] = $0.89\pm0.30$ for low-mass starless cores in the Taurus, Perseus, and Orion A molecular clouds.
Although HCN is more stable than HNC, HCN can be isomerized into HNC through reactions with H$_3^+$ followed by the dissociative recombination of HCNH$^+$ under conditions where atomic carbon and CO are depleted \citep{2014Loison_HCN-HNC}.

In our study, the [HCN]/[HNC] is \vlr{2.8}{0.5}{0.5}, slightly higher than one. 
However, this value may not be robust due to that HCN and HNC $J=1-0$ transitions are often reported to be optically thick (see Sections~\ref{sec:isotope:$^{12}$C$^{13}$C} and \ref{sec:isotope:14N15N}). 
We, instead, estimate their abundances from each of their rare isotopologues. 
Adopting standard isotopic ratios at the local ISM \citep[$^{12}$C/$^{13}$C$=77$ and $^{14}$N/$^{15}$N $=450$][]{1994Wilson_isotope}, we derive [HCN]/[HNC] $=$ ([HC$^{15}$N] $\times450$)/([HN$^{13}$C]$\times77$)$=$\vlr{1.2}{0.4}{0.4}, consistent with the expected value of unity. 



\subsection{Isotopologues}
\label{sec:iso}


Our observations cover a broad set of isotopologues, enabling us to derive isotope ratios including $^{12}$C/$^{13}$C, $^{14}$N/$^{15}$N, $^{16}$O/$^{18}$O, $^{32}$S/$^{34}$S, and D/H.
In this section, we present and discuss these ratios, which are shown in Table~\ref{tab:isotope}.

\begin{deluxetable*}{lccccccccccc}
\setlength{\tabcolsep}{3pt}
\tablecaption{\label{tab:isotope} Summary of the isotopologue column density ratios in this study. 
$\mu$ and $\mu'$ denote the mean values calculated from all data points and from the data after excluding outliers, respectively.
Outliers are defined as data points lying beyond 1.5 times the IQR from the first or third quartile.
}
\tablehead{\colhead{source} & \colhead{$\mathrm{\frac{[CS]}{[^{13}CS]}}$} & \colhead{$\mathrm{\frac{[HNC]}{[HN^{13}C]}}$} & \colhead{$\mathrm{\frac{[HCN]}{[HC^{15}N]}}$} & \colhead{$\mathrm{\frac{77\times[HN^{13}C]}{[HC^{15}N]}}$} & \colhead{$\mathrm{\frac{[HCO^+]}{[HC^{18}O^+]}}$} & \colhead{$\mathrm{\frac{[CS]}{[C^{34}S]}}$} & \colhead{$\mathrm{\frac{77\times[^{13}CS]}{[C^{34}S]}}$} & \colhead{$\mathrm{\frac{[DNC]}{[HNC]}}$} & \colhead{$\mathrm{\frac{[DNC]}{77\times[HN^{13}C]}}$} & \colhead{$\mathrm{\frac{[c-C_3HD]}{[\it{c}\text{-}\text{C}_3\text{H}_2]}}$} & \colhead{$\mathrm{\frac{[DC_3N]}{[HC_3N]}}$}
\\ \colhead{} 
& \colhead{$\frac{^{12}\text{C}}{^{13}\text{C}}$} & \colhead{$\frac{^{12}\text{C}}{^{13}\text{C}}$} 
& \colhead{$\frac{^{14}\text{N}}{^{15}\text{N}}$} & \colhead{$\frac{^{14}\text{N}}{^{15}\text{N}}$} 
& \colhead{$\frac{^{16}\text{O}}{^{18}\text{O}}$} 
& \colhead{$\frac{^{32}\text{S}}{^{34}\text{S}}$} & \colhead{$\frac{^{32}\text{S}}{^{34}\text{S}}$} 
& \colhead{$\frac{\text{D}}{\text{H}}$} & \colhead{$\frac{\text{D}}{\text{H}}$} & \colhead{$\frac{\text{D}}{\text{H}}$} & \colhead{$\frac{\text{D}}{\text{H}}$} 
}
\colnumbers
\startdata
G198.69N1 & 63$^{+31}_{-37}$ & \nodata & \nodata & \nodata & \nodata & 11$^{+5}_{-5}$ & 14$^{+9}_{-8}$ & 0.11$^{+0.08}_{-0.08}$ & \nodata & \nodata & \nodata \\
G198.69N2 & 25$^{+12}_{-13}$ & 13$^{+10}_{-10}$ & \nodata & \nodata & 53$^{+39}_{-45}$ & 13$^{+6}_{-6}$ & 40$^{+22}_{-22}$ & 0.39$^{+0.26}_{-0.29}$ & 0.065$^{+0.043}_{-0.050}$ & 0.26$^{+0.38}_{-0.18}$ & \nodata \\
G203.21E1 & 16$^{+7}_{-7}$ & 4$^{+3}_{-3}$ & \nodata & \nodata & 4$^{+3}_{-3}$ & 5$^{+2}_{-2}$ & 25$^{+11}_{-11}$ & 1.37$^{+0.91}_{-1.03}$ & 0.068$^{+0.045}_{-0.051}$ & 0.31$^{+0.04}_{-0.05}$ & 0.096$^{+0.029}_{-0.023}$ \\
G203.21E2 & 15$^{+7}_{-7}$ & 5$^{+4}_{-4}$ & \nodata & \nodata & 3$^{+3}_{-2}$ & 5$^{+2}_{-2}$ & 24$^{+11}_{-11}$ & 1.00$^{+0.65}_{-0.75}$ & 0.059$^{+0.039}_{-0.044}$ & 0.39$^{+0.07}_{-0.09}$ & 0.074$^{+0.010}_{-0.009}$ \\
G205.46M3 & \nodata & 13$^{+11}_{-10}$ & 119$^{+119}_{-103}$ & 330$^{+249}_{-262}$ & 69$^{+53}_{-55}$ & 14$^{+6}_{-7}$ & \nodata & 0.21$^{+0.14}_{-0.16}$ & 0.037$^{+0.025}_{-0.028}$ & \nodata & 0.033$^{+0.072}_{-0.024}$ \\
G206.21N & \nodata & 28$^{+21}_{-23}$ & \nodata & \nodata & 75$^{+55}_{-62}$ & \nodata & \nodata & 0.25$^{+0.17}_{-0.19}$ & 0.093$^{+0.060}_{-0.074}$ & \nodata & \nodata \\
G206.21S & \nodata & 14$^{+10}_{-11}$ & \nodata & \nodata & \nodata & \nodata & \nodata & 0.32$^{+0.22}_{-0.25}$ & 0.057$^{+0.038}_{-0.045}$ & \nodata & \nodata \\
G207.36N4 & 38$^{+17}_{-19}$ & 15$^{+12}_{-12}$ & \nodata & \nodata & 48$^{+36}_{-39}$ & 10$^{+4}_{-4}$ & 20$^{+10}_{-9}$ & 0.36$^{+0.24}_{-0.27}$ & 0.069$^{+0.046}_{-0.052}$ & \nodata & \nodata \\
G208.68N2 & 33$^{+15}_{-15}$ & 11$^{+8}_{-8}$ & 96$^{+91}_{-79}$ & 214$^{+163}_{-165}$ & 41$^{+31}_{-31}$ & 11$^{+5}_{-5}$ & 26$^{+12}_{-12}$ & 0.31$^{+0.21}_{-0.23}$ & 0.043$^{+0.029}_{-0.032}$ & \nodata & \nodata \\
G209.29N1 & \nodata & 16$^{+13}_{-13}$ & 159$^{+143}_{-142}$ & 265$^{+197}_{-226}$ & \nodata & 15$^{+7}_{-7}$ & \nodata & 0.22$^{+0.14}_{-0.16}$ & 0.045$^{+0.030}_{-0.034}$ & \nodata & \nodata \\
G209.29S1 & 38$^{+18}_{-19}$ & 7$^{+5}_{-5}$ & \nodata & \nodata & 29$^{+23}_{-23}$ & \nodata & \nodata & 0.36$^{+0.24}_{-0.27}$ & 0.031$^{+0.021}_{-0.023}$ & \nodata & \nodata \\
G209.29S2 & 42$^{+20}_{-21}$ & 9$^{+7}_{-7}$ & \nodata & \nodata & 41$^{+32}_{-32}$ & 10$^{+5}_{-5}$ & 19$^{+10}_{-10}$ & 0.32$^{+0.21}_{-0.24}$ & 0.038$^{+0.025}_{-0.028}$ & \nodata & \nodata \\
G209.55N2 & 77$^{+38}_{-48}$ & 9$^{+7}_{-7}$ & \nodata & \nodata & 61$^{+46}_{-50}$ & 8$^{+4}_{-4}$ & 8$^{+6}_{-5}$ & 0.33$^{+0.22}_{-0.25}$ & 0.038$^{+0.025}_{-0.028}$ & \nodata & \nodata \\
G209.77E3 & 28$^{+13}_{-13}$ & 10$^{+8}_{-8}$ & \nodata & \nodata & \nodata & 8$^{+4}_{-4}$ & 22$^{+11}_{-11}$ & 0.40$^{+0.27}_{-0.30}$ & 0.053$^{+0.036}_{-0.040}$ & \nodata & \nodata \\
G209.79W & \nodata & 9$^{+7}_{-7}$ & 101$^{+100}_{-94}$ & 543$^{+396}_{-465}$ & 53$^{+40}_{-42}$ & 26$^{+13}_{-15}$ & \nodata & 0.37$^{+0.25}_{-0.28}$ & 0.044$^{+0.030}_{-0.033}$ & 0.24$^{+0.07}_{-0.06}$ & \nodata \\
G209.94N & \nodata & 5$^{+4}_{-4}$ & 178$^{+156}_{-163}$ & 939$^{+675}_{-816}$ & 48$^{+36}_{-40}$ & 10$^{+4}_{-5}$ & \nodata & 0.81$^{+0.53}_{-0.61}$ & 0.049$^{+0.032}_{-0.036}$ & 0.33$^{+0.06}_{-0.08}$ & 0.101$^{+0.026}_{-0.020}$ \\
G209.94S1 & 40$^{+18}_{-19}$ & 5$^{+4}_{-4}$ & \nodata & \nodata & 35$^{+27}_{-29}$ & 10$^{+5}_{-5}$ & 20$^{+10}_{-9}$ & 0.61$^{+0.40}_{-0.46}$ & 0.039$^{+0.026}_{-0.029}$ & 0.18$^{+0.06}_{-0.06}$ & 0.087$^{+0.132}_{-0.053}$ \\
G210.37N & \nodata & 11$^{+8}_{-9}$ & \nodata & \nodata & \nodata & \nodata & \nodata & 0.38$^{+0.26}_{-0.29}$ & 0.053$^{+0.035}_{-0.041}$ & 0.57$^{+0.72}_{-0.43}$ & \nodata \\
G210.82N2 & \nodata & 11$^{+9}_{-9}$ & 150$^{+139}_{-141}$ & 490$^{+360}_{-435}$ & 40$^{+30}_{-33}$ & 15$^{+7}_{-7}$ & \nodata & 0.51$^{+0.34}_{-0.39}$ & 0.074$^{+0.049}_{-0.056}$ & 0.37$^{+0.56}_{-0.23}$ & \nodata \\
G211.16N4 & \nodata & 7$^{+5}_{-5}$ & \nodata & \nodata & 25$^{+20}_{-20}$ & 8$^{+4}_{-4}$ & \nodata & 0.51$^{+0.34}_{-0.39}$ & 0.046$^{+0.030}_{-0.034}$ & 0.29$^{+0.07}_{-0.07}$ & 0.065$^{+0.101}_{-0.040}$ \\
G211.16N5 & \nodata & 6$^{+5}_{-5}$ & \nodata & \nodata & 94$^{+64}_{-86}$ & 15$^{+7}_{-9}$ & \nodata & 0.58$^{+0.39}_{-0.44}$ & 0.049$^{+0.033}_{-0.037}$ & \nodata & \nodata \\
G211.72S1 & 14$^{+6}_{-7}$ & 2$^{+2}_{-2}$ & 33$^{+27}_{-28}$ & 289$^{+221}_{-240}$ & 12$^{+9}_{-10}$ & 7$^{+3}_{-3}$ & 36$^{+19}_{-19}$ & 1.97$^{+1.29}_{-1.49}$ & 0.055$^{+0.036}_{-0.041}$ & 0.31$^{+0.04}_{-0.05}$ & 0.074$^{+0.032}_{-0.023}$ \\
G212.10N1 & 13$^{+6}_{-6}$ & 4$^{+3}_{-3}$ & 70$^{+66}_{-61}$ & 450$^{+334}_{-364}$ & 26$^{+19}_{-21}$ & 5$^{+2}_{-3}$ & 32$^{+16}_{-16}$ & 1.41$^{+0.92}_{-1.06}$ & 0.069$^{+0.046}_{-0.051}$ & 0.29$^{+0.38}_{-0.18}$ & 0.092$^{+0.132}_{-0.055}$ \\
$\mu$ & 34$^{+5}_{-6}$ & 10$^{+2}_{-2}$ & 113$^{+40}_{-39}$ & 440$^{+127}_{-148}$ & 42$^{+8}_{-9}$ & 11$^{+1}_{-1}$ & 24$^{+4}_{-4}$ & 0.57$^{+0.10}_{-0.11}$ & 0.053$^{+0.008}_{-0.009}$ & 0.32$^{+0.10}_{-0.05}$ & 0.078$^{+0.029}_{-0.012}$ \\
$\mu^{'}$ & 30$^{+5}_{-5}$ & 9$^{+2}_{-2}$ & 113$^{+40}_{-39}$ & 369$^{+108}_{-123}$ & 42$^{+8}_{-9}$ & 10$^{+1}_{-1}$ & 22$^{+4}_{-3}$ & 0.42$^{+0.07}_{-0.08}$ & 0.053$^{+0.008}_{-0.009}$ & 0.30$^{+0.08}_{-0.04}$ & 0.084$^{+0.031}_{-0.014}$
\enddata
\tablecomments{Column (5) presented the estimated $\frac{^{14}\text{N}}{^{15}\text{N}}$ of HCN, assuming [HCN]$=$[HNC]$=$77$\times$[HN$^{13}$C]. }
\end{deluxetable*}

\subsubsection{$^{12}$C and $^{13}$C Isotopologues}
\label{sec:isotope:$^{12}$C$^{13}$C}

In our observation, two $^{12}$C/$^{13}$C ratios are measured, CS/$^{13}$CS and HNC/HN$^{13}$C. 
The measured $^{12}$C/$^{13}$C ratio of CS and HNC are $34\pm19$ and $10\pm6$, respectively. 
The relatively low $^{12}$C/$^{13}$C ratio compared to the typical ISM value of $\sim77$ \citep{1994Wilson_isotope} indicates that the emission of their main isotopologues, CS $J=1-0$ and HNC $J=1-0$, are most likely optically thick or the molecules are centrally depleted \citep[e.g., ][]{2002Tafalla_starless_chem,2020Kim_starless_CS}. 

\subsubsection{$^{14}$N and $^{15}$N Isotopologues}
\label{sec:isotope:14N15N}

We detected three $^{15}$N-substituted molecules: HC$^{15}$N, N$^{15}$NH$^+$, and $^{15}$NNH$^+$. 
HC$^{15}$N is detected in eight sources, and N$^{15}$NH$^+$ and $^{15}$NNH$^+$ are detected in the same and only source, G208.68N2. 

From the column density ratio of HCN to HC$^{15}$N, the $^{14}$N/$^{15}$N ratio is \vlr{113}{40}{39}.
This value is lower than the $240\pm200$ reported by \citet{2025Tasa-Chaveli_GEMS}, though their ratio decreases to $163\pm49$ when two outliers (560 and 850) are excluded.
This directly derived $^{14}$N/$^{15}$N ratio for HCN is also much lower than the local ISM value of $\sim450$ \citep{1994Wilson_isotope}. 
As discussed for $^{12}$C/$^{13}$C ratio, this value is likely a lower limit since the main isotopologue emission, HCN $J=1-0$, is most likely optically thick \citep[e.g.,][]{2021Rodriguez-Baras_GEMS_statistics,2024Jensen_HCN_isotope}.

Since we do not have observations of the $^{13}$C isotopologue of HCN (i.e., H$^{13}$CN), we cannot directly estimate the HCN column density from isotopologue scaling.
An alternative approach is to approximate the HCN column density using its isomer, HNC, given that surveys have shown their abundances to be broadly comparable. 
This is an inverse approach of what we have used in Section~\ref{sec:isomer:HCN}. 
Assuming identical column densities and $^{12}$C/$^{13}$C ratios \citep[77, ][]{1994Wilson_isotope} of HCN and HNC, we estimate the column density of HCN by [HCN]$=$[HNC]$=77\times$[HN$^{13}$C]. 
The resulting $^{14}$N/$^{15}$N ratio is \vlr{440}{127}{148} (or \vlr{369}{108}{123} if excluding the outliers), appearing to be consistent with the local ISM value of $\sim450$ \citep{1994Wilson_isotope}.

For the N$_2$H$^+$/N$^{15}$NH$^+$ column density ratio, values of $1050 \pm 220$, \vlr{400}{100}{65}, and \vlr{300}{170}{100} have been reported for L1544 \citep{2013Bizzocchi_L1544_14N15N}, Barnard 1b \citep{2013Daniel_Barnard1_N}, and IRAS 16293E \citep{2016Daniel_16293E_14N15N}, respectively.
\citet{2018Redaelli_survey_14N15N} investigated three starless cores, L183, L429, and L694-2, and their reported values are \vlr{670}{150}{230}, \vlr{740}{250}{250}, and \vlr{580}{140}{110}, respectively. 

We detected N$^{15}$NH$^+$ and $^{15}$NNH$^+$ in one source, G208.68N2. 
Since N$_2$H$^+$ is not detected in our observations, we cannot directly use it to derive the column density ratios.
As a result, we estimate the N$_2$H$^+$ column density from that of N$_2$D$^+$, adopting the D/H ratio of N$_2$H$^+$ in G208.68N2, $0.11\pm0.01$, reported by \citet{2020Kim_PGCC_45m}. 
We do not directly adopting the column density of N$_2$H$^+$ from \citet{2020Kim_PGCC_45m} since the value of column density is highly depending on the methodology for the measurement. 
The corresponding $^{14}$N/$^{15}$N ratio, derived from the N$_2$H$^+$/N$^{15}$NH$^+$ column density ratio, is \vlr{596}{348}{401}, which is consistent with general literature values.

On the other hand, \citet{2013Bizzocchi_L1544_14N15N} found that the N$^{15}$NH$^+$ is slightly more abundant than $^{15}$NNH$^+$ with a [N$^{15}$NH$^+$]/[$^{15}$NNH$^+$] column density ratio of $1.1\pm0.3$ in L1544. 
Similarly, an upper limit of $3.05^{+3.35}_{-0.35}$ can be derived for Barnard 1b based on the values in Table~2 of \citet{2013Daniel_Barnard1_N}. 
For L694-2, this ratio is $1.24^{+0.38}_{-0.30}$, reported by \citet{2018Redaelli_survey_14N15N}. 
Although there exist great uncertainties, our estimated N$^{15}$NH$^+$ and $^{15}$NNH$^+$ column densities agree with the above tentative column density difference, with a [N$^{15}$NH$^+$]/[$^{15}$NNH$^+$] of $1.63^{+1.22}_{-1.10}$.

\subsubsection{$^{16}$O and $^{18}$O Isotopologues}
\label{sec:isotope:16O18O}

We detected two HCO$^+$ isotopologues, HCO$^+$ and HC$^{18}$O$^+$, in all 23 and 18 sources, respectively. 
The directly derived $^{16}$O/$^{18}$O ratio of HCO$^+$ is \vlr{42}{8}{9}.
This value is likely a lower limit because the main isotopologue emission, HCO$^+$ $J=1-0$, is often optically thick \citep[e.g.,][]{2021Rodriguez-Baras_GEMS_statistics}.
Consistently, the derived ratio is far lower than the local ISM value of $\sim$560 \citep{1994Wilson_isotope}.

\subsubsection{$^{32}$S and $^{34}$S Isotopologues}
\label{sec:isotope:32S34S}

We detected one $^{34}$S species, C$^{34}$S, in 19 sources. 
Directly deriving isotopologue ratios from these detections gives statistics of $^{32}$S/$^{34}$S = $11\pm5$.
Adopting [CS]$=77\times$[$^{13}$CS], the $^{32}$S/$^{34}$S ratio becomes \vlr{24}{4}{4}, consistent with the local ISM value of $\sim$19 \citep{1998Lucas_isotope}.

\subsubsection{Deuterium}
\label{sec:isotope:DH}

Deuterium enrichment in starless cores is driven primarily by the exothermic reaction
\begin{equation}
    \mathrm{H}_3^+ + \mathrm{HD} \rightleftharpoons \mathrm{H}_2\mathrm{D}^+ + \mathrm{H}_2 + 230 \mathrm{K},
\end{equation}
which proceeds efficiently below 20–30 K. 
Thus, deuteration is enhanced in cold, dense gas. 
Deuterium fractionation is widely used as the indicator of the evolutionary stage of starless cores, although the general range of the D/H ratios differs from species to species \citep[e.g., ][]{2013Pagani_prestellar_ortho-H2,2015Kong_deuteration,2022Esplugues_GEMS_sulfur,2023Navarro-Almaida_chem}. 

We detected five deuterated species, DNC, c-C$_3$HD, DC$_3$N, NH$_2$D, and N$_2$D$^{+}$. 
In this section, we discuss the column density ratios of DNC, c-C$_3$HD, and DC$_3$N with respect to their main isotopologues in order to derive the D/H ratios.
For NH$_2$D, we did not observe any other isotopologue. 
For N$_2$D$^{+}$, we only detected its isotopologue (N$^{15}$NH$^+$ and $^{15}$NNH$^+$) in one source (G208.68N2.)

We detected DNC and HNC in all 23 sources. 
Direct derivation of the D/H ratio from these detections yields \vlr{0.57}{0.10}{0.11}. 
Using [HN$^{13}$C] and adopting $^{12}$C/$^{13}$C = 77, the D/H ratio for HNC is \vlr{0.053}{0.008}{0.009}. 
As a pioneering survey of this project, \citet{2020Kim_PGCC_45m} measured [DNC]/[HN$^{13}$C] with the Nobeyama 45-m telescope toward most of our targets using the same transitions of this study. 
Our [DNC]/[HN$^{13}$C] ratios agree with their reported values within a factor of two. 
c-C$_3$H$_2$ and c-C$_3$HD are simultaneously detected in 11 sources, and the D/H ratio of c-C$_3$H$_2$ is \vlr{0.32}{0.10}{0.05}. 
HC$_3$N and DC$_3$N are simultaneously detected in eight sources, and the D/H ratio of HC$_3$N is $0.078\pm0.021$. 

We compare our results with those of \citet{2001Turner_deuteration}, who observed TMC-1 (CP) and reported D/H ratios for the same three species. 
Their derived D/H ratios are 0.0198 for HNC (assuming a $^{12}$C/$^{13}$C = 77), 0.068 for $c$-C$_3$H$_2$, and 0.016 for HC$_3$N. 
These values are consistently smaller than our averaged ratios by factors of $\sim$3--5.

\begin{figure*}[htb!]
\centering
\includegraphics[width=\linewidth]{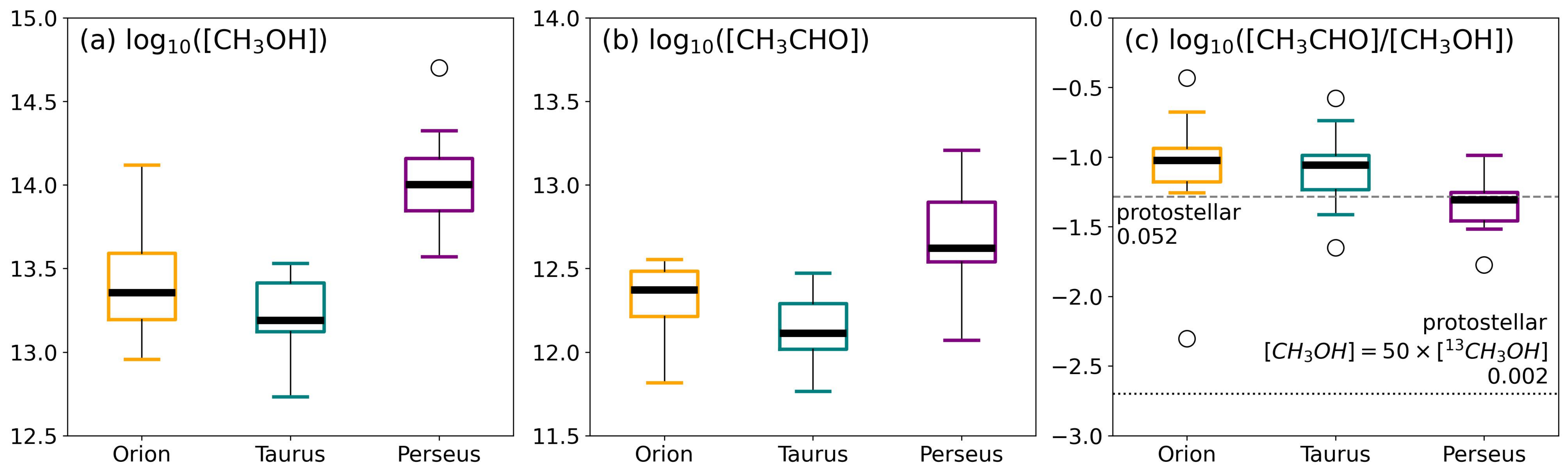}
\caption{\label{fig:boxplot_COM} The boxplots of the CH$_3$OH column densities (a), CH$_3$CHO column densities (b), and the ratio between them (c) in starless cores from different clouds. 
The statistics was achieved in logarithmic space. 
The data of Taurus and Perseus clouds were adopted from \citet{2020Scibelli_COM_Taurus} and \citet{2024Scibelli_COM_Perseus}, respectively. 
In panel (c), the protostellar core values are adopted from \citet{2022Hsu_ALMASOP}. 
}
\end{figure*}

\begin{figure*}[htb!]
\centering
\includegraphics[width=\linewidth]{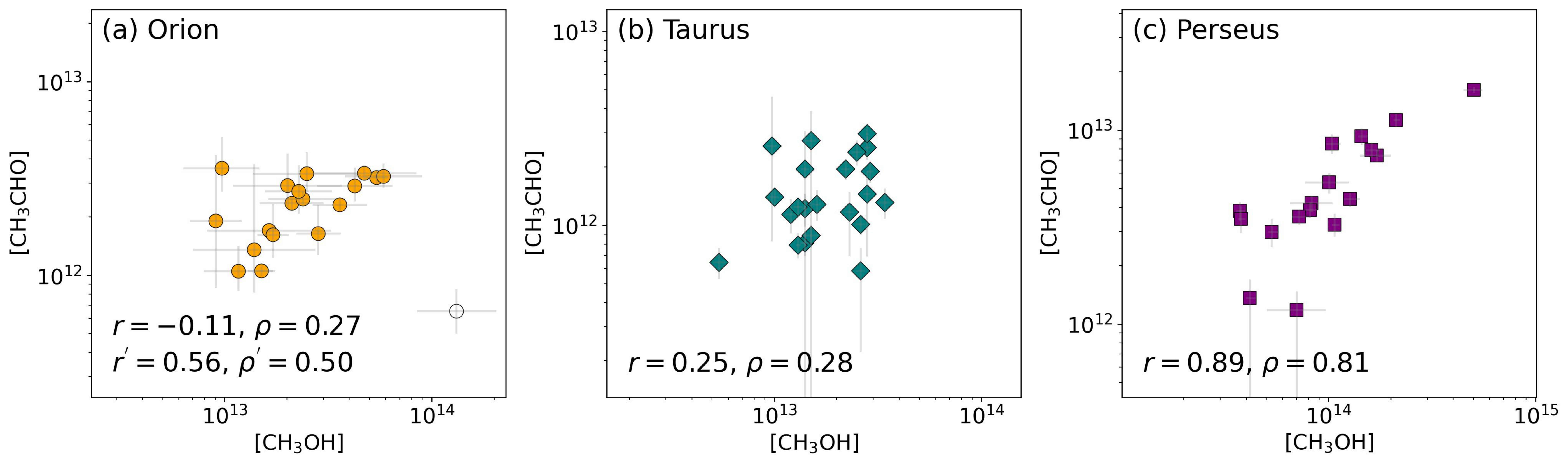}
\caption{\label{fig:scatter_COM} The scatter plots of the CH$_3$OH and CH$_3$CHO column densities in starless cores from different clouds. 
The data of Taurus and Perseus clouds were adopted from \citet{2020Scibelli_COM_Taurus} and \citet{2024Scibelli_COM_Perseus}, respectively. 
The $r$ and $\rho$ represent the Pearson and Spearman correlation coefficients, respectively. 
In panel (a), the $r{'}$ and $\rho{'}$ are the coefficients excluding the outlier illustrated by the white color. 
}
\end{figure*}

\subsection{Saturated Complex Organics: CH$_3$OH and CH$_3$CHO}
\label{sec:COM}

COMs in star-forming regions are of great interest because of their potential connection to prebiotic chemistry.
Gas-phase COMs have been found in a wide range of environments in low mass young stellar objects \citep[e.g.,][]{2008Arce_L1157-B1_COMs,2019Jacobsen_L483_COM,2019Lee_HH212,2024Hsu_HOPS87,2024DeSimone_IRAS4A2_COMdiskwind,2025Hsu_G192}.
The ubiquitous of COMs found in protostellar envelopes \citep[e.g.,][]{2004Ceccarelli_HotCorino,2023Hsu_ALMASOP} appears to be in full consistency with the prevailing chemical model, in which COMs are primarily form on the icy mantles of dust grains via thermally driven diffusive grain chemistry during the warm-up phase \citep[e.g., ][]{2008Garrod_grain-surface} and subsequently desorb into the gas phase through processes such as ice sublimation and ice destruction \citep[see][and references therein]{2009Herbst_COM_review}.
The detection of COMs in cold starless cores, however, challenges this above grain-surface warm-up scenario \citep[e.g., ][]{1985Matthews_TMC1_L134N_COM,1988Friberg_TMC1_L134N_B335_COM,2012Bacmann_L1689B_COM,2006Tafalla_L1498_1517B}. 
One possible scenario is to incorporate formation mechanisms that operate under cold conditions \citep{2022Garrod_non-diffusive} together with non-thermal desorption processes such as cosmic-ray-induced sputtering \citep{2020Dartois_CR-sputtering,2021Wakelam_CR_sputtering} and grain–grain collision shocks \citep[e.g.,][]{2001Dickens_TMC1,2015Soma_TMC1,2020Harju_H-MM1_CH3OH,2022Lin_L1544_CH3OH_HNCO,2022Kalvans_desorption,2025Hsu_ALMASOP_starless}.
A different scenario couples formation and desorption through so-called reactive desorption \citep[e.g., ][]{2007Garrod_reactive-desorption,2013Vasyunin_reactive-desorption,2017Vasyunin_reactive-desorption,2018Chuang_reactive-desorption,2020Jin_COM_reactive_desorption,2025Borshcheva_COM_reactive_desorption}.
Other alternatives involve gas-phase formation pathways for COMs \citep[e.g.,][]{2020Vazart_acetaldehyde_gas-phase}.
Investigating the occurrence, abundance of COMs and their correlations is therefore essential for constraining both their formation and desorption mechanisms in starless cores.

\subsubsection{Detection Rates and Ubiquity of COMs}
\label{sec:COM:rate}

In this study, we detected two (saturated) COMs, CH$_3$OH and CH$_3$CHO. 
In surveys toward starless cores in the Taurus cloud \citep{2020Scibelli_COM_Taurus} and the Perseus cloud \citep{2024Scibelli_COM_Perseus}, the detection rates of CH$_3$OH were 100\%, suggesting the prevalence of CH$_3$OH in starless cores. 
As a pioneering project of this study, \citet{2026Hsu_starless_core} observed 16 cores among the sample of this study and found also a  detection rate of 100\% for CH$_3$OH. 

In this survey of the Orion cloud, the detection rates of CH$_3$OH is 87\% (20/23).
The three sources without CH$_3$OH detections are G206.21N, G206.21S and G210.37N.
For G206.21N and G206.21S, \citet{2026Hsu_starless_core} reported the CH$_3$OH detection with the ACA array, suggesting that their non-detections with Yebes are due to limited sensitivity rather than different chemistry. 
Given the ubiquity of CH$_3$OH detections in surveys such as \citet{2020Scibelli_COM_Taurus}, \citet{2024Scibelli_COM_Perseus}, and \citet{2026Hsu_starless_core}, we argue that G210.37N is also not devoid of CH$_3$OH, but instead its emission is too weak to be detected. 

For CH$_3$CHO, in our survey of the Orion cloud, the detection rate is 83\% (19/23).
Such a high detection rate suggests that gaseous CH$_3$CHO, similar with CH$_3$OH, could also be ubiquitous in starless cores. 
We note that this detection rate is higher than those reported for the Taurus cloud (70\%, 21/31; \citealt{2020Scibelli_COM_Taurus}) and the Perseus cloud (49\%, 17/35; \citealt{2024Scibelli_COM_Perseus}).
In both literature surveys, the CH$_3$CHO detection rates were obtained with the ARO 12-m telescope at 3~mm (95~GHz), where the targeted transitions have higher upper-state energies (5.0-13.9~K). 
In contrast, our observations covered the CH$_3$CHO transitions with upper energies of around 3~K. 
If we restrict our Orion analysis to the CH$_3$CHO transitions with $E_u \sim 5$~K only, the detection rate drops to $\sim$60\%, consistent with the values reported in Taurus and Perseus. 
The higher detection rate of the lower upper energy transition, moreover, implies the cold nature of the CH$_3$CHO gas in starless environments.

\subsubsection{Comparisons of Column Densities between Clouds}
\label{sec:COM:colDens}

In Figures~\ref{fig:boxplot_COM}(a) and (b), we present the boxplots of the CH$_3$OH and CH$_3$CHO column densities, respectively, measured toward the Orion (this study), the Taurus \citep{2020Scibelli_COM_Taurus} and the Perseus \citep{2024Scibelli_COM_Perseus} molecular clouds.
Since the CH$_3$CHO column densities reported by \citet{2024Scibelli_COM_Perseus} include only the {\textit{A}}-type isomer, we multiplied them by a factor of two, assuming an {\textit{A}}-to-{\textit{E}} ratio of unity.
The same as in Figure~\ref{fig:bar_box_Ntot}, the box represents Q1 and Q3, the line inside the box represents Q2 (i.e., median), the two whiskers represent Q1-1.5$\times$IQR and Q3+1.5$\times$IQR, and the circles represent the outliers. 
Statistics were derived in logarithmic space. 
The box (from Q1 to Q3) of the CH$_3$OH column density spans approximately from $10^{13.25}$ to $10^{13.50}$\persqcm, and the box of the CH$_3$CHO column density spans from $10^{12.2}$ to $10^{12.5}$\persqcm. 


As shown in Figures~\ref{fig:boxplot_COM}(a) and (b), for both CH$_3$OH and CH$_3$CHO, the locations of the boxes of the Taurus sample are comparable to those of our Orion sample, whereas the Perseus sample shows boxes that are systematically higher by about an order of magnitude.
Judging from their H$_2$ gas column densities, one finds that the difference of a factor $\sim10^{0.5}\sim3$ in the apparent column densities of the two COMs between the Taurus and Perseus clouds is not resulting from the molecular fractional abundance. 
Instead, the fractional abundances of these two COMs (i.e., [COM]/[H$_2$]) in the Perseus sample are higher than those in the Taurus sample by $\sim10^{0.5}\sim3$ (see the upper left panel of Fig. 17 in \citet{2024Scibelli_COM_Perseus}). 
For CH$_3$OH, the difference between the fractional abundances (and column densities) could be further attributed to differences in the adopted analysis methods, particularly the adopted source sizes \citep{2024Scibelli_COM_Perseus}.
Still, there remains a residual enhancement of a factor of $\sim$1.6–1.8 in Perseus \citep{2024Scibelli_COM_Perseus}. 

\subsubsection{Column Density Ratios between CH$_3$OH and CH$_3$CHO}
\label{sec:COM:ratio}

CH$_3$OH is the most commonly detected COM and therefore often serve as a reference to normalize other COM column densities when comparing chemical compositions across COM-rich regions \citep[e.g.,][]{2021Yang_PEACHES,2022Hsu_ALMASOP,2024Scibelli_COM_Perseus}.
In Figure~\ref{fig:boxplot_COM}(c), we made the boxplots of the [CH$_3$CHO]/[CH$_3$OH] column density ratio for the starless cores in the three clouds. 
The statistics were, same as Figures~\ref{fig:boxplot_COM}(a) and (b), derived in the logarithmic space. 
The median value of [CH$_3$CHO]/[CH$_3$OH] in the Orion and the Taurus clouds are around $10^{-1}$. 
The median of this ratio is slightly lower in the Perseus cloud, which is around $10^{-1.3}$.

To inspect whether or how the [CH$_3$CHO]/[CH$_3$OH] column density ratio varies with the evolution of star formation, we compare our derived ratios with those in hot corinos (i.e., localized warm regions rich in COMs surrounding protostars). 
In Figure~\ref{fig:boxplot_COM}(c), we label the [CH$_3$CHO]/[CH$_3$OH] column density ratios inferred from the survey of hot corinos in Orion {it under the ALMASOP project} \citep{2022Hsu_ALMASOP}. 
The value 0.052 illustrated by the dashed line in Figure~\ref{fig:boxplot_COM}(c) was directly derived from the slope of the proportional relation between [CH$_3$CHO] and [CH$_3$OH].
The other value 0.002 illustrated by the dotted line was derived from [$^{13}$CH$_3$OH] with an assuming $^{12}$C/$^{13}$C ratio of 50 (i.e., [CH$_3$OH]$=50\times$[$^{13}$CH$_3$OH]) adopted by \citet{2022Hsu_ALMASOP}.
Such indirect derivation was motivated by the commonly suggested optically thick CH$_3$OH emission in hot corinos, based on the observed low $^{12}$C/$^{13}$C ratios \citep[e.g.,][]{2013Zapata_IRAS16293B_depth,2019Jacobsen_L483_COM,2019Lee_HH212,2020Hsu_ALMASOP,2020Manigand_IRAS16293A_COM,2022Hsu_ALMASOP,2026Hsu_HOPS-288}. 

We see, in Figure~\ref{fig:boxplot_COM}(c), that the directly derived value ($\sim$0.05) is comparable to the values for the starless cores, while the indirectly derived value ($\sim$0.002) is lower by a factor of 25. 
This yields several scenarios. 
One straightforward scenario is that the [CH$_3$CHO]/[CH$_3$OH] ratio indeed varies with the formation of stars at their early stages.
In other words, the [CH$_3$CHO]/[CH$_3$OH] ratio is $\sim0.05$ in the starless phase and then decrease to $\sim 0.002$ in the protostellar phase.
This could imply relatively enhanced production of gas-phase CH$_3$OH in the protostellar stage compared to the starless stage, possibly due to more efficient desorption mechanisms such as thermal desorption. 

In contrast, assuming that the [CH$_3$CHO]/[CH$_3$OH] ratio remains similar between the starless and protostellar stages would imply either an underestimation of [CH$_3$OH] in starless cores or an overestimation of [CH$_3$OH] in protostellar cores.
For the former, the [CH$_3$CHO]/[CH$_3$OH] ratios may be $\sim 0.002$ in both the starless and protostellar stages. 
In this case, the CH$_3$OH emission in starless cores would generally be optically thick, similar to that in protostellar stage, leading to an underestimation of the CH$_3$OH column density by about an order of magnitude.
However, \citet{2024Scibelli_COM_Perseus} applied non-LTE method for CH$_3$OH in the Perseus starless cores and found that the optical depths are generally consistent with optically thin emission. 
For the latter, the [CH$_3$CHO]/[CH$_3$OH] ratios may instead be $\sim 0.05$ in both evolutionary stages. 
This would imply that the CH$_3$OH emission in hot corinos is not severely optically thick and, therefore, the directly derived column densities of CH$_3$OH in protostellar cores are not underestimated severely as well. 
In other words, the $^{12}$C/$^{13}$C ratio in hot corinos could be lower than the typical ISM value ($\sim$50--80), possibly reflecting enhanced formation of $^{13}$C-COMs on dust grains during star formation \citep[e.g., ][]{2024Ichimura_SFR_12C13C,2025Ichimura_COM_12C13C}. 

\subsubsection{Column Density Correlations between CH$_3$OH and CH$_3$CHO}
\label{sec:COM:correlation}


To further investigate the chemical connection between CH$_3$OH and CH$_3$CHO, we show in Figures~\ref{fig:scatter_COM}(a), (b), and (c) the [CH$_3$CHO] versus [CH$_3$OH] scatter plots of the Orion (this study), Taurus \citep{2020Scibelli_COM_Taurus}, and Perseus \citep{2024Scibelli_COM_Perseus} starless cores, respectively. 
We label their Pearson correlation coefficients ($r$) and Spearman correlation coefficients ($\rho$) at the lower left of the panels. 
The Spearman correlation coefficient is computed as the Pearson correlation coefficient applied to the ranks of the data rather than to their raw values.
As a result, the Spearman coefficient is sensitive not only to linear relationships but also to any monotonic correlation between variables. 

As indicated in the panels of Figure~\ref{fig:scatter_COM}, among the three clouds, only the Perseus sample appears to have a clear positive correlation between [CH$_3$OH] and [CH$_3$CHO], with $r=0.89$ and $\rho=0.81$. 
In contrast, the Orion and Taurus samples do not exhibit a clear correlation between [CH$_3$OH] and [CH$_3$CHO], with $r$ and $\rho$ less than 0.3. 
Note that the Spearman correlation coefficient for the Taurus cloud derived here (0.28) differs from the value of 0.54 reported by \citet{2020Scibelli_COM_Taurus}, which was calculated by including upper limits on the CH$_3$CHO column density (private communication). 
For the Orion cloud, even if we manually remove the outlier data, the white marker at the bottom right in Figures~\ref{fig:scatter_COM} (c), the updated coefficients ($r{'}=0.56$ and $\rho{'}=0.50$) do not imply at any statistically significant level a linear correlation between [CH$_3$OH] and [CH$_3$CHO] in the Orion starless cores. 
While both CH$_3$OH and CH$_3$CHO are assumed to be of grain surface origin, more observations, presumably, toward a larger sample of starless cores in different clouds, would be desired to reveal quantitatively if and how CH$_3$OH and CH$_3$CHO are chemically linked in starless core environments.


\section{Conclusions}
\label{sec:Conclusions}

Our Q/W-band Observations toward Starless Cores in Orion (QWOSCO) conducted wide-band chemical surveys with Yebes 40-m observatory toward 23 starless cores in the Orion cloud. 
This paper presents the first results of the QWOSCO project:

\begin{enumerate}
  \item \textit{Overview}:
  We detect $\sim40$ molecular species and present the estimated column densities of them, if detected, in each of the 23 starless cores. 
  The non-detections of the molecules are likely due to limited sensitivity rather than true chemical absence, while chemical segregation between sources can still be discerned. 
  The column density of each molecule has an interquartile range (IQR) spanning approximately half an order of magnitude, indicating a well-defined typical value.
  
  \item \textit{Isomers}: 
  The averaged {\textit{A}}-to-{\textit{E}} ratio of CH$_3$CHO is $1.1^{+0.2}_{−0.1}$, consistent with the expected value of unity. 
  The averaged {\textit{ortho}}-to-{\textit{para}} ratio of $c$-C$_3$H$_2$ is $2.0^{+0.7}_{−0.7}$, in agreement with the expectation that this ratio should be less than three in cold environments.
  The averaged {\textit{cyclic}}-to-{\textit{linear}} ratio of C$_3$H$_2$ derived from the {\textit{ortho}} isomers is $5.3^{+0.6}_{−0.5}$, which is lower than values reported in the literature.
  The averaged column density ratio between HCN and HNC deriving from their rare isotopologues is $1.2^{+0.4}_{−0.4}$. 
  Their column densities are expected to be comparable in cold environments. 

  \item \textit{Isotopologues}:
    The isotope ratios $^{12}$C/$^{13}$C, $^{14}$N/$^{15}$N, $^{16}$O/$^{18}$O, and $^{32}$S/$^{34}$S derived directly from the corresponding molecular species, are all smaller than the typical local ISM values. 
    The comparatively low ratios are probably resulting from the optically thick lines of their main isotopologues. 
    We detected N$^{15}$NH$^+$ and $^{15}$NNH$^+$ in one source, G208.68–19.20N2 (G206.68N2).
    Using its D/H ratio of N$_2$H$^+$ reported in the literature, the derived column density ratio [N$_2$H$^+$]/[N$^{15}$NH$^+$] is consistent with the literature.
    In addition, the estimated column densities of N$^{15}$NH$^+$ and $^{15}$NNH$^+$ show a tentative difference, although the associated uncertainties are large.

  \item \textit{Complex Organics}:
  \begin{enumerate}
      \item \textit{Ubiquity:} Both CH$_3$OH and CH$_3$CHO are widely detected in our sample, with detection rates of 87\% and 83\%, respectively. 
      The non-detections of both CH$_3$OH and CH$_3$CHO are most likely attributed to the limited sensitivity. 
      As a result, CH$_3$CHO could also be ubiquitous in starless cores, similar to CH$_3$OH. 
    
      \item{Chemical Correlation between CH$_3$OH and CH$_3$CHO: }
      The column density ratio [CH$_3$CHO]/[CH$_3$OH] in starless cores across Taurus, Perseus, and Orion is consistent ($\sim0.05$), and is comparable to the directly measured value inferred for hot corinos (i.e., protostellar stage). 
      However, if CH$_3$OH in protostellar regions is heavily optically thick, as many studies suggest, the true ratio there may be $10^{-3}$, much lower than in starless cores.
      We discuss several possible explanations for this discrepancy. 

      The CH$_3$OH and CH$_3$CHO column densities show a weak linearity (Pearson coefficient $r=0.56$ and Spearman coefficient $\rho=0.50$).
      More observations toward a larger sample of starless cores in different clouds would be desired to reveal quantitatively if and how CH$_3$OH and CH$_3$CHO are chemically linked in starless environments.

  \end{enumerate}
  
\end{enumerate}


\acknowledgments
Based on observations carried out with the Yebes 40 m telescope (22A010). The 40 m radio telescope at Yebes Observatory is operated by the Spanish Geographic Institute (IGN; Ministerio de Transportes y Movilidad Sostenible). 
X.-C. Liu and T. Liu acknowledges the supports by the National Key R\&D Program of China (No. 2022YFA1603101). 
S.-Y. Hsu acknowledges supports from the Academia Sinica of Taiwan (grant No. AS-PD-1142-M02-2) and National Science and Technology Council of Taiwan (grant No. 112-2112-M-001- 039-MY3).  
S.-Y. Liu acknowledges supports from National Science and Technology Council of Taiwan (grant No. 113-2112-M-001-004- and 114-2112-M-001-035-MY3).
MJ acknowledges the support of the Research Council of Finland Grant No. 348342.
S.L. acknowledges support from the National SKA Program of China with No. 2025SKA0140100, “Double First-Class” Funding with No. 14912217, and National Natural Science Foundation of China (NSFC) grant with No. 13004007. 

\software{
astropy \citep{astropy:2013, astropy:2018, astropy:2022},
\texttt{CASA} \citep{casa:2007,casa:2022},
\texttt{CARTA}  \citep{2021Comrie_CARTA}, 
\texttt{GILDAS} \citep{2005GILDAS}.
}

\clearpage
\appendix

\resetapptablenumbers
\section{Sample Literature Review \label{appx:sample}}

In Table~\ref{tab:lit} we show the survey-type studies having targets significantly overlapping with this study. 
They include \citet{2018Yi_PGCC_SCUBA2_II}, \citet{2020Kim_PGCC_45m}, \citet{2020Dutta_ALMASOP},\citet{2021Sahu_ALMASOP_presstellar}, \citet{2021Yi_PGCC_chem}, \citet{2021Tatematsu_SCOPE}, \citet{2022Tatematsu_ALMASOP_inward}, \citet{2023Sahu_ALMASOP_density}, \citet{2025Hsu_ALMASOP_starless}, and finally \citet{2026Hsu_starless_core}, which briefly summarizes in its Sect. B1 all the results of the papers listed above.
\citet{2026Hsu_starless_core} observed 16 starless cores (including prestellar ones) with ACA at Band~3 (3~mm) and presented the ubiquity of CH$_3$OH and the related chemical segregation with N$_2$H$^+$, CCS, and $c$-C$_3$HD.

\begin{deluxetable}{llcccccccccc}
\caption{\label{tab:lit} Summary of sources included in literature.}
\tablehead{\colhead{Name} & \colhead{Short Name} & \colhead{Prestellar} & \colhead{Yi+18} & \colhead{Kim+20} & \colhead{Dutta+20} & \colhead{Yi+21} & \colhead{Tatematsu+21} & \colhead{Tatematsu+22} & \colhead{Hsu+26}}
\startdata
G198.69-09.12N1 & G198.69N1 &            & \checkmark & \checkmark & \checkmark$^a$ & \checkmark &            & \checkmark & \\
G198.69-09.12N2 & G198.69N2 &            & \checkmark & \checkmark & \checkmark$^a$ & \checkmark &            & \checkmark & \\
G203.21-11.20E1 & G203.21E1 &            & \checkmark & \checkmark & \checkmark & \checkmark & \checkmark & \checkmark & \checkmark \\
G203.21-11.20E2 & G203.21E2 &            & \checkmark & \checkmark & \checkmark$^a$ & \checkmark & \checkmark & \checkmark & \\
G205.46-14.56M3 & G205.46M3 & \checkmark & \checkmark & \checkmark & \checkmark &            &            & \checkmark & \checkmark \\
G206.21-16.17N  & G206.21N  &            & \checkmark & \checkmark & \checkmark & \checkmark &            & \checkmark & \checkmark \\
G206.21-16.17S  & G206.21S  &            & \checkmark &            & \checkmark & \checkmark &            & \checkmark & \checkmark \\
G207.36-19.82N4 & G207.36N4 &            & \checkmark & \checkmark & \checkmark & \checkmark & \checkmark & \checkmark & \\
G208.68-19.20N2 & G208.68N2 & \checkmark & \checkmark & \checkmark & \checkmark & \checkmark & \checkmark & \checkmark & \checkmark \\
G209.29-19.65N1 & G209.29N1 &            & \checkmark & \checkmark & \checkmark &            &            & \checkmark & \checkmark \\
G209.29-19.65S1 & G209.29S1 & \checkmark & \checkmark & \checkmark & \checkmark &            &            & \checkmark & \checkmark \\
G209.29-19.65S2 & G209.29S2 &            & \checkmark & \checkmark & \checkmark &            & \checkmark & \checkmark & \checkmark \\
G209.55-19.68N2 & G209.55N2 &            & \checkmark & \checkmark & \checkmark & \checkmark &            & \checkmark & \checkmark \\
G209.77-19.40E3 & G209.77E3 &            & \checkmark & \checkmark & \checkmark & \checkmark &            & \checkmark & \checkmark \\
G209.79-19.80W  & G209.79W  &            & \checkmark & \checkmark & \checkmark$^a$ & \checkmark &            & \checkmark & \\
G209.94-19.52N  & G209.94N  & \checkmark & \checkmark & \checkmark & \checkmark & \checkmark & \checkmark & \checkmark & \checkmark \\
G209.94-19.52S1 & G209.94S1 &            & \checkmark & \checkmark & \checkmark & \checkmark &            & \checkmark & \\
G210.37-19.53N  & G210.37N  &            & \checkmark & \checkmark & \checkmark &            &            & \checkmark & \checkmark \\
G210.82-19.47N2 & G210.82N2 &            & \checkmark & \checkmark & \checkmark &            &            & \checkmark & \checkmark \\
G211.16-19.33N4 & G211.16N4 &            & \checkmark & \checkmark & \checkmark & \checkmark &            & \checkmark & \checkmark \\
G211.16-19.33N5 & G211.16N5 &            & \checkmark & \checkmark & \checkmark & \checkmark & \checkmark & \checkmark & \checkmark \\
G211.72-19.25S1 & G211.72S1 &            &   $^b$     & \checkmark & \checkmark$^a$ &           & \checkmark & \checkmark & \\
G212.10-19.15N1 & G212.10N1 & \checkmark & \checkmark & \checkmark & \checkmark &            & \checkmark & \checkmark & \checkmark \\
\enddata
\tablecomments{
The ``\checkmark'' labels the the sources included in the corresponding literature. 
The five ``Prestellar'' sources were also be studied by \citet{2021Sahu_ALMASOP_presstellar}, \citet{2023Sahu_ALMASOP_density}, and \citet{2025Hsu_ALMASOP_starless}. 
$^a$: Observed but either non- or weakly detected. 
$^b$: A different source with a similar name is included in the literature. 
}
\tablerefs{Yi+18: \citet{2018Yi_PGCC_SCUBA2_II}; Kim+20: \citet{2020Kim_PGCC_45m}; Dutta+20: \citet{2020Dutta_ALMASOP}; Sahu+21: \citet{2021Sahu_ALMASOP_presstellar}; Yi+21: \citet{2021Yi_PGCC_chem}; Tatematsu+21: \citet{2021Tatematsu_SCOPE};  Tatematsu+22: \citet{2022Tatematsu_ALMASOP_inward}; Sahu+23: \citet{2023Sahu_ALMASOP_density}; Hsu+25: \citet{2025Hsu_ALMASOP_starless}; Hsu+26: \citet{2026Hsu_starless_core}}
\end{deluxetable}

\clearpage
\resetapptablenumbers
\section{Transition Parameters \label{appx:trans}}
Table~\ref{tab:trans} lists the parameters of the transitions detected in this study. 

\startlongtable
\begin{deluxetable*}{lrrrrl}
\tablecaption{\label{tab:trans} {Detected transitions of this report. }}
\tablehead{\colhead{Formula} & \colhead{$f_\mathrm{rest}$} & \colhead{\Eu} & \colhead{\gu} & \colhead{$\log_{10}\frac{A_\mathrm{ij}}{\mathrm{Hz}}$} & \colhead{QNs} \\
\colhead{} & \colhead{(MHz)} & \colhead{(K)} & \colhead{} & \colhead{} & \colhead{}
}
\startdata
CS & 48990.95 & 2.4 & 3 & -5.7572 & $J=1-0$ \\
$^{13}$CS & 46247.56 & 2.2 & 6 & -5.8323 & $J=1-0$ \\
C$^{34}$S & 48206.94 & 2.3 & 3 & -5.7783 & $J=1-0$ \\
C$_2$H & 87328.59 & 4.2 & 3 & -5.8956 & $S=0.5$, $J=1.5-0.5$, $N=1-0$, $F_H=1-0$ \\
 & 87446.47 & 4.2 & 3 & -6.5828 & $S=0.5$, $J=0.5$, $N=1-0$, $F_H=1-0$ \\
 & 87284.10 & 4.2 & 3 & -6.5852 & $S=0.5$, $J=1.5-0.5$, $N=1-0$, $F_H=1$ \\
 & 87316.90 & 4.2 & 5 & -5.8149 & $S=0.5$, $J=1.5-0.5$, $N=1-0$, $F_H=2-1$ \\
HCO & 86708.36 & 4.2 & 3 & -5.3376 & $S=0.5$, $J=1.5-0.5$, $N=1-0$, $K_a=0$, $K_c=1-0$, $F_H=1-0$ \\
 & 86670.76 & 4.2 & 5 & -5.3288 & $S=0.5$, $J=1.5-0.5$, $N=1-0$, $K_a=0$, $K_c=1-0$, $F_H=2-1$ \\
 & 86777.46 & 4.2 & 3 & -5.3366 & $S=0.5$, $J=0.5$, $N=1-0$, $K_a=0$, $K_c=1-0$, $F_H=1$ \\
 & 86805.78 & 4.2 & 1 & -5.3268 & $S=0.5$, $J=0.5$, $N=1-0$, $K_a=0$, $K_c=1-0$, $F_H=0-1$ \\
HCO$^+$ & 89188.52 & 4.3 & 3 & -4.3781 & $J=1-0$ \\
HC$^{18}$O$^+$ & 85162.22 & 4.1 & 3 & -4.4383 & $J=1-0$ \\
N$_2$D$^+$ & 77109.24 & 3.7 & 27 & -4.9077 & $J=1-0$ \\
$^{15}$NNH$^+$ & 90263.84 & 4.3 & 9 & -4.4817 & $J=1-0$ \\
N$^{15}$NH$^+$ & 91205.70 & 4.4 & 9 & -4.4681 & $J=1-0$ \\
HCN & 88631.60 & 4.2 & 9 & -4.6184 & $J=1-0$ \\
HC$^{15}$N & 86054.97 & 4.1 & 3 & -4.6569 & $J=1-0$ \\
HNC & 90663.57 & 4.3 & 3 & -4.5703 & $J=1-0$ \\
DNC & 76305.70 & 3.7 & 3 & -4.7949 & $J=1-0$ \\
HN$^{13}$C & 87090.83 & 4.2 & 3 & -4.6227 & $J=1-0$ \\
HNO & 81477.49 & 3.9 & 3 & -5.6524 & $J=1-0$, $K_a=0$, $K_c=1-0$, $F\_N=1$ \\
C$_2$S & 33751.37 & 3.2 & 7 & -5.8019 & $S=1$, $J=3-2$, $N=2-1$ \\
 & 45379.05 & 5.4 & 9 & -5.3999 & $S=1$, $J=4-3$, $N=3-2$ \\
 & 38866.42 & 12.4 & 7 & -5.6656 & $S=1$, $J=3-2$, $N=3-2$ \\
 & 43981.02 & 12.9 & 7 & -5.5042 & $S=1$, $J=3-2$, $N=4-3$ \\
 & 81505.17 & 15.4 & 15 & -4.6145 & $S=1$, $J=7-6$, $N=6-5$ \\
HCS$^+$ & 42674.20 & 2.0 & 3 & -5.9370 & $J=1-0$ \\
 & 85347.89 & 6.1 & 5 & -4.9548 & $J=2-1$ \\
OCS & 36488.81 & 3.5 & 7 & -6.9067 & $J=3-2$ \\
 & 48651.60 & 5.8 & 9 & -6.5161 & $J=4-3$ \\
 & 85139.10 & 16.3 & 15 & -5.7658 & $J=7-6$ \\
C$_3$H & 32627.30 & 1.6 & 5 & -5.7709 & $S=0.5$, $J=1.5-0.5$, $N=2-1$, $F_H=2-1$ \\
 & 32660.65 & 1.6 & 5 & -5.7696 & $S=0.5$, $J=1.5-0.5$, $N=2-1$, $F_H=2-1$ \\
 & 32663.36 & 1.6 & 3 & -5.9456 & $S=0.5$, $J=1.5-0.5$, $N=2-1$, $F_H=1-0$ \\
H$_2$CO & 72837.95 & 3.5 & 3 & -5.0887 & $J=1-0$, $K_a=0$, $K_c=1-0$ \\
HOCO$^+$ & 42766.19 & 3.1 & 5 & -5.5759 & $J=2-1$, $K_a=0$, $K_c=2-1$ \\
NH$_2$D & 85926.28 & 20.7 & 27 & -5.1067 & $J=1$, $K_a=1-0$, $K_c=1$, $vibInv=s-a$ \\
HNCO & 43963.04 & 3.2 & 5 & -6.0055 & $J=2-1$, $K_a=0$, $K_c=2-1$ \\
 & 87925.24 & 10.6 & 9 & -5.0565 & $J=4-3$, $K_a=0$, $K_c=4-3$ \\
C$_3$S & 34684.37 & 5.8 & 13 & -5.5368 & $J=6-5$ \\
 & 40465.01 & 7.8 & 15 & -5.3312 & $J=7-6$ \\
 & 46245.62 & 10.0 & 17 & -5.1536 & $J=8-7$ \\
C$_3$H & 32634.39 & 1.6 & 3 & -5.9469 & $S=0.5$, $J=1.5-0.5$, $N=2-1$, $F_H=1-0$ \\
H$_2$CCO & 40417.95 & 2.9 & 5 & -6.2064 & $J=2-1$, $K_a=0$, $K_c=2-1$ \\
 & 80832.12 & 9.7 & 9 & -5.2576 & $J=4-3$, $K_a=0$, $K_c=4-3$ \\
 & 40039.02 & 15.9 & 15 & -6.3435 & $J=2-1$, $K_a=1$, $K_c=2-1$ \\
 & 40793.83 & 16.0 & 15 & -6.3193 & $J=2-1$, $K_a=1$, $K_c=1-0$ \\
 & 80076.65 & 22.7 & 27 & -5.2978 & $J=4-3$, $K_a=1$, $K_c=4-3$ \\
 & 81586.23 & 22.8 & 27 & -5.2735 & $J=4-3$, $K_a=1$, $K_c=3-2$ \\
$\it{c}\text{-}\text{C}_3\text{H}_2$; $ortho$ & 85338.89 & 4.1 & 15 & -4.6341 & $J=2-1$, $K_a=1-0$, $K_c=2-1$ \\
$\it{c}\text{-}\text{C}_3\text{H}_2$; $ortho$ & 82966.20 & 13.7 & 21 & -5.0035 & $J=3$, $K_a=1-0$, $K_c=2-3$ \\
$\it{c}\text{-}\text{C}_3\text{H}_2$; $ortho$ & 44104.78 & 15.8 & 21 & -5.4953 & $J=3$, $K_a=2-1$, $K_c=1-2$ \\
$\it{c}\text{-}\text{C}_3\text{H}_2$; $para$ & 82093.54 & 6.4 & 5 & -4.7246 & $J=2-1$, $K_a=0-1$, $K_c=2-1$ \\
$\it{c}\text{-}\text{C}_3\text{H}_2$; $para$ & 46755.61 & 8.7 & 5 & -5.5717 & $J=2$, $K_a=1-0$, $K_c=1-2$ \\
$\it{c}\text{-}\text{C}_3\text{H}_2$; $para$ & 84727.69 & 16.1 & 7 & -4.9820 & $J=3$, $K_a=2-1$, $K_c=2-3$ \\
$\it{l}\text{-}\text{C}_3\text{H}_2$; $ortho$ & 41198.34 & 2.0 & 15 & -5.3868 & $J=2-1$, $K_a=1$, $K_c=2-1$ \\
$\it{l}\text{-}\text{C}_3\text{H}_2$; $ortho$ & 41967.67 & 2.0 & 15 & -5.3627 & $J=2-1$, $K_a=1$, $K_c=1-0$ \\
$\it{l}\text{-}\text{C}_3\text{H}_2$; $para$ & 41584.68 & 3.0 & 5 & -5.2496 & $J=2-1$, $K_a=0$, $K_c=2-1$ \\
$\it{c}\text{-}\text{C}_3\text{HD}$ & 49615.86 & 2.4 & 9 & -5.3585 & $J=1-0$, $K_a=1-0$, $K_c=1-0$ \\
 & 79812.33 & 5.8 & 15 & -4.7850 & $J=2-1$, $K_a=1-0$, $K_c=2-1$ \\
 & 38224.44 & 7.6 & 15 & -5.7575 & $J=2$, $K_a=1-0$, $K_c=1-2$ \\
C$_4$H & 38049.69 & 4.6 & 11 & -5.9008 & $S=0.5$, $J=4.5-3.5$, $J=4.5-3.5$, $N=4-3$, $F_H=5-4$, $F=5-4$ \\
 & 38049.62 & 4.6 & 9 & -5.9140 & $S=0.5$, $J=4.5-3.5$, $J=4.5-3.5$, $N=4-3$, $F_H=4-3$, $F=4-3$ \\
 & 38088.48 & 4.6 & 7 & -5.9364 & $S=0.5$, $J=3.5-2.5$, $J=3.5-2.5$, $N=4-3$, $F_H=3-2$, $F=3-2$ \\
 & 38088.44 & 4.6 & 9 & -5.9139 & $S=0.5$, $J=3.5-2.5$, $J=3.5-2.5$, $N=4-3$, $F_H=4-3$, $F=4-3$ \\
 & 47566.81 & 6.8 & 13 & -5.6001 & $S=0.5$, $J=5.5-4.5$, $J=5.5-4.5$, $N=5-4$, $F_H=6-5$, $F=6-5$ \\
 & 47566.77 & 6.8 & 11 & -5.6087 & $S=0.5$, $J=5.5-4.5$, $J=5.5-4.5$, $N=5-4$, $F_H=5-4$, $F=5-4$ \\
 & 47605.50 & 6.9 & 9 & -5.6210 & $S=0.5$, $J=4.5-3.5$, $J=4.5-3.5$, $N=5-4$, $F_H=4-3$, $F=4-3$ \\
 & 47605.49 & 6.9 & 11 & -5.6082 & $S=0.5$, $J=4.5-3.5$, $J=4.5-3.5$, $N=5-4$, $F_H=5-4$, $F=5-4$ \\
 & 76117.45 & 16.4 & 19 & -4.9725 & $S=0.5$, $J=8.5-7.5$, $J=8.5-7.5$, $N=8-7$, $F_H=9-8$, $F=9-8$ \\
 & 76117.43 & 16.4 & 17 & -4.9758 & $S=0.5$, $J=8.5-7.5$, $J=8.5-7.5$, $N=8-7$, $F_H=8-7$, $F=8-7$ \\
 & 76156.03 & 16.4 & 15 & -4.9796 & $S=0.5$, $J=7.5-6.5$, $J=7.5-6.5$, $N=8-7$, $F_H=7-6$, $F=7-6$ \\
 & 76156.03 & 16.4 & 17 & -4.9753 & $S=0.5$, $J=7.5-6.5$, $J=7.5-6.5$, $N=8-7$, $F_H=8-7$, $F=8-7$ \\
 & 85634.02 & 20.6 & 21 & -4.8162 & $S=0.5$, $J=9.5-8.5$, $J=9.5-8.5$, $N=9-8$, $F_H=10-9$, $F=10-9$ \\
 & 85634.00 & 20.6 & 19 & -4.8189 & $S=0.5$, $J=9.5-8.5$, $J=9.5-8.5$, $N=9-8$, $F_H=9-8$, $F=9-8$ \\
 & 85672.58 & 20.6 & 17 & -4.8216 & $S=0.5$, $J=8.5-7.5$, $J=8.5-7.5$, $N=9-8$, $F_H=8-7$, $F=8-7$ \\
 & 85672.58 & 20.6 & 19 & -4.8183 & $S=0.5$, $J=8.5-7.5$, $J=8.5-7.5$, $N=9-8$, $F_H=9-8$, $F=9-8$ \\
CH$_3$OH & 48372.46 & 2.3 & 12 & -6.4498 & $J=1-0$, $K_a=0$, $K_c=1-0$, $rovib=A1$ \\
CH$_3$OH & 48376.89 & 15.4 & 12 & -6.4497 & $J=1-0$, $K_a=0$, $K_c=1-0$, $rovib=E$ \\
CH$_3$OH & 36169.26 & 28.8 & 36 & -6.8121 & $J=4-3$, $K_a=1-0$, $K_c=4-3$, $rovib=E$ \\
CH$_3$OH & 84521.21 & 40.4 & 44 & -5.7055 & $J=5-4$, $K_a=1-0$, $K_c=5-4$, $rovib=E$ \\
t-HCOOH & 89579.18 & 10.8 & 9 & -5.1244 & $J=4-3$, $K_a=0$, $K_c=4-3$ \\
HC$_3$N & 36392.32 & 4.4 & 9 & -5.4593 & $J=4-3$ \\
 & 45490.31 & 6.5 & 11 & -5.1589 & $J=5-4$ \\
DC$_3$N & 33772.53 & 4.0 & 9 & -5.5546 & $J=4-3$ \\
 & 42215.58 & 6.1 & 11 & -5.2541 & $J=5-4$ \\
l-C$_4$H$_2$ & 35727.38 & 4.3 & 9 & -5.4017 & $J=4-3$, $K_a=0$, $K_c=4-3$ \\
 & 35577.01 & 17.8 & 27 & -5.4352 & $J=4-3$, $K_a=1$, $K_c=4-3$ \\
 & 35875.77 & 17.8 & 27 & -5.4243 & $J=4-3$, $K_a=1$, $K_c=3-2$ \\
 & 44471.14 & 19.9 & 33 & -5.1244 & $J=5-4$, $K_a=1$, $K_c=5-4$ \\
 & 44844.59 & 20.0 & 33 & -5.1135 & $J=5-4$, $K_a=1$, $K_c=4-3$ \\
CH$_3$CCH & 34183.41 & 2.5 & 10 & -6.9420 & $J=2-1$, $K=0$, $rovib=A1-A2$ \\
CH$_3$CCH & 85457.30 & 12.3 & 22 & -5.6927 & $J=5-4$, $K=0$, $rovib=A2-A1$ \\
CH$_3$CHO; $A$ & 38512.08 & 2.8 & 10 & -5.7742 & $J=2-1$, $K_a=0$, $K_c=2-1$, $rovib=A$ \\
CH$_3$CHO; $E$ & 38506.03 & 2.9 & 10 & -5.7742 & $J=2-1$, $K_a=0$, $K_c=2-1$, $rovib=E$ \\
CH$_3$CHO; $A$ & 37464.20 & 5.0 & 10 & -5.9351 & $J=2-1$, $K_a=1$, $K_c=2-1$, $rovib=A$ \\
CH$_3$CHO; $A$ & 39594.29 & 5.1 & 10 & -5.8630 & $J=2-1$, $K_a=1$, $K_c=1-0$, $rovib=A$ \\
CH$_3$CHO; $E$ & 37686.93 & 5.0 & 10 & -5.9565 & $J=2-1$, $K_a=1$, $K_c=2-1$, $rovib=E$ \\
CH$_3$CHO; $E$ & 39362.54 & 5.2 & 10 & -5.9000 & $J=2-1$, $K_a=1$, $K_c=1-0$, $rovib=E$ \\
CH$_3$CHO; $A$ & 76878.95 & 9.2 & 18 & -4.8281 & $J=4-3$, $K_a=0$, $K_c=4-3$, $rovib=A$ \\
CH$_3$CHO; $E$ & 76866.44 & 9.3 & 18 & -4.8280 & $J=4-3$, $K_a=0$, $K_c=4-3$, $rovib=E$ \\
CH$_3$CHO; $E$ & 79099.31 & 11.8 & 18 & -4.8195 & $J=4-3$, $K_a=1$, $K_c=3-2$, $rovib=E$ \\
HC$_5$N & 31951.77 & 10.0 & 25 & -5.4664 & $J=12-11$ \\
 & 34614.39 & 11.6 & 27 & -5.3607 & $J=13-12$ \\
 & 37276.99 & 13.4 & 29 & -5.2630 & $J=14-13$ \\
 & 39939.59 & 15.3 & 31 & -5.1721 & $J=15-14$ \\
 & 42602.15 & 17.4 & 33 & -5.0872 & $J=16-15$ \\
 & 45264.72 & 19.6 & 35 & -5.0074 & $J=17-16$
\enddata
\end{deluxetable*}

\clearpage
\resetapptablenumbers
\section{Molecular Spectra \label{appx:spec}}
\begin{figure*}[h]\centering\includegraphics[width=.99\textwidth]{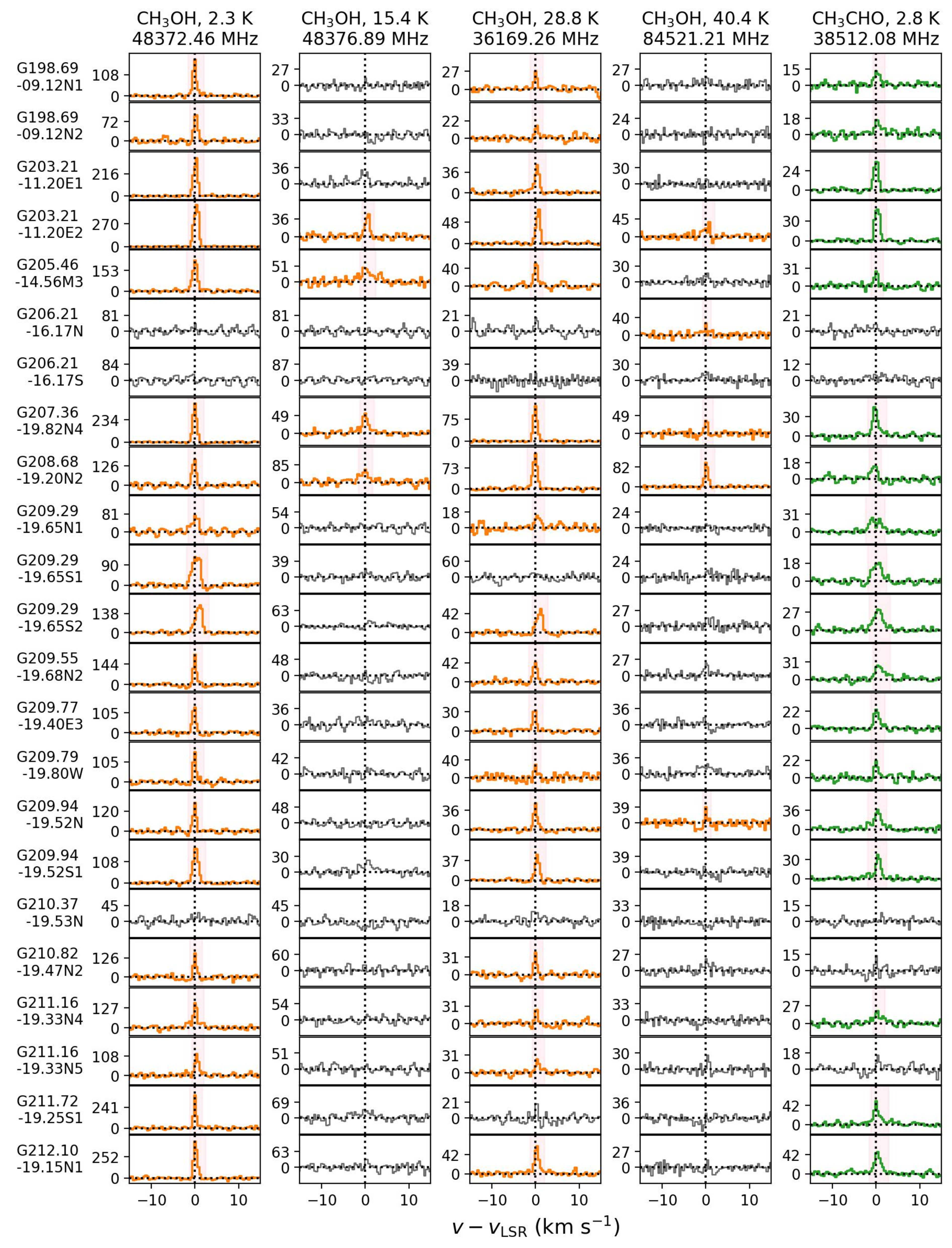}\caption{\label{fig:appx:spec_all_trans_0}Molecular spectra. Please see Appendix \ref{appx:spec} for the captions.}\end{figure*}

Figure~\ref{fig:appx:spec_all_trans_0} shows the spectra of selected molecular transitions. 
The complete figure set is available online. 

\clearpage
\resetapptablenumbers
\section{Integrated intensities and column densities \label{appx:Ntot}}
Table~\ref{tab:W} lists the integrated intensities of each transition in each source. 
Table~\ref{tab:Ntot} presents the estimated column densities with three different rotational temperatures. 
The molecular species having more than one transition The value having the lowest $\chi^2$-value are illustrated in bold font. 

\begin{splitdeluxetable*}{rcccccccBccccccccBcccccccc}
\tabletypesize{\scriptsize}
\tablecaption{\label{tab:W} Example of the integrated intensity of each transition. The complete table is provided in the format of machine readable table (MRT.) 
}
\tablehead{
\colhead{Formula ($f$)}
& \colhead{G198.69-09.12N1} 
& \colhead{G198.69-09.12N2} 
& \colhead{G203.21-11.20E1} 
& \colhead{G203.21-11.20E2} 
& \colhead{G205.46-14.56M3} 
& \colhead{G206.21-16.17N } 
& \colhead{G206.21-16.17S } 
& \colhead{G207.36-19.82N4} 
& \colhead{G208.68-19.20N2} 
& \colhead{G209.29-19.65N1} 
& \colhead{G209.29-19.65S1} 
& \colhead{G209.29-19.65S2} 
& \colhead{G209.55-19.68N2} 
& \colhead{G209.77-19.40E3} 
& \colhead{G209.79-19.80W } 
& \colhead{G209.94-19.52N } 
& \colhead{G209.94-19.52S1} 
& \colhead{G210.37-19.53N } 
& \colhead{G210.82-19.47N2} 
& \colhead{G211.16-19.33N4} 
& \colhead{G211.16-19.33N5} 
& \colhead{G211.72-19.25S1} 
& \colhead{G212.10-19.15N1} \\
\colhead{(MHz)} & \colhead{(mK km s$^{-1}$)}  & \colhead{(mK km s$^{-1}$)}  & \colhead{(mK km s$^{-1}$)}  & \colhead{(mK km s$^{-1}$)}  & \colhead{(mK km s$^{-1}$)}  & \colhead{(mK km s$^{-1}$)}  & \colhead{(mK km s$^{-1}$)}  & \colhead{(mK km s$^{-1}$)}  & \colhead{(mK km s$^{-1}$)}  & \colhead{(mK km s$^{-1}$)}  & \colhead{(mK km s$^{-1}$)}  & \colhead{(mK km s$^{-1}$)}  & \colhead{(mK km s$^{-1}$)}  & \colhead{(mK km s$^{-1}$)}  & \colhead{(mK km s$^{-1}$)}  & \colhead{(mK km s$^{-1}$)}  & \colhead{(mK km s$^{-1}$)}  & \colhead{(mK km s$^{-1}$)}  & \colhead{(mK km s$^{-1}$)}  & \colhead{(mK km s$^{-1}$)}  & \colhead{(mK km s$^{-1}$)}  & \colhead{(mK km s$^{-1}$)}  & \colhead{(mK km s$^{-1}$)} 
}
\startdata
CH$_3$OH (48372.46) & 190$\pm$6 & 106$\pm$8 & 414$\pm$8 & 673$\pm$7 & 314$\pm$12 & \nodata & \nodata & 437$\pm$8 & 160$\pm$15 & 127$\pm$13 & 275$\pm$9 & 377$\pm$18 & 181$\pm$10 & 126$\pm$10 & 132$\pm$9 & 133$\pm$11 & 265$\pm$7 & \nodata & 100$\pm$14 & 210$\pm$12 & 128$\pm$10 & 298$\pm$16 & 437$\pm$16 \\
CH$_3$OH (48376.89) & \nodata & \nodata & \nodata & 56$\pm$5 & 101$\pm$12 & \nodata & \nodata & 82$\pm$7 & 127$\pm$13 & \nodata & \nodata & \nodata & \nodata & \nodata & \nodata & \nodata & \nodata & \nodata & \nodata & \nodata & \nodata & \nodata & \nodata \\
CH$_3$OH (36169.26) & 24$\pm$3 & 16$\pm$3 & 62$\pm$3 & 83$\pm$2 & 56$\pm$5 & \nodata & \nodata & 135$\pm$3 & 154$\pm$5 & \nodata & \nodata & 81$\pm$4 & 46$\pm$4 & 35$\pm$3 & 27$\pm$5 & 49$\pm$4 & 51$\pm$3 & \nodata & 26$\pm$4 & 23$\pm$3 & 27$\pm$4 & \nodata & 76$\pm$5 \\
CH$_3$OH (84521.21) & \nodata & \nodata & \nodata & 28$\pm$5 & \nodata & 18$\pm$4 & \nodata & 29$\pm$5 & 97$\pm$7 & \nodata & \nodata & \nodata & \nodata & \nodata & \nodata & 25$\pm$6 & \nodata & \nodata & \nodata & \nodata & \nodata & \nodata & \nodata \\
CH$_3$CHO (38512.08) & 18$\pm$3 & \nodata & 48$\pm$2 & 70$\pm$2 & 28$\pm$4 & \nodata & \nodata & 73$\pm$4 & \nodata & 62$\pm$5 & 47$\pm$3 & 81$\pm$5 & 60$\pm$5 & 35$\pm$3 & 23$\pm$3 & 73$\pm$5 & 54$\pm$3 & \nodata & \nodata & 29$\pm$3 & \nodata & 72$\pm$5 & 86$\pm$6 \\
CH$_3$CHO (38506.03) & 25$\pm$3 & \nodata & 43$\pm$3 & 77$\pm$5 & 23$\pm$5 & \nodata & \nodata & 53$\pm$4 & 23$\pm$4 & 86$\pm$6 & 52$\pm$3 & 70$\pm$5 & 60$\pm$6 & 35$\pm$3 & 17$\pm$3 & 66$\pm$5 & 54$\pm$3 & \nodata & 24$\pm$4 & 24$\pm$4 & \nodata & 39$\pm$5 & 57$\pm$6 \\
CH$_3$CHO (37464.20) & \nodata & 43$\pm$5 & 33$\pm$3 & 35$\pm$3 & \nodata & \nodata & \nodata & 34$\pm$3 & \nodata & \nodata & \nodata & \nodata & \nodata & \nodata & \nodata & \nodata & 18$\pm$3 & \nodata & \nodata & \nodata & \nodata & \nodata & \nodata \\
CH$_3$CHO (39594.29) & 13$\pm$3 & \nodata & 27$\pm$2 & 41$\pm$2 & \nodata & \nodata & \nodata & 29$\pm$3 & \nodata & \nodata & 30$\pm$3 & 35$\pm$4 & \nodata & 19$\pm$3 & \nodata & \nodata & \nodata & \nodata & \nodata & \nodata & \nodata & \nodata & 38$\pm$4 \\
CH$_3$CHO (37686.93) & 12$\pm$3 & \nodata & 27$\pm$3 & 28$\pm$3 & \nodata & \nodata & \nodata & 11$\pm$3 & \nodata & \nodata & 18$\pm$3 & \nodata & \nodata & \nodata & \nodata & 33$\pm$4 & \nodata & \nodata & \nodata & 30$\pm$4 & \nodata & \nodata & \nodata \\
CH$_3$CHO (39362.54) & \nodata & \nodata & 25$\pm$2 & 35$\pm$2 & 31$\pm$4 & \nodata & \nodata & 57$\pm$4 & \nodata & \nodata & \nodata & 28$\pm$3 & \nodata & 20$\pm$3 & \nodata & \nodata & 26$\pm$3 & \nodata & \nodata & \nodata & \nodata & 23$\pm$4 & 29$\pm$3 \\
CH$_3$CHO (76878.95) & \nodata & \nodata & 21$\pm$7 & 57$\pm$8 & \nodata & \nodata & \nodata & \nodata & \nodata & \nodata & \nodata & \nodata & \nodata & \nodata & 59$\pm$10 & \nodata & \nodata & \nodata & 54$\pm$7 & \nodata & \nodata & \nodata & 35$\pm$7 \\
CH$_3$CHO (76866.44) & \nodata & \nodata & \nodata & 55$\pm$7 & \nodata & \nodata & \nodata & 46$\pm$7 & 29$\pm$5 & \nodata & \nodata & \nodata & \nodata & \nodata & \nodata & \nodata & \nodata & \nodata & \nodata & \nodata & \nodata & \nodata & 39$\pm$6 \\
CH$_3$CHO (79099.31) & \nodata & \nodata & \nodata & \nodata & 24$\pm$5 & \nodata & \nodata & 56$\pm$7 & 46$\pm$7 & \nodata & \nodata & \nodata & \nodata & \nodata & \nodata & \nodata & \nodata & \nodata & \nodata & \nodata & \nodata & \nodata & \nodata \\
CS (48990.95) & 1343$\pm$9 & 1088$\pm$8 & 1767$\pm$8 & 2135$\pm$11 & 2637$\pm$18 & 519$\pm$12 & 319$\pm$13 & 2468$\pm$14 & 4646$\pm$39 & 3065$\pm$71 & 2906$\pm$72 & 3210$\pm$107 & 1868$\pm$20 & 2088$\pm$12 & 1499$\pm$33 & 2036$\pm$22 & 2693$\pm$25 & 882$\pm$34 & 1547$\pm$18 & 1509$\pm$18 & 1432$\pm$18 & 1220$\pm$16 & 1653$\pm$28 \\
$^{13}$CS (46247.56) & 19$\pm$4 & 40$\pm$5 & 103$\pm$4 & 126$\pm$4 & \nodata & \nodata & \nodata & 59$\pm$5 & 129$\pm$8 & \nodata & 70$\pm$6 & 70$\pm$6 & 22$\pm$5 & 67$\pm$5 & \nodata & \nodata & 62$\pm$4 & \nodata & \nodata & \nodata & \nodata & 80$\pm$7 & 116$\pm$9 \\
C$^{34}$S (48206.94) & 114$\pm$7 & 81$\pm$7 & 338$\pm$7 & 430$\pm$6 & 180$\pm$11 & \nodata & \nodata & 248$\pm$6 & 414$\pm$15 & 204$\pm$17 & \nodata & 299$\pm$15 & 217$\pm$12 & 256$\pm$14 & 55$\pm$8 & 206$\pm$14 & 258$\pm$6 & \nodata & 103$\pm$10 & 184$\pm$14 & 90$\pm$14 & 182$\pm$15 & 297$\pm$20  \\
\enddata
\end{splitdeluxetable*}

\clearpage

\begin{deluxetable*}{ccccccccc}
\caption{\label{tab:Ntot} Example of the estimated column density of CS in each source. The complete table exhibiting the column density of each molecular species in each source is provided in the format of machine readable table (MRT.) }
\tablehead{\colhead{Source} & \colhead{Formula} & \colhead{Flag} & \colhead{$N$(5K)} & \colhead{$\chi^2$(5K)} & \colhead{$N$(7K5)} & \colhead{$\chi^2$(7K5)} & \colhead{$N$(10K)} & \colhead{$\chi^2$(10K)}}
\startdata
G198.69-09.12N1 & CS & O & 8.8$^{+5.4}_{-3.3}\,\text{E+}{12}$ & 0.00E+00 & 1.1$^{+0.4}_{-0.3}\,\text{E+}{13}$ & 0.00E+00 & 1.3$^{+0.4}_{-0.3}\,\text{E+}{13}$ & 0.00E+00 \\
G198.69-09.12N2 & CS & O & 7.1$^{+4.4}_{-2.7}\,\text{E+}{12}$ & 0.00E+00 & 8.9$^{+3.4}_{-2.4}\,\text{E+}{12}$ & 0.00E+00 & 1.1$^{+0.3}_{-0.2}\,\text{E+}{13}$ & 0.00E+00 \\
G203.21-11.20E1 & CS & O & 1.2$^{+0.7}_{-0.4}\,\text{E+}{13}$ & 0.00E+00 & 1.4$^{+0.5}_{-0.4}\,\text{E+}{13}$ & 0.00E+00 & 1.8$^{+0.5}_{-0.4}\,\text{E+}{13}$ & 0.00E+00 \\
G203.21-11.20E2 & CS & O & 1.4$^{+0.8}_{-0.5}\,\text{E+}{13}$ & 0.00E+00 & 1.8$^{+0.7}_{-0.5}\,\text{E+}{13}$ & 0.00E+00 & 2.1$^{+0.6}_{-0.5}\,\text{E+}{13}$ & 0.00E+00 \\
G205.46-14.56M3 & CS & O & 1.7$^{+1.1}_{-0.7}\,\text{E+}{13}$ & 0.00E+00 & 2.2$^{+0.8}_{-0.6}\,\text{E+}{13}$ & 0.00E+00 & 2.6$^{+0.7}_{-0.6}\,\text{E+}{13}$ & 0.00E+00 \\
G206.21-16.17N & CS & O & 3.4$^{+2.2}_{-1.3}\,\text{E+}{12}$ & 0.00E+00 & 4.2$^{+1.7}_{-1.2}\,\text{E+}{12}$ & 0.00E+00 & 5.2$^{+1.5}_{-1.2}\,\text{E+}{12}$ & 0.00E+00 \\
G206.21-16.17S & CS & O & 2.1$^{+1.4}_{-0.8}\,\text{E+}{12}$ & 0.00E+00 & 2.6$^{+1.1}_{-0.8}\,\text{E+}{12}$ & 0.00E+00 & 3.2$^{+1.0}_{-0.8}\,\text{E+}{12}$ & 0.00E+00 \\
G207.36-19.82N4 & CS & O & 1.6$^{+1.0}_{-0.6}\,\text{E+}{13}$ & 0.00E+00 & 2.0$^{+0.8}_{-0.6}\,\text{E+}{13}$ & 0.00E+00 & 2.5$^{+0.7}_{-0.5}\,\text{E+}{13}$ & 0.00E+00 \\
G208.68-19.20N2 & CS & O & 3.0$^{+1.9}_{-1.2}\,\text{E+}{13}$ & 0.00E+00 & 3.8$^{+1.4}_{-1.1}\,\text{E+}{13}$ & 0.00E+00 & 4.6$^{+1.3}_{-1.0}\,\text{E+}{13}$ & 0.00E+00 \\
G209.29-19.65N1 & CS & O & 2.0$^{+1.3}_{-0.8}\,\text{E+}{13}$ & 0.00E+00 & 2.5$^{+1.0}_{-0.7}\,\text{E+}{13}$ & 0.00E+00 & 3.0$^{+0.9}_{-0.7}\,\text{E+}{13}$ & 0.00E+00 \\
G209.29-19.65S1 & CS & O & 1.9$^{+1.2}_{-0.7}\,\text{E+}{13}$ & 0.00E+00 & 2.4$^{+1.0}_{-0.7}\,\text{E+}{13}$ & 0.00E+00 & 2.9$^{+0.9}_{-0.7}\,\text{E+}{13}$ & 0.00E+00 \\
G209.29-19.65S2 & CS & O & 2.1$^{+1.4}_{-0.8}\,\text{E+}{13}$ & 0.00E+00 & 2.6$^{+1.1}_{-0.8}\,\text{E+}{13}$ & 0.00E+00 & 3.2$^{+1.0}_{-0.8}\,\text{E+}{13}$ & 0.00E+00 \\
G209.55-19.68N2 & CS & O & 1.2$^{+0.8}_{-0.5}\,\text{E+}{13}$ & 0.00E+00 & 1.5$^{+0.6}_{-0.4}\,\text{E+}{13}$ & 0.00E+00 & 1.9$^{+0.5}_{-0.4}\,\text{E+}{13}$ & 0.00E+00 \\
G209.77-19.40E3 & CS & O & 1.4$^{+0.8}_{-0.5}\,\text{E+}{13}$ & 0.00E+00 & 1.7$^{+0.6}_{-0.5}\,\text{E+}{13}$ & 0.00E+00 & 2.1$^{+0.6}_{-0.4}\,\text{E+}{13}$ & 0.00E+00 \\
G209.79-19.80W & CS & O & 9.8$^{+6.2}_{-3.8}\,\text{E+}{12}$ & 0.00E+00 & 1.2$^{+0.5}_{-0.3}\,\text{E+}{13}$ & 0.00E+00 & 1.5$^{+0.4}_{-0.3}\,\text{E+}{13}$ & 0.00E+00 \\
G209.94-19.52N & CS & O & 1.3$^{+0.8}_{-0.5}\,\text{E+}{13}$ & 0.00E+00 & 1.7$^{+0.6}_{-0.5}\,\text{E+}{13}$ & 0.00E+00 & 2.0$^{+0.6}_{-0.4}\,\text{E+}{13}$ & 0.00E+00 \\
G209.94-19.52S1 & CS & O & 1.8$^{+1.1}_{-0.7}\,\text{E+}{13}$ & 0.00E+00 & 2.2$^{+0.8}_{-0.6}\,\text{E+}{13}$ & 0.00E+00 & 2.7$^{+0.7}_{-0.6}\,\text{E+}{13}$ & 0.00E+00 \\
G210.37-19.53N & CS & O & 5.8$^{+3.8}_{-2.3}\,\text{E+}{12}$ & 0.00E+00 & 7.2$^{+3.0}_{-2.1}\,\text{E+}{12}$ & 0.00E+00 & 8.8$^{+2.8}_{-2.1}\,\text{E+}{12}$ & 0.00E+00 \\
G210.82-19.47N2 & CS & O & 1.0$^{+0.6}_{-0.4}\,\text{E+}{13}$ & 0.00E+00 & 1.3$^{+0.5}_{-0.4}\,\text{E+}{13}$ & 0.00E+00 & 1.5$^{+0.4}_{-0.3}\,\text{E+}{13}$ & 0.00E+00 \\
G211.16-19.33N4 & CS & O & 9.9$^{+6.1}_{-3.8}\,\text{E+}{12}$ & 0.00E+00 & 1.2$^{+0.5}_{-0.3}\,\text{E+}{13}$ & 0.00E+00 & 1.5$^{+0.4}_{-0.3}\,\text{E+}{13}$ & 0.00E+00 \\
G211.16-19.33N5 & CS & O & 9.4$^{+5.8}_{-3.6}\,\text{E+}{12}$ & 0.00E+00 & 1.2$^{+0.5}_{-0.3}\,\text{E+}{13}$ & 0.00E+00 & 1.4$^{+0.4}_{-0.3}\,\text{E+}{13}$ & 0.00E+00 \\
G211.72-19.25S1 & CS & O & 8.0$^{+5.0}_{-3.1}\,\text{E+}{12}$ & 0.00E+00 & 10.0$^{+3.9}_{-2.8}\,\text{E+}{12}$ & 0.00E+00 & 1.2$^{+0.3}_{-0.3}\,\text{E+}{13}$ & 0.00E+00 \\
G212.10-19.15N1 & CS & O & 1.1$^{+0.7}_{-0.4}\,\text{E+}{13}$ & 0.00E+00 & 1.4$^{+0.5}_{-0.4}\,\text{E+}{13}$ & 0.00E+00 & 1.6$^{+0.5}_{-0.4}\,\text{E+}{13}$ & 0.00E+00
\enddata
\tablecomments{Flag: "C," "M," or "W" if an assumed temperature of 5 K, 7.5 K, or 10 K, respectively, yields the lowest χ2 value; "O" if only one transition is detected, resulting in χ2 values of zero; "U" if the line is not detected and the column density is an upper limit.}
\end{deluxetable*}


\bibliography{REFERENCE.bib}{}

@article{astropy:2013,
Adsurl = {http://adsabs.harvard.edu/abs/2013A%26A...558A..33A},
Archiveprefix = {arXiv},
Author = {{Astropy Collaboration} and {Robitaille}, T.~P. and {Tollerud}, E.~J. and {Greenfield}, P. and {Droettboom}, M. and {Bray}, E. and {Aldcroft}, T. and {Davis}, M. and {Ginsburg}, A. and {Price-Whelan}, A.~M. and {Kerzendorf}, W.~E. and {Conley}, A. and {Crighton}, N. and {Barbary}, K. and {Muna}, D. and {Ferguson}, H. and {Grollier}, F. and {Parikh}, M.~M. and {Nair}, P.~H. and {Unther}, H.~M. and {Deil}, C. and {Woillez}, J. and {Conseil}, S. and {Kramer}, R. and {Turner}, J.~E.~H. and {Singer}, L. and {Fox}, R. and {Weaver}, B.~A. and {Zabalza}, V. and {Edwards}, Z.~I. and {Azalee Bostroem}, K. and {Burke}, D.~J. and {Casey}, A.~R. and {Crawford}, S.~M. and {Dencheva}, N. and {Ely}, J. and {Jenness}, T. and {Labrie}, K. and {Lim}, P.~L. and {Pierfederici}, F. and {Pontzen}, A. and {Ptak}, A. and {Refsdal}, B. and {Servillat}, M. and {Streicher}, O.},
Doi = {10.1051/0004-6361/201322068},
Eid = {A33},
Eprint = {1307.6212},
Journal = {\aap},
Month = oct,
Pages = {A33},
Primaryclass = {astro-ph.IM},
Title = {{Astropy: A community Python package for astronomy}},
Volume = 558,
Year = 2013}

@ARTICLE{astropy:2018,
       author = {{Astropy Collaboration} and {Price-Whelan}, A.~M. and
         {Sip{\H{o}}cz}, B.~M. and {G{\"u}nther}, H.~M. and {Lim}, P.~L. and
         {Crawford}, S.~M. and {Conseil}, S. and {Shupe}, D.~L. and
         {Craig}, M.~W. and {Dencheva}, N. and {Ginsburg}, A. and {Vand
        erPlas}, J.~T. and {Bradley}, L.~D. and {P{\'e}rez-Su{\'a}rez}, D. and
         {de Val-Borro}, M. and {Aldcroft}, T.~L. and {Cruz}, K.~L. and
         {Robitaille}, T.~P. and {Tollerud}, E.~J. and {Ardelean}, C. and
         {Babej}, T. and {Bach}, Y.~P. and {Bachetti}, M. and {Bakanov}, A.~V. and
         {Bamford}, S.~P. and {Barentsen}, G. and {Barmby}, P. and
         {Baumbach}, A. and {Berry}, K.~L. and {Biscani}, F. and {Boquien}, M. and
         {Bostroem}, K.~A. and {Bouma}, L.~G. and {Brammer}, G.~B. and
         {Bray}, E.~M. and {Breytenbach}, H. and {Buddelmeijer}, H. and
         {Burke}, D.~J. and {Calderone}, G. and {Cano Rodr{\'\i}guez}, J.~L. and
         {Cara}, M. and {Cardoso}, J.~V.~M. and {Cheedella}, S. and {Copin}, Y. and
         {Corrales}, L. and {Crichton}, D. and {D'Avella}, D. and {Deil}, C. and
         {Depagne}, {\'E}. and {Dietrich}, J.~P. and {Donath}, A. and
         {Droettboom}, M. and {Earl}, N. and {Erben}, T. and {Fabbro}, S. and
         {Ferreira}, L.~A. and {Finethy}, T. and {Fox}, R.~T. and
         {Garrison}, L.~H. and {Gibbons}, S.~L.~J. and {Goldstein}, D.~A. and
         {Gommers}, R. and {Greco}, J.~P. and {Greenfield}, P. and
         {Groener}, A.~M. and {Grollier}, F. and {Hagen}, A. and {Hirst}, P. and
         {Homeier}, D. and {Horton}, A.~J. and {Hosseinzadeh}, G. and {Hu}, L. and
         {Hunkeler}, J.~S. and {Ivezi{\'c}}, {\v{Z}}. and {Jain}, A. and
         {Jenness}, T. and {Kanarek}, G. and {Kendrew}, S. and {Kern}, N.~S. and
         {Kerzendorf}, W.~E. and {Khvalko}, A. and {King}, J. and {Kirkby}, D. and
         {Kulkarni}, A.~M. and {Kumar}, A. and {Lee}, A. and {Lenz}, D. and
         {Littlefair}, S.~P. and {Ma}, Z. and {Macleod}, D.~M. and
         {Mastropietro}, M. and {McCully}, C. and {Montagnac}, S. and
         {Morris}, B.~M. and {Mueller}, M. and {Mumford}, S.~J. and {Muna}, D. and
         {Murphy}, N.~A. and {Nelson}, S. and {Nguyen}, G.~H. and
         {Ninan}, J.~P. and {N{\"o}the}, M. and {Ogaz}, S. and {Oh}, S. and
         {Parejko}, J.~K. and {Parley}, N. and {Pascual}, S. and {Patil}, R. and
         {Patil}, A.~A. and {Plunkett}, A.~L. and {Prochaska}, J.~X. and
         {Rastogi}, T. and {Reddy Janga}, V. and {Sabater}, J. and
         {Sakurikar}, P. and {Seifert}, M. and {Sherbert}, L.~E. and
         {Sherwood-Taylor}, H. and {Shih}, A.~Y. and {Sick}, J. and
         {Silbiger}, M.~T. and {Singanamalla}, S. and {Singer}, L.~P. and
         {Sladen}, P.~H. and {Sooley}, K.~A. and {Sornarajah}, S. and
         {Streicher}, O. and {Teuben}, P. and {Thomas}, S.~W. and
         {Tremblay}, G.~R. and {Turner}, J.~E.~H. and {Terr{\'o}n}, V. and
         {van Kerkwijk}, M.~H. and {de la Vega}, A. and {Watkins}, L.~L. and
         {Weaver}, B.~A. and {Whitmore}, J.~B. and {Woillez}, J. and
         {Zabalza}, V. and {Astropy Contributors}},
        title = "{The Astropy Project: Building an Open-science Project and Status of the v2.0 Core Package}",
      journal = {\aj},
         year = 2018,
        month = sep,
       volume = {156},
       number = {3},
          eid = {123},
        pages = {123},
          doi = {10.3847/1538-3881/aabc4f},
archivePrefix = {arXiv},
       eprint = {1801.02634},
 primaryClass = {astro-ph.IM},
       adsurl = {https://ui.adsabs.harvard.edu/abs/2018AJ....156..123A}
}

@ARTICLE{astropy:2022,
       author = {{Astropy Collaboration} and {Price-Whelan}, Adrian M. and {Lim}, Pey Lian and {Earl}, Nicholas and {Starkman}, Nathaniel and {Bradley}, Larry and {Shupe}, David L. and {Patil}, Aarya A. and {Corrales}, Lia and {Brasseur}, C.~E. and {N{"o}the}, Maximilian and {Donath}, Axel and {Tollerud}, Erik and {Morris}, Brett M. and {Ginsburg}, Adam and {Vaher}, Eero and {Weaver}, Benjamin A. and {Tocknell}, James and {Jamieson}, William and {van Kerkwijk}, Marten H. and {Robitaille}, Thomas P. and {Merry}, Bruce and {Bachetti}, Matteo and {G{"u}nther}, H. Moritz and {Aldcroft}, Thomas L. and {Alvarado-Montes}, Jaime A. and {Archibald}, Anne M. and {B{'o}di}, Attila and {Bapat}, Shreyas and {Barentsen}, Geert and {Baz{'a}n}, Juanjo and {Biswas}, Manish and {Boquien}, M{'e}d{'e}ric and {Burke}, D.~J. and {Cara}, Daria and {Cara}, Mihai and {Conroy}, Kyle E. and {Conseil}, Simon and {Craig}, Matthew W. and {Cross}, Robert M. and {Cruz}, Kelle L. and {D'Eugenio}, Francesco and {Dencheva}, Nadia and {Devillepoix}, Hadrien A.~R. and {Dietrich}, J{"o}rg P. and {Eigenbrot}, Arthur Davis and {Erben}, Thomas and {Ferreira}, Leonardo and {Foreman-Mackey}, Daniel and {Fox}, Ryan and {Freij}, Nabil and {Garg}, Suyog and {Geda}, Robel and {Glattly}, Lauren and {Gondhalekar}, Yash and {Gordon}, Karl D. and {Grant}, David and {Greenfield}, Perry and {Groener}, Austen M. and {Guest}, Steve and {Gurovich}, Sebastian and {Handberg}, Rasmus and {Hart}, Akeem and {Hatfield-Dodds}, Zac and {Homeier}, Derek and {Hosseinzadeh}, Griffin and {Jenness}, Tim and {Jones}, Craig K. and {Joseph}, Prajwel and {Kalmbach}, J. Bryce and {Karamehmetoglu}, Emir and {Ka{l}uszy{'n}ski}, Miko{l}aj and {Kelley}, Michael S.~P. and {Kern}, Nicholas and {Kerzendorf}, Wolfgang E. and {Koch}, Eric W. and {Kulumani}, Shankar and {Lee}, Antony and {Ly}, Chun and {Ma}, Zhiyuan and {MacBride}, Conor and {Maljaars}, Jakob M. and {Muna}, Demitri and {Murphy}, N.~A. and {Norman}, Henrik and {O'Steen}, Richard and {Oman}, Kyle A. and {Pacifici}, Camilla and {Pascual}, Sergio and {Pascual-Granado}, J. and {Patil}, Rohit R. and {Perren}, Gabriel I. and {Pickering}, Timothy E. and {Rastogi}, Tanuj and {Roulston}, Benjamin R. and {Ryan}, Daniel F. and {Rykoff}, Eli S. and {Sabater}, Jose and {Sakurikar}, Parikshit and {Salgado}, Jes{'u}s and {Sanghi}, Aniket and {Saunders}, Nicholas and {Savchenko}, Volodymyr and {Schwardt}, Ludwig and {Seifert-Eckert}, Michael and {Shih}, Albert Y. and {Jain}, Anany Shrey and {Shukla}, Gyanendra and {Sick}, Jonathan and {Simpson}, Chris and {Singanamalla}, Sudheesh and {Singer}, Leo P. and {Singhal}, Jaladh and {Sinha}, Manodeep and {Sip{H{o}}cz}, Brigitta M. and {Spitler}, Lee R. and {Stansby}, David and {Streicher}, Ole and {{{S}}umak}, Jani and {Swinbank}, John D. and {Taranu}, Dan S. and {Tewary}, Nikita and {Tremblay}, Grant R. and {Val-Borro}, Miguel de and {Van Kooten}, Samuel J. and {Vasovi{'c}}, Zlatan and {Verma}, Shresth and {de Miranda Cardoso}, Jos{'e} Vin{'i}cius and {Williams}, Peter K.~G. and {Wilson}, Tom J. and {Winkel}, Benjamin and {Wood-Vasey}, W.~M. and {Xue}, Rui and {Yoachim}, Peter and {Zhang}, Chen and {Zonca}, Andrea and {Astropy Project Contributors}},
        title = "{The Astropy Project: Sustaining and Growing a Community-oriented Open-source Project and the Latest Major Release (v5.0) of the Core Package}",
      journal = {\apj},
         year = 2022,
        month = aug,
       volume = {935},
       number = {2},
          eid = {167},
        pages = {167},
          doi = {10.3847/1538-4357/ac7c74},
archivePrefix = {arXiv},
       eprint = {2206.14220},
 primaryClass = {astro-ph.IM},
       adsurl = {https://ui.adsabs.harvard.edu/abs/2022ApJ...935..167A}
}

@conference{casa:2007,
    Author = {{McMullin}, J. P. and Waters, B. and Schiebel, D. and Young, W. and Golap, K.},
    Title = {CASA Architecture and Applications},
    Page = "127",
    Volume = "376",
    Publisher = {ASP Conference Series},
    Organization = {San Francisco, Astronomical Society of the Pacific},
    Editor = {R. A. Shaw and F. Hill and D. J. Bell},
    Booktitle = {Astronomical Data Analysis Software and Systems XVI},
    Year = "2007"
}

@article{casa:2022,
   author = {{CASA Team} and Bean, Ben and Bhatnagar, Sanjay and Castro, Sandra and Donovan Meyer, Jennifer and Emonts, Bjorn and Garcia, Enrique and Garwood, Robert and Golap, Kumar and Gonzalez Villalba, Justo and Harris, Pamela and Hayashi, Yohei and Hoskins, Josh and Hsieh, Mingyu and Jagannathan, Preshanth and Kawasaki, Wataru and Keimpema, Aard and Kettenis, Mark and Lopez, Jorge and Marvil, Joshua and Masters, Joseph and McNichols, Andrew and Mehringer, David and Miel, Renaud and Moellenbrock, George and Montesino, Federico and Nakazato, Takeshi and Ott, Juergen and Petry, Dirk and Pokorny, Martin and Raba, Ryan and Rau, Urvashi and Schiebel, Darrell and Schweighart, Neal and Sekhar, Srikrishna and Shimada, Kazuhiko and Small, Des and Steeb, Jan-Willem and Sugimoto, Kanako and Suoranta, Ville and Tsutsumi, Takahiro and van Bemmel, Ilse M. and Verkouter, Marjolein and Wells, Akeem and Xiong, Wei and Szomoru, Arpad and Griffith, Morgan and Glendenning, Brian and Kern, Jeff},
   title = {CASA, the Common Astronomy Software Applications for Radio Astronomy},
   journal = {Publications of the Astronomical Society of the Pacific},
   volume = {134},
   pages = {114501},
   ISSN = {0004-6280},
   DOI = {10.1088/1538-3873/ac9642},
   url = {https://ui.adsabs.harvard.edu/abs/2022PASP..134k4501C},
   year = {2022},
   type = {Journal Article}
}

@article{2021Comrie_CARTA,
   author = {Comrie, Angus and Wang, Kuo-Song and Hsu, Shou-Chieh and Moraghan, Anthony and Harris, Pamela and Pang, Qi and Pińska, Adrianna and Chiang, Cheng-Chin and Simmonds, Rob and Chang, Tien-Hao and Jan, Hengtai and Lin, Ming-Yi},
   title = {CARTA: Cube Analysis and Rendering Tool for Astronomy},
   journal = {Astrophysics Source Code Library},
   pages = {ascl:2103.031},
   url = {https://ui.adsabs.harvard.edu/abs/2021ascl.soft03031C},
   year = {2021},
   type = {Journal Article}
}

@INPROCEEDINGS{2005GILDAS,
       author = {{Pety}, J.},
        title = "{Successes of and Challenges to GILDAS, a State-of-the-Art Radioastronomy Toolkit}",
    booktitle = {SF2A-2005: Semaine de l'Astrophysique Francaise},
         year = 2005,
       editor = {{Casoli}, F. and {Contini}, T. and {Hameury}, J.~M. and {Pagani}, L.},
        month = dec,
        pages = {721},
       adsurl = {https://ui.adsabs.harvard.edu/abs/2005sf2a.conf..721P}
}

@article{1985Matthews_TMC1_L134N_COM,
   author = {Matthews, H. E. and Friberg, P. and Irvine, W. M.},
   title = {The detection of acetaldehyde in cold dust clouds},
   journal = {The Astrophysical Journal},
   volume = {290},
   pages = {609-614},
   ISSN = {0004-637X},
   DOI = {10.1086/163018},
   url = {https://ui.adsabs.harvard.edu/abs/1985ApJ...290..609M},
   year = {1985},
   type = {Journal Article}
}

@article{1988JPL,
title = "SUBMILLIMETER, MILLIMETER, AND MICROWAVE SPECTRAL LINE CATALOG",
journal = "Journal of Quantitative Spectroscopy and Radiative Transfer",
volume = "60",
number = "5",
pages = "883 - 890",
year = "1998",
issn = "0022-4073",
doi = "https://doi.org/10.1016/S0022-4073(98)00091-0",
url = "http://www.sciencedirect.com/science/article/pii/S0022407398000910",
author = "H.M. Pickett and R.L. Poynter and E.A. Cohen and M.L. Delitsky and J.C. Pearson and H.S.P. Muller"
}

@article{1988Friberg_TMC1_L134N_B335_COM,
   author = {Friberg, P. and Madden, S. C. and Hjalmarson, A. and Irvine, W. M.},
   title = {Methanol in dark clouds},
   journal = {Astron Astrophys},
   volume = {195},
   pages = {281-9},
   ISSN = {0004-6361 (Print) 0004-6361 (Linking)},
   url = {https://www.ncbi.nlm.nih.gov/pubmed/11540080},
   year = {1988},
   type = {Journal Article}
}

@ARTICLE{1992Suzuki_chem_survey,
       author = {{Suzuki}, Hiroko and {Yamamoto}, Satoshi and {Ohishi}, Masatoshi and {Kaifu}, Norio and {Ishikawa}, Shin-Ichi and {Hirahara}, Yasuhiro and {Takano}, Shuro},
        title = "{A Survey of CCS, HC 3N, HC 5N, and NH 3 toward Dark Cloud Cores and Their Production Chemistry}",
      journal = {\apj},
         year = 1992,
        month = jun,
       volume = {392},
        pages = {551},
          doi = {10.1086/171456},
       adsurl = {https://ui.adsabs.harvard.edu/abs/1992ApJ...392..551S}
}

@article{1994Wilson_isotope,
   author = {Wilson, T. L. and Rood, R. T.},
   title = {Abundances in the Interstellar-Medium},
   journal = {Annual Review of Astronomy and Astrophysics},
   volume = {32},
   pages = {191-226},
   ISSN = {0066-4146},
   DOI = {10.1146/annurev.aa.32.090194.001203},
   year = {1994},
   type = {Journal Article}
}

@ARTICLE{1998Lucas_isotope,
       author = {{Lucas}, R. and {Liszt}, H.},
        title = "{Interstellar isotope ratios from mm-wave molecular absorption spectra}",
      journal = {\aap},
         year = 1998,
        month = sep,
       volume = {337},
        pages = {246-252},
       adsurl = {https://ui.adsabs.harvard.edu/abs/1998A&A...337..246L}
}

@ARTICLE{1999Goldsmith_popdiagram,
       author = {{Goldsmith}, Paul F. and {Langer}, William D.},
        title = "{Population Diagram Analysis of Molecular Line Emission}",
      journal = {\apj},
         year = 1999,
        month = may,
       volume = {517},
       number = {1},
        pages = {209-225},
          doi = {10.1086/307195},
       adsurl = {https://ui.adsabs.harvard.edu/abs/1999ApJ...517..209G}
}

@article{2001Dickens_TMC1,
   author = {Dickens, J. E. and Langer, W. D. and Velusamy, T.},
   title = {Small-Scale Abundance Variations in TMC-1: Dynamics and Hydrocarbon Chemistry},
   journal = {The Astrophysical Journal},
   volume = {558},
   pages = {693-701},
   ISSN = {0004-637X},
   DOI = {10.1086/322292},
   url = {https://ui.adsabs.harvard.edu/abs/2001ApJ...558..693D
https://iopscience.iop.org/article/10.1086/322292},
   year = {2001},
   type = {Journal Article}
}

@ARTICLE{2001Takakuwa_OPR_C3H2,
       author = {{Takakuwa}, Shigehisa and {Kawaguchi}, Kentarou and {Mikami}, Hitomi and {Saito}, Masao},
        title = "{The Ortho-to-Para Ratio and the Chemical Properties of C$_{3}$ H$_{2}$ in Dark Cloud Cores}",
      journal = {\pasj},
         year = 2001,
        month = apr,
       volume = {53},
       number = {2},
        pages = {251-257},
          doi = {10.1093/pasj/53.2.251},
       adsurl = {https://ui.adsabs.harvard.edu/abs/2001PASJ...53..251T}
}

@article{2001Turner_deuteration,
   author = {Turner, B. E.},
   title = {Deuterated molecules in translucent and dark clouds},
   journal = {Astrophysical Journal Supplement Series},
   volume = {136},
   number = {2},
   pages = {579-629},
   ISSN = {0067-0049},
   DOI = {10.1086/322536},
   year = {2001},
   type = {Journal Article}
}

@ARTICLE{2002Tafalla_starless_chem,
       author = {{Tafalla}, M. and {Myers}, P.~C. and {Caselli}, P. and {Walmsley}, C.~M. and {Comito}, C.},
        title = "{Systematic Molecular Differentiation in Starless Cores}",
      journal = {\apj},
         year = 2002,
        month = apr,
       volume = {569},
       number = {2},
        pages = {815-835},
          doi = {10.1086/339321},
archivePrefix = {arXiv},
       eprint = {astro-ph/0112487},
 primaryClass = {astro-ph},
       adsurl = {https://ui.adsabs.harvard.edu/abs/2002ApJ...569..815T}
}

@conference{2004Ceccarelli_HotCorino,
    Author = {C. Ceccarelli},
    Title = {The Hot Corinos of Solar Type Protostars},
    Page = 195,
    Volume = 323,
    Publisher = {ASP Conference Proceedings},
    Organization = {San Francisco, Astronomical Society of the Pacific},
    Editor = {D. Johnstone and F.C. Adams and D.N.C. Lin and D.A. Neufeld and E.C. Ostriker},
    Booktitle = {Star Formation in the Interstellar Medium: In Honor of David Hollenbach, Chris McKee and Frank Shu, ASP Conference Proceeding},
    Year = 2004
}

@article{2004Kaifu_TMC1_chem,
   author = {Kaifu, Norio and Ohishi, Masatoshi and Kawaguchi, Kentarou and Saito, Shuji and Yamamoto, Satoshi and Miyaji, Takeshi and Miyazawa, Keisuke and Ishikawa, Shin-Ichi and Noumaru, Chiaki and Harasawa, Sumiko and Okuda, Michiko and Suzuki, Hiroko},
   title = {A 8.8–50GHz Complete Spectral Line Survey toward TMC-1 I. Survey Data},
   journal = {Publications of the Astronomical Society of Japan},
   volume = {56},
   pages = {69-173},
   ISSN = {0004-6264},
   DOI = {10.1093/pasj/56.1.69},
   url = {https://ui.adsabs.harvard.edu/abs/2004PASJ...56...69K},
   year = {2004},
   type = {Journal Article}
}

@article{2005CDMS,
title = "The Cologne Database for Molecular Spectroscopy, CDMS: a useful tool for astronomers and spectroscopists",
journal = "Journal of Molecular Structure",
volume = "742",
number = "1",
pages = "215 - 227",
year = "2005",
note = "MOLECULAR SPECTROSCOPY AND STRUCTURE",
issn = "0022-2860",
doi = "https://doi.org/10.1016/j.molstruc.2005.01.027",
url = "http://www.sciencedirect.com/science/article/pii/S0022286005000888",
author = "Holger S.P. Müller and Frank Schlöder and Jürgen Stutzki and Gisbert Winnewisser"
}

@ARTICLE{2006Flower_OPR_H2,
       author = {{Flower}, D.~R. and {Pineau Des For{\^e}ts}, G. and {Walmsley}, C.~M.},
        title = "{The importance of the ortho:para H$_{2}$ ratio for the deuteration of molecules during pre-protostellar collapse}",
      journal = {\aap},
         year = 2006,
        month = apr,
       volume = {449},
       number = {2},
        pages = {621-629},
          doi = {10.1051/0004-6361:20054246},
archivePrefix = {arXiv},
       eprint = {astro-ph/0601429},
 primaryClass = {astro-ph},
       adsurl = {https://ui.adsabs.harvard.edu/abs/2006A&A...449..621F}
}

@ARTICLE{2006Tafalla_L1498_1517B,
       author = {{Tafalla}, M. and {Santiago-Garc{\'\i}a}, J. and {Myers}, P.~C. and {Caselli}, P. and {Walmsley}, C.~M. and {Crapsi}, A.},
        title = "{On the internal structure of starless cores. II. A molecular survey of L1498 and L1517B}",
      journal = {\aap},
         year = 2006,
        month = aug,
       volume = {455},
       number = {2},
        pages = {577-593},
          doi = {10.1051/0004-6361:20065311},
archivePrefix = {arXiv},
       eprint = {astro-ph/0605513},
 primaryClass = {astro-ph},
       adsurl = {https://ui.adsabs.harvard.edu/abs/2006A&A...455..577T}
}

@inproceedings{2007DiFrancesco_review,
    author = "Di Francesco, James and Evans, II, Neal J. and Caselli, Paola and Myers, Philip C. and Shirley, Yancy and Aikawa, Yuri and Tafalla, Mario",
    title = "{An observational perspective of low-mass dense cores. 1. internal physical and chemical properties}",
    booktitle = "{Protostars and Planets V}",
    eprint = "astro-ph/0602379",
    archivePrefix = "arXiv",
    month = "2",
    year = "2006"
}

@article{2007Garrod_reactive-desorption,
   author = {Garrod, R. T. and Wakelam, V. and Herbst, E.},
   title = {Non-thermal desorption from interstellar dust grains via exothermic surface reactions},
   journal = {Astronomy and Astrophysics},
   volume = {467},
   pages = {1103-1115},
   ISSN = {0004-6361},
   DOI = {10.1051/0004-6361:20066704},
   url = {https://ui.adsabs.harvard.edu/abs/2007A&A...467.1103G},
   year = {2007},
   type = {Journal Article}
}

@article{2008Arce_L1157-B1_COMs,
   author = {Arce, H. G. and Santiago-Garcia, J. and Jorgensen, J. K. and Tafalla, M. and Bachiller, R.},
   title = {Complex molecules in the L1157 molecular outflow},
   journal = {Astrophysical Journal Letters},
   volume = {681},
   number = {1},
   pages = {L21-L24},
   ISSN = {2041-8205},
   DOI = {10.1086/590110},
   year = {2008},
   type = {Journal Article}
}

@article{2008Garrod_grain-surface,
   author = {Garrod, R. T. and Weaver, S. L. W. and Herbst, E.},
   title = {Complex chemistry in star-forming regions: An expanded gas-grain warm-up chemical model},
   journal = {Astrophysical Journal},
   volume = {682},
   number = {1},
   pages = {283-302},
   ISSN = {0004-637x},
   DOI = {10.1086/588035},
   year = {2008},
   type = {Journal Article}
}

@article{2009Herbst_COM_review,
   author = {Herbst, E. and van Dishoeck, E. F.},
   title = {Complex Organic Interstellar Molecules},
   journal = {Annual Review of Astronomy and Astrophysics, Vol 47},
   volume = {47},
   pages = {427-480},
   ISSN = {0066-4146},
   DOI = {10.1146/annurev-astro-082708-101654},
   year = {2009},
   type = {Journal Article}
}

@article{2012Bacmann_L1689B_COM,
   author = {Bacmann, A. and Taquet, V. and Faure, A. and Kahane, C. and Ceccarelli, C.},
   title = {Detection of complex organic molecules in a prestellar core: a new challenge for astrochemical models},
   journal = {Astronomy \& Astrophysics},
   volume = {541},
   pages = {L12},
   ISSN = {0004-6361},
   DOI = {10.1051/0004-6361/201219207},
   url = {https://www.aanda.org/articles/aa/pdf/2012/05/aa19207-12.pdf},
   year = {2012},
   type = {Journal Article}
}

@article{2013Bizzocchi_L1544_14N15N,
   author = {Bizzocchi, L. and Caselli, P. and Leonardo, E. and Dore, L.},
   title = {Detection of <SUP>15</SUP>NNH<SUP>+</SUP> in L1544: non-LTE modelling of dyazenilium hyperfine line emission and accurate <SUP>14</SUP>N/<SUP>15</SUP>N values},
   journal = {Astronomy and Astrophysics},
   volume = {555},
   pages = {A109},
   ISSN = {0004-6361},
   DOI = {10.1051/0004-6361/201321276},
   url = {https://ui.adsabs.harvard.edu/abs/2013A&A...555A.109B},
   year = {2013},
   type = {Journal Article}
}

@ARTICLE{2013Daniel_Barnard1_N,
       author = {{Daniel}, F. and {G{\'e}rin}, M. and {Roueff}, E. and {Cernicharo}, J. and {Marcelino}, N. and {Lique}, F. and {Lis}, D.~C. and {Teyssier}, D. and {Biver}, N. and {Bockel{\'e}e-Morvan}, D.},
        title = "{Nitrogen isotopic ratios in Barnard 1: a consistent study of the N$_{2}$H$^{+}$, NH$_{3}$, CN, HCN, and HNC isotopologues}",
      journal = {\aap},
         year = 2013,
        month = dec,
       volume = {560},
          eid = {A3},
        pages = {A3},
          doi = {10.1051/0004-6361/201321939},
archivePrefix = {arXiv},
       eprint = {1309.5782},
 primaryClass = {astro-ph.GA},
       adsurl = {https://ui.adsabs.harvard.edu/abs/2013A&A...560A...3D}
}

@ARTICLE{2013Pagani_prestellar_ortho-H2,
       author = {{Pagani}, L. and {Lesaffre}, P. and {Jorfi}, M. and {Honvault}, P. and {Gonz{\'a}lez-Lezana}, T. and {Faure}, A.},
        title = "{Ortho-H$_{2}$ and the age of prestellar cores}",
      journal = {\aap},
         year = 2013,
        month = mar,
       volume = {551},
          eid = {A38},
        pages = {A38},
          doi = {10.1051/0004-6361/201117161},
       adsurl = {https://ui.adsabs.harvard.edu/abs/2013A&A...551A..38P}
}

@article{2013Sakai_review_carbon-chain,
   author = {Sakai, Nami and Yamamoto, Satoshi},
   title = {Warm Carbon-Chain Chemistry},
   journal = {Chemical Reviews},
   volume = {113},
   pages = {8981-9015},
   DOI = {10.1021/cr4001308},
   url = {https://ui.adsabs.harvard.edu/abs/2013ChRv..113.8981S
https://pubs.acs.org/doi/pdf/10.1021/cr4001308?ref=article_openPDF},
   year = {2013},
   type = {Journal Article}
}

@article{2013Vasyunin_reactive-desorption,
   author = {Vasyunin, A. I. and Herbst, Eric},
   title = {Reactive Desorption and Radiative Association as Possible Drivers of Complex Molecule Formation in the Cold Interstellar Medium},
   journal = {The Astrophysical Journal},
   volume = {769},
   pages = {34},
   ISSN = {0004-637X},
   DOI = {10.1088/0004-637X/769/1/34},
   url = {https://ui.adsabs.harvard.edu/abs/2013ApJ...769...34V
https://iopscience.iop.org/article/10.1088/0004-637X/769/1/34},
   year = {2013},
   type = {Journal Article}
}

@article{2013Zapata_IRAS16293B_depth,
   author = {Zapata, Luis A. and Loinard, Laurent and Rodríguez, Luis F. and Hernández-Hernández, Vicente and Takahashi, Satoko and Trejo, Alfonso and Parise, Bérengère},
   title = {ALMA 690 GHz Observations of IRAS 16293-2422B: Infall in a Highly Optically Thick Disk},
   journal = {The Astrophysical Journal},
   volume = {764},
   pages = {L14},
   ISSN = {0004-637X},
   DOI = {10.1088/2041-8205/764/1/L14},
   url = {https://ui.adsabs.harvard.edu/abs/2013ApJ...764L..14Z},
   year = {2013},
   type = {Journal Article}
}

@ARTICLE{2014Loison_HCN-HNC,
       author = {{Loison}, Jean-Christophe and {Wakelam}, Valentine and {Hickson}, Kevin M.},
        title = "{The interstellar gas-phase chemistry of HCN and HNC}",
      journal = {\mnras},
         year = 2014,
        month = sep,
       volume = {443},
       number = {1},
        pages = {398-410},
          doi = {10.1093/mnras/stu1089},
archivePrefix = {arXiv},
       eprint = {1406.1696},
 primaryClass = {astro-ph.GA},
       adsurl = {https://ui.adsabs.harvard.edu/abs/2014MNRAS.443..398L}
}

@article{2014Tatematsu_OrionA,
   author = {Tatematsu, Ken'ichi and Ohashi, Satoshi and Umemoto, Tomofumi and Lee, Jeong-Eun and Hirota, Tomoya and Yamamoto, Satoshi and Choi, Minho and Kandori, Ryo and Mizuno, Norikazu},
   title = {Chemical variation in molecular cloud cores in the Orion A cloud. II},
   journal = {Publications of the Astronomical Society of Japan},
   volume = {66},
   pages = {16},
   ISSN = {0004-6264},
   DOI = {10.1093/pasj/pst016},
   url = {https://ui.adsabs.harvard.edu/abs/2014PASJ...66...16T},
   year = {2014},
   type = {Journal Article}
}

@ARTICLE{2015Kong_deuteration,
       author = {{Kong}, Shuo and {Caselli}, Paola and {Tan}, Jonathan C. and {Wakelam}, Valentine and {Sipil{\"a}}, Olli},
        title = "{The Deuterium Fractionation Timescale in Dense Cloud Cores: A Parameter Space Exploration}",
      journal = {\apj},
         year = 2015,
        month = may,
       volume = {804},
       number = {2},
          eid = {98},
        pages = {98},
          doi = {10.1088/0004-637X/804/2/98},
archivePrefix = {arXiv},
       eprint = {1312.0971},
 primaryClass = {astro-ph.SR},
       adsurl = {https://ui.adsabs.harvard.edu/abs/2015ApJ...804...98K}
}

@article{2015Mangum_colDens,
   author = {Mangum, Jeffrey G. and Shirley, Yancy L.},
   title = {How to Calculate Molecular Column Density},
   journal = {Publications of the Astronomical Society of the Pacific},
   volume = {127},
   pages = {266},
   ISSN = {0004-6280},
   DOI = {10.1086/680323},
   url = {https://ui.adsabs.harvard.edu/abs/2015PASP..127..266M},
   year = {2015},
   type = {Journal Article}
}

@article{2015Soma_TMC1,
   author = {Soma, Tatsuya and Sakai, Nami and Watanabe, Yoshimasa and Yamamoto, Satoshi},
   title = {Methanol in the Starless Core, Taurus Molecular Cloud-1},
   journal = {The Astrophysical Journal},
   volume = {802},
   pages = {74},
   ISSN = {0004-637X},
   DOI = {10.1088/0004-637x/802/2/74},
   url = {https://ui.adsabs.harvard.edu/abs/2015ApJ...802...74S
https://iopscience.iop.org/article/10.1088/0004-637X/802/2/74},
   year = {2015},
   type = {Journal Article}
}

@ARTICLE{2016Daniel_16293E_14N15N,
       author = {{Daniel}, F. and {Faure}, A. and {Pagani}, L. and {Lique}, F. and {G{\'e}rin}, M. and {Lis}, D. and {Hily-Blant}, P. and {Bacmann}, A. and {Roueff}, E.},
        title = "{N$_{2}$H$^{+}$ and N$^{15}$NH$^{+}$ toward the prestellar core 16293E in L1689N}",
      journal = {\aap},
         year = 2016,
        month = jul,
       volume = {592},
          eid = {A45},
        pages = {A45},
          doi = {10.1051/0004-6361/201628192},
archivePrefix = {arXiv},
       eprint = {1603.07128},
 primaryClass = {astro-ph.GA},
       adsurl = {https://ui.adsabs.harvard.edu/abs/2016A&A...592A..45D}
}

@article{2016Planck_PGCC,
   author = {Planck, Collaboration and Ade, P. A. R. and Aghanim, N. and Arnaud, M. and Ashdown, M. and Aumont, J. and Baccigalupi, C. and Banday, A. J. and Barreiro, R. B. and Bartolo, N. and Battaner, E. and Benabed, K. and Benoît, A. and Benoit-Lévy, A. and Bernard, J. P. and Bersanelli, M. and Bielewicz, P. and Bonaldi, A. and Bonavera, L. and Bond, J. R. and Borrill, J. and Bouchet, F. R. and Boulanger, F. and Bucher, M. and Burigana, C. and Butler, R. C. and Calabrese, E. and Catalano, A. and Chamballu, A. and Chiang, H. C. and Christensen, P. R. and Clements, D. L. and Colombi, S. and Colombo, L. P. L. and Combet, C. and Couchot, F. and Coulais, A. and Crill, B. P. and Curto, A. and Cuttaia, F. and Danese, L. and Davies, R. D. and Davis, R. J. and de Bernardis, P. and de Rosa, A. and de Zotti, G. and Delabrouille, J. and Désert, F. X. and Dickinson, C. and Diego, J. M. and Dole, H. and Donzelli, S. and Doré, O. and Douspis, M. and Ducout, A. and Dupac, X. and Efstathiou, G. and Elsner, F. and Enßlin, T. A. and Eriksen, H. K. and Falgarone, E. and Fergusson, J. and Finelli, F. and Forni, O. and Frailis, M. and Fraisse, A. A. and Franceschi, E. and Frejsel, A. and Galeotta, S. and Galli, S. and Ganga, K. and Giard, M. and Giraud-Héraud, Y. and Gjerløw, E. and González-Nuevo, J. and Górski, K. M. and Gratton, S. and Gregorio, A. and Gruppuso, A. and Gudmundsson, J. E. and Hansen, F. K. and Hanson, D. and Harrison, D. L. and Helou, G. and Henrot-Versillé, S. and Hernández-Monteagudo, C. and Herranz, D. and Hildebrandt, S. R. and Hivon, E. and Hobson, M. and Holmes, W. A. and Hornstrup, A. and Hovest, W. and Huffenberger, K. M. and Hurier, G. and Jaffe, A. H. and Jaffe, T. R. and Jones, W. C. and Juvela, M. and Keihänen, E. and others },
   title = {Planck 2015 results. XXVIII. The Planck Catalogue of Galactic cold clumps},
   journal = {Astronomy and Astrophysics},
   volume = {594},
   pages = {A28},
   ISSN = {0004-6361},
   DOI = {10.1051/0004-6361/201525819},
   url = {https://ui.adsabs.harvard.edu/abs/2016A&A...594A..28P},
   year = {2016},
   type = {Journal Article}
}

@ARTICLE{2016Sipila_L1544_C3H2,
       author = {{Sipil{\"a}}, O. and {Spezzano}, S. and {Caselli}, P.},
        title = "{Understanding the C$_{3}$H$_{2}$ cyclic-to-linear ratio in L1544}",
      journal = {\aap},
         year = 2016,
        month = jun,
       volume = {591},
          eid = {L1},
        pages = {L1},
          doi = {10.1051/0004-6361/201628689},
archivePrefix = {arXiv},
       eprint = {1605.07379},
 primaryClass = {astro-ph.GA},
       adsurl = {https://ui.adsabs.harvard.edu/abs/2016A&A...591L...1S}
}

@ARTICLE{2016Spezzano_L1544_TMC1C_C3H2,
       author = {{Spezzano}, S. and {Gupta}, H. and {Br{\"u}nken}, S. and {Gottlieb}, C.~A. and {Caselli}, P. and {Menten}, K.~M. and {M{\"u}ller}, H.~S.~P. and {Bizzocchi}, L. and {Schilke}, P. and {McCarthy}, M.~C. and {Schlemmer}, S.},
        title = "{A study of the C$_{3}$H$_{2}$ isomers and isotopologues: first interstellar detection of HDCCC}",
      journal = {\aap},
         year = 2016,
        month = feb,
       volume = {586},
          eid = {A110},
        pages = {A110},
          doi = {10.1051/0004-6361/201527460},
archivePrefix = {arXiv},
       eprint = {1511.04878},
 primaryClass = {astro-ph.GA},
       adsurl = {https://ui.adsabs.harvard.edu/abs/2016A&A...586A.110S}
}

@ARTICLE{2017Dapra_L1498_methanolAE,
       author = {{Dapr{\`a}}, M. and {Henkel}, C. and {Levshakov}, S.~A. and {Menten}, K.~M. and {Muller}, S. and {Bethlem}, H.~L. and {Leurini}, S. and {Lapinov}, A.~V. and {Ubachs}, W.},
        title = "{Testing the variability of the proton-to-electron mass ratio from observations of methanol in the dark cloud core L1498}",
      journal = {\mnras},
         year = 2017,
        month = dec,
       volume = {472},
       number = {4},
        pages = {4434-4443},
          doi = {10.1093/mnras/stx2308},
archivePrefix = {arXiv},
       eprint = {1709.03103},
 primaryClass = {astro-ph.CO},
       adsurl = {https://ui.adsabs.harvard.edu/abs/2017MNRAS.472.4434D}
}

@ARTICLE{2017Loison_C3H_C3H2,
       author = {{Loison}, Jean-Christophe and {Ag{\'u}ndez}, Marcelino and {Wakelam}, Valentine and {Roueff}, Evelyne and {Gratier}, Pierre and {Marcelino}, N{\'u}ria and {Reyes}, Dianailys Nu{\~n}ez and {Cernicharo}, Jos{\'e} and {Gerin}, Maryvonne},
        title = "{The interstellar chemistry of C$_{3}$H and C$_{3}$H$_{2}$ isomers}",
      journal = {\mnras},
         year = 2017,
        month = oct,
       volume = {470},
       number = {4},
        pages = {4075-4088},
          doi = {10.1093/mnras/stx1265},
archivePrefix = {arXiv},
       eprint = {1707.07926},
 primaryClass = {astro-ph.GA},
       adsurl = {https://ui.adsabs.harvard.edu/abs/2017MNRAS.470.4075L}
}

@article{2017Vasyunin_reactive-desorption,
   author = {Vasyunin, A. I. and Caselli, P. and Dulieu, F. and Jiménez-Serra, I.},
   title = {Formation of Complex Molecules in Prestellar Cores: A Multilayer Approach},
   journal = {The Astrophysical Journal},
   volume = {842},
   pages = {33},
   ISSN = {0004-637X},
   DOI = {10.3847/1538-4357/aa72ec},
   url = {https://ui.adsabs.harvard.edu/abs/2017ApJ...842...33V
https://iopscience.iop.org/article/10.3847/1538-4357/aa72ec/pdf},
   year = {2017},
   type = {Journal Article}
}

@article{2018Chuang_reactive-desorption,
   author = {Chuang, K. J. and Fedoseev, G. and Qasim, D. and Ioppolo, S. and van Dishoeck, E. F. and Linnartz, H.},
   title = {Reactive Desorption of CO Hydrogenation Products under Cold Pre-stellar Core Conditions},
   journal = {The Astrophysical Journal},
   volume = {853},
   pages = {102},
   ISSN = {0004-637X},
   DOI = {10.3847/1538-4357/aaa24e},
   url = {https://ui.adsabs.harvard.edu/abs/2018ApJ...853..102C},
   year = {2018},
   type = {Journal Article}
}

@article{2018Yi_PGCC_SCUBA2_II,
   author = {Yi, Hee-Weon and Lee, Jeong-Eun and Liu, Tie and Kim, Kee-Tae and Choi, Minho and Eden, David and Evans, Neal J., II and Di Francesco, James and Fuller, Gary and Hirano, N. and Juvela, Mika and Kang, Sung-ju and Kim, Gwanjeong and Koch, Patrick M. and Lee, Chang Won and Li, Di and Liu, H. Y. B. and Liu, Hong-Li and Liu, Sheng-Yuan and Rawlings, Mark G. and Ristorcelli, I. and Sanhueza, Patrico and Soam, Archana and Tatematsu, Ken'ichi and Thompson, Mark and Toth, L. V. and Wang, Ke and White, Glenn J. and Wu, Yuefang and Yang, Yao-Lun and Collaboration, Jcmt Large Program “SCOPE” and Collaboration, Trao Key Science Program “TOP”},
   title = {Planck Cold Clumps in the λ Orionis Complex. II. Environmental Effects on Core Formation},
   journal = {\apjs},
   volume = {236},
   pages = {51},
   ISSN = {0067-0049},
   DOI = {10.3847/1538-4365/aac2e0},
   url = {https://ui.adsabs.harvard.edu/abs/2018ApJS..236...51Y},
   year = {2018},
   type = {Journal Article}
}

@ARTICLE{2018Redaelli_survey_14N15N,
       author = {{Redaelli}, E. and {Bizzocchi}, L. and {Caselli}, P. and {Harju}, J. and {Chac{\'o}n-Tanarro}, A. and {Leonardo}, E. and {Dore}, L.},
        title = "{$^{14}$N/$^{15}$N ratio measurements in prestellar cores with N$_{2}$H$^{+}$: new evidence of $^{15}$N-antifractionation}",
      journal = {\aap},
         year = 2018,
        month = sep,
       volume = {617},
          eid = {A7},
        pages = {A7},
          doi = {10.1051/0004-6361/201833065},
archivePrefix = {arXiv},
       eprint = {1806.01088},
 primaryClass = {astro-ph.GA},
       adsurl = {https://ui.adsabs.harvard.edu/abs/2018A&A...617A...7R}
}

@article{2019Jacobsen_L483_COM,
   author = {Jacobsen, Steffen K. and Jørgensen, Jes K. and Di Francesco, James and Evans, Neal J. and Choi, Minho and Lee, Jeong-Eun},
   title = {Organic chemistry in the innermost, infalling envelope of the Class 0 protostar L483},
   journal = {Astronomy and Astrophysics},
   volume = {629},
   pages = {A29},
   ISSN = {0004-6361},
   DOI = {10.1051/0004-6361/201833214},
   url = {https://ui.adsabs.harvard.edu/abs/2019A&A...629A..29J},
   year = {2019},
   type = {Journal Article}
}

@article{2019Lee_HH212,
   author = {Lee, Chin-Fei and Codella, Claudio and Li, Zhi-Yun and Liu, Sheng-Yuan},
   title = {First Abundance Measurement of Organic Molecules in the Atmosphere of HH 212 Protostellar Disk},
   journal = {The Astrophysical Journal},
   volume = {876},
   pages = {63},
   ISSN = {0004-637X},
   DOI = {10.3847/1538-4357/ab15db},
   url = {https://ui.adsabs.harvard.edu/abs/2019ApJ...876...63L},
   year = {2019},
   type = {Journal Article}
}

@article{2020Hsu_ALMASOP,
   author = {Hsu, Shih-Ying and Liu, Sheng-Yuan and Liu, Tie and Sahu, Dipen and Hirano, Naomi and Lee, Chin-Fei and Tatematsu, Ken’ichi and Kim, Gwanjeong and Juvela, Mika and Sanhueza, Patricio and He, Jinhua and Johnstone, Doug and Qin, Sheng-Li and Bronfman, Leonardo and Chen, Huei-Ru Vivien and Dutta, Somnath and Eden, David J. and Jhan, Kai-Syun and Kim, Kee-Tae and Kuan, Yi-Jehng and Kwon, Woojin and Lee, Chang Won and Lee, Jeong-Eun and Moraghan, Anthony and Rawlings, M. G. and Shang, Hsien and Soam, Archana and Thompson, M. A. and Traficante, Alessio and Wu, Yuefang and Yang, Yao-Lun and Zhang, Qizhou},
   title = {ALMA Survey of Orion Planck Galactic Cold Clumps (ALMASOP). I. Detection of New Hot Corinos with the ACA},
   journal = {\apj},
   volume = {898},
   number = {2},
   pages = {107},
   ISSN = {1538-4357},
   DOI = {10.3847/1538-4357/ab9f3a},
   url = {http://dx.doi.org/10.3847/1538-4357/ab9f3a},
   year = {2020},
   type = {Journal Article}
}

@article{2020Dartois_CR-sputtering,
   author = {Dartois, E. and Chabot, M. and Bacmann, A. and Boduch, P. and Domaracka, A. and Rothard, H.},
   title = {Non-thermal desorption of complex organic molecules. Cosmic-ray sputtering of CH<SUB>3</SUB>OH embedded in CO<SUB>2</SUB> ice},
   journal = {Astronomy and Astrophysics},
   volume = {634},
   pages = {A103},
   ISSN = {0004-6361},
   DOI = {10.1051/0004-6361/201936934},
   url = {https://ui.adsabs.harvard.edu/abs/2020A&A...634A.103D
https://www.aanda.org/articles/aa/pdf/2020/02/aa36934-19.pdf},
   year = {2020},
   type = {Journal Article}
}

@article{2020Dutta_ALMASOP,
   author = {Dutta, Somnath and Lee, Chin-Fei and Liu, Tie and Hirano, Naomi and Liu, Sheng-Yuan and Tatematsu, Ken'ichi and Kim, Kee-Tae and Shang, Hsien and Sahu, Dipen and Kim, Gwanjeong and Moraghan, Anthony and Jhan, Kai-Syun and Hsu, Shih-Ying and Evans, Neal J. and Johnstone, Doug and Ward-Thompson, Derek and Kuan, Yi-Jehng and Lee, Chang Won and Lee, Jeong-Eun and Traficante, Alessio and Juvela, Mika and Vastel, Charlotte and Zhang, Qizhou and Sanhueza, Patricio and Soam, Archana and Kwon, Woojin and Bronfman, Leonardo and Eden, David and Goldsmith, Paul F. and He, Jinhua and Wu, Yuefang and Pelkonen, Veli-Matti and Qin, Sheng-Li and Li, Shanghuo and Li, Di},
   title = {ALMA Survey of Orion Planck Galactic Cold Clumps (ALMASOP). II. Survey Overview: A First Look at 1.3 mm Continuum Maps and Molecular Outflows},
   journal = {\apjs},
   volume = {251},
   pages = {20},
   ISSN = {0067-0049},
   DOI = {10.3847/1538-4365/abba26},
   url = {https://ui.adsabs.harvard.edu/abs/2020ApJS..251...20D},
   year = {2020},
   type = {Journal Article}
}

@article{2020Harju_H-MM1_CH3OH,
   author = {Harju, Jorma and Pineda, Jaime E. and Vasyunin, Anton I. and Caselli, Paola and Offner, Stella S. R. and Goodman, Alyssa A. and Juvela, Mika and Sipilä, Olli and Faure, Alexandre and Le Gal, Romane and Hily-Blant, Pierre and Alves, João and Bizzocchi, Luca and Burkert, Andreas and Chen, Hope and Friesen, Rachel K. and Güsten, Rolf and Myers, Philip C. and Punanova, Anna and Rist, Claire and Rosolowsky, Erik and Schlemmer, Stephan and Shirley, Yancy and Spezzano, Silvia and Vastel, Charlotte and Wiesenfeld, Laurent},
   title = {Efficient Methanol Production on the Dark Side of a Prestellar Core},
   journal = {The Astrophysical Journal},
   volume = {895},
   pages = {101},
   ISSN = {0004-637X},
   DOI = {10.3847/1538-4357/ab8f93},
   url = {https://ui.adsabs.harvard.edu/abs/2020ApJ...895..101H
https://iopscience.iop.org/article/10.3847/1538-4357/ab8f93/pdf},
   year = {2020},
   type = {Journal Article}
}

@ARTICLE{2020Jin_COM_reactive_desorption,
       author = {{Jin}, Mihwa and {Garrod}, Robin T.},
        title = "{Formation of Complex Organic Molecules in Cold Interstellar Environments through Nondiffusive Grain-surface and Ice-mantle Chemistry}",
      journal = {\apjs},
         year = 2020,
        month = aug,
       volume = {249},
       number = {2},
          eid = {26},
        pages = {26},
          doi = {10.3847/1538-4365/ab9ec8},
archivePrefix = {arXiv},
       eprint = {2006.11127},
 primaryClass = {astro-ph.GA},
       adsurl = {https://ui.adsabs.harvard.edu/abs/2020ApJS..249...26J}
}

@article{2020Kim_PGCC_45m,
   author = {Kim, Gwanjeong and Tatematsu, Ken'ichi and Liu, Tie and Yi, Hee-Weon and He, Jinhua and Hirano, Naomi and Liu, Sheng-Yuan and Choi, Minho and Sanhueza, Patricio and Tóth, L. Viktor and Evans, I. I. Neal J. and Feng, Siyi and Juvela, Mika and Kim, Kee-Tae and Vastel, Charlotte and Lee, Jeong-Eun and Nguyễn Lu'o'ng, Quang and Kang, Miju and Ristorcelli, Isabelle and Fehér, Orsolya and Wu, Yuefang and Ohashi, Satoshi and Wang, Ke and Kandori, Ryo and Hirota, Tomoya and Sakai, Takeshi and Lu, Xing and Thompson, Mark A. and Fuller, Gary A. and Li, Di and Shinnaga, Hiroko and Kim, Jungha},
   title = {Molecular Cloud Cores with a High Deuterium Fraction: Nobeyama Single-pointing Survey},
   journal = {The Astrophysical Journal Supplement Series},
   volume = {249},
   pages = {33},
   ISSN = {0067-0049},
   DOI = {10.3847/1538-4365/aba746},
   url = {https://ui.adsabs.harvard.edu/abs/2020ApJS..249...33K
https://iopscience.iop.org/article/10.3847/1538-4365/aba746/pdf},
   year = {2020},
   type = {Journal Article}
}

@ARTICLE{2020Kim_starless_CS,
       author = {{Kim}, Shinyoung and {Lee}, Chang Won and {Gopinathan}, Maheswar and {Tafalla}, Mario and {Sohn}, Jungjoo and {Kim}, Gwanjeong and {Kim}, Mi-Ryang and {Soam}, Archana and {Myers}, Philip C.},
        title = "{CS Depletion in Prestellar Cores}",
      journal = {\apj},
         year = 2020,
        month = mar,
       volume = {891},
       number = {2},
          eid = {169},
        pages = {169},
          doi = {10.3847/1538-4357/ab774d},
archivePrefix = {arXiv},
       eprint = {2002.06857},
 primaryClass = {astro-ph.GA},
       adsurl = {https://ui.adsabs.harvard.edu/abs/2020ApJ...891..169K}
}

@article{2020Manigand_IRAS16293A_COM,
   author = {Manigand, S. and Jørgensen, J. K. and Calcutt, H. and Müller, H. S. P. and Ligterink, N. F. W. and Coutens, A. and Drozdovskaya, M. N. and van Dishoeck, E. F. and Wampfler, S. F.},
   title = {The ALMA-PILS survey: inventory of complex organic molecules towards IRAS 16293-2422 A},
   journal = {Astronomy and Astrophysics},
   volume = {635},
   pages = {A48},
   ISSN = {0004-6361},
   DOI = {10.1051/0004-6361/201936299},
   url = {https://ui.adsabs.harvard.edu/abs/2020A&A...635A..48M},
   year = {2020},
   type = {Journal Article}
}

@ARTICLE{2020McGuire_TMC-1_GOTHAM,
       author = {{McGuire}, Brett A. and {Burkhardt}, Andrew M. and {Loomis}, Ryan A. and {Shingledecker}, Christopher N. and {Kelvin Lee}, Kin Long and {Charnley}, Steven B. and {Cordiner}, Martin A. and {Herbst}, Eric and {Kalenskii}, Sergei and {Momjian}, Emmanuel and {Willis}, Eric R. and {Xue}, Ci and {Remijan}, Anthony J. and {McCarthy}, Michael C.},
        title = "{Early Science from GOTHAM: Project Overview, Methods, and the Detection of Interstellar Propargyl Cyanide (HCCCH$_{2}$CN) in TMC-1}",
      journal = {\apjl},
         year = 2020,
        month = sep,
       volume = {900},
       number = {1},
          eid = {L10},
        pages = {L10},
          doi = {10.3847/2041-8213/aba632},
archivePrefix = {arXiv},
       eprint = {2008.12349},
 primaryClass = {astro-ph.GA},
       adsurl = {https://ui.adsabs.harvard.edu/abs/2020ApJ...900L..10M}
}

@article{2020Scibelli_COM_Taurus,
   author = {Scibelli, Samantha and Shirley, Yancy},
   title = {Prevalence of Complex Organic Molecules in Starless and Prestellar Cores within the Taurus Molecular Cloud},
   journal = {The Astrophysical Journal},
   volume = {891},
   pages = {73},
   ISSN = {0004-637X},
   DOI = {10.3847/1538-4357/ab7375},
   url = {https://ui.adsabs.harvard.edu/abs/2020ApJ...891...73S
https://iopscience.iop.org/article/10.3847/1538-4357/ab7375/pdf},
   year = {2020},
   type = {Journal Article}
}

@ARTICLE{2020Vazart_acetaldehyde_gas-phase,
       author = {{Vazart}, Fanny and {Ceccarelli}, Cecilia and {Balucani}, Nadia and {Bianchi}, Eleonora and {Skouteris}, Dimitrios},
        title = "{Gas-phase formation of acetaldehyde: review and new theoretical computations}",
      journal = {\mnras},
         year = 2020,
        month = dec,
       volume = {499},
       number = {4},
        pages = {5547-5561},
          doi = {10.1093/mnras/staa3060},
archivePrefix = {arXiv},
       eprint = {2010.02718},
 primaryClass = {astro-ph.GA},
       adsurl = {https://ui.adsabs.harvard.edu/abs/2020MNRAS.499.5547V}
}

@ARTICLE{2021Rodriguez-Baras_GEMS_statistics,
       author = {{Rodr{\'\i}guez-Baras}, M. and {Fuente}, A. and {Rivi{\'e}re-Marichalar}, P. and {Navarro-Almaida}, D. and {Caselli}, P. and {Gerin}, M. and {Kramer}, C. and {Roueff}, E. and {Wakelam}, V. and {Esplugues}, G. and {Garc{\'\i}a-Burillo}, S. and {Le Gal}, R. and {Spezzano}, S. and {Alonso-Albi}, T. and {Bachiller}, R. and {Cazaux}, S. and {Commercon}, B. and {Goicoechea}, J.~R. and {Loison}, J.~C. and {Trevi{\~n}o-Morales}, S.~P. and {Roncero}, O. and {Jim{\'e}nez-Serra}, I. and {Laas}, J. and {Hacar}, A. and {Kirk}, J. and {Lattanzi}, V. and {Mart{\'\i}n-Dom{\'e}nech}, R. and {Mu{\~n}oz-Caro}, G. and {Pineda}, J.~E. and {Tercero}, B. and {Ward-Thompson}, D. and {Tafalla}, M. and {Marcelino}, N. and {Malinen}, J. and {Friesen}, R. and {Giuliano}, B.~M.},
        title = "{Gas phase Elemental abundances in Molecular cloudS (GEMS). IV. Observational results and statistical trends}",
      journal = {\aap},
         year = 2021,
        month = apr,
       volume = {648},
          eid = {A120},
        pages = {A120},
          doi = {10.1051/0004-6361/202040112},
archivePrefix = {arXiv},
       eprint = {2102.13153},
 primaryClass = {astro-ph.GA},
       adsurl = {https://ui.adsabs.harvard.edu/abs/2021A&A...648A.120R}
}

@article{2021Sahu_ALMASOP_presstellar,
   author = {Sahu, Dipen and Liu, Sheng-Yuan and Liu, Tie and Evans, I. I. Neal J. and Hirano, Naomi and Tatematsu, Ken'ichi and Lee, Chin-Fei and Kim, Kee-Tae and Dutta, Somnath and Alina, Dana and Bronfman, Leonardo and Cunningham, Maria and Eden, David J. and Garay, Guido and Goldsmith, Paul F. and He, Jinhua and Hsu, Shih-Ying and Jhan, Kai-Syun and Johnstone, Doug and Juvela, Mika and Kim, Gwanjeong and Kuan, Yi-Jehng and Kwon, Woojin and Lee, Chang Won and Lee, Jeong-Eun and Li, Di and Li, Pak Shing and Li, Shanghuo and Luo, Qiu-Yi and Montillaud, Julien and Moraghan, Anthony and Pelkonen, Veli-Matti and Qin, Sheng-Li and Ristorcelli, Isabelle and Sanhueza, Patricio and Shang, Hsien and Shen, Zhi-Qiang and Soam, Archana and Wu, Yuefang and Zhang, Qizhou and Zhou, Jianjun},
   title = {ALMA Survey of Orion Planck Galactic Cold Clumps (ALMASOP): Detection of Extremely High-density Compact Structure of Prestellar Cores and Multiple Substructures Within},
   journal = {The Astrophysical Journal},
   volume = {907},
   pages = {L15},
   ISSN = {0004-637X},
   DOI = {10.3847/2041-8213/abd3aa},
   url = {https://ui.adsabs.harvard.edu/abs/2021ApJ...907L..15S
https://iopscience.iop.org/article/10.3847/2041-8213/abd3aa},
   year = {2021},
   type = {Journal Article}
}

@article{2021Tatematsu_SCOPE,
   author = {Tatematsu, Ken'ichi and Kim, Gwanjeong and Liu, Tie and Evans, Neal J., II and Yi, Hee-Weon and Lee, Jeong-Eun and Wu, Yuefang and Hirano, Naomi and Liu, Sheng-Yuan and Dutta, Somnath and Sahu, Dipen and Sanhueza, Patricio and Kim, Kee-Tae and Juvela, Mika and Tóth, L. Viktor and Fehér, Orsolya and He, Jinhua and Ge, Jixing and Feng, Siyi and Choi, Minho and Kang, Miju and Thompson, Mark A. and Fuller, Gary A. and Li, Di and Ristorcelli, Isabelle and Wang, Ke and di Francesco, James and Eden, David and Ohashi, Satoshi and Kandori, Ryo and Vastel, Charlotte and Hirota, Tomoya and Sakai, Takeshi and Lu, Xing and Nguyên Lu'O'Ng, Quang and Shinnaga, Hiroko and Kim, Jungha and Scope, Collaboration and Jcmt Large, Program},
   title = {Molecular Cloud Cores with High Deuterium Fractions: Nobeyama Mapping Survey},
   journal = {The Astrophysical Journal Supplement Series},
   volume = {256},
   pages = {25},
   ISSN = {0067-0049},
   DOI = {10.3847/1538-4365/ac0978},
   url = {https://ui.adsabs.harvard.edu/abs/2021ApJS..256...25T},
   year = {2021},
   type = {Journal Article}
}

@article{2021Yang_PEACHES,
   author = {Yang, Yao-Lun and Sakai, Nami and Zhang, Yichen and Murillo, Nadia M. and Zhang, Ziwei E. and Higuchi, Aya E. and Zeng, Shaoshan and López-Sepulcre, Ana and Yamamoto, Satoshi and Lefloch, Bertrand and Bouvier, Mathilde and Ceccarelli, Cecilia and Hirota, Tomoya and Imai, Muneaki and Oya, Yoko and Sakai, Takeshi and Watanabe, Yoshimasa},
   title = {The Perseus ALMA Chemistry Survey (PEACHES). I. The Complex Organic Molecules in Perseus Embedded Protostars},
   journal = {\apj},
   volume = {910},
   number = {1},
   pages = {20},
   ISSN = {0004-637X 1538-4357},
   DOI = {10.3847/1538-4357/abdfd6},
   url = {https://iopscience.iop.org/article/10.3847/1538-4357/abdfd6/pdf},
   year = {2021},
   type = {Journal Article}
}

@article{2021Yi_PGCC_chem,
   author = {Yi, Hee-Weon and Lee, Jeong-Eun and Kim, Kee-Tae and Liu, Tie and Lim, Beomdu and Tatematsu, Ken’ichi},
   title = {Planck Cold Clumps in the λ Orionis Complex. III. A Chemical Probe of Stellar Feedback on Cores in the λ Orionis Cloud},
   journal = {The Astrophysical Journal Supplement Series},
   volume = {254},
   number = {1},
   pages = {14},
   ISSN = {0067-0049 1538-4365},
   DOI = {10.3847/1538-4365/abec4a},
   url = {https://ui.adsabs.harvard.edu/abs/2021ApJS..254...14Y},
   year = {2021},
   type = {Journal Article}
}

@article{2021Wakelam_CR_sputtering,
   author = {Wakelam, V. and Dartois, E. and Chabot, M. and Spezzano, S. and Navarro-Almaida, D. and Loison, J. C. and Fuente, A.},
   title = {Efficiency of non-thermal desorptions in cold-core conditions. Testing the sputtering of grain mantles induced by cosmic rays},
   journal = {Astronomy and Astrophysics},
   volume = {652},
   pages = {A63},
   ISSN = {0004-6361},
   DOI = {10.1051/0004-6361/202039855},
   url = {https://ui.adsabs.harvard.edu/abs/2021A&A...652A..63W
https://www.aanda.org/articles/aa/pdf/2021/08/aa39855-20.pdf},
   year = {2021},
   type = {Journal Article}
}

@article{2022Hsu_ALMASOP,
   author = {Hsu, Shih-Ying and Liu, Sheng-Yuan and Liu, Tie and Sahu, Dipen and Lee, Chin-Fei and Tatematsu, Kenichi and Kim, Kee-Tae and Hirano, Naomi and Yang, Yao-Lun and Johnstone, Doug and Liu, Hongli and Juvela, Mika and Bronfman, Leonardo and Chen, Huei-Ru Vivien and Dutta, Somnath and Eden, David J. and Jhan, Kai-Syun and Kuan, Yi-Jehng and Lee, Chang Won and Lee, Jeong-Eun and Li, Shanghuo and Liu, Chun-Fan and Qin, Sheng-Li and Sanhueza, Patricio and Shang, Hsien and Soam, Archana and Traficante, Alessio and Zhou, Jianjun},
   title = {ALMA Survey of Orion Planck Galactic Cold Clumps (ALMASOP): A Hot Corino Survey toward Protostellar Cores in the Orion Cloud},
   journal = {\apj},
   volume = {927},
   pages = {218},
   ISSN = {0004-637X},
   DOI = {10.3847/1538-4357/ac49e0},
   url = {https://ui.adsabs.harvard.edu/abs/2022ApJ...927..218H},
   year = {2022},
   type = {Journal Article}
}

@INPROCEEDINGS{2022Cernicharo_TMC-1_QUIJOTE,
       author = {{Cernicharo}, Jos{\'e} and {Ag{\'u}ndez}, Marcelino and {Cabezas}, Carlos and {Marcelino}, Nuria and {Tercero}, Bel{\'e}n and {Pardo}, Juan Ram{\'o}n and {Fuentetaja}, Ra{\'u}l and {de Vicente}, Pablo},
        title = "{The QUIJOTE1 line survey of TMC-1}",
    booktitle = {European Physical Journal Web of Conferences},
         year = 2022,
       series = {European Physical Journal Web of Conferences},
       volume = {265},
        month = apr,
    publisher = {EDP},
          eid = {00041},
        pages = {00041},
          doi = {10.1051/epjconf/202226500041},
       adsurl = {https://ui.adsabs.harvard.edu/abs/2022EPJWC.26500041C}
}

@ARTICLE{2022Esplugues_GEMS_sulfur,
       author = {{Esplugues}, G. and {Fuente}, A. and {Navarro-Almaida}, D. and {Rodr{\'\i}guez-Baras}, M. and {Majumdar}, L. and {Caselli}, P. and {Wakelam}, V. and {Roueff}, E. and {Bachiller}, R. and {Spezzano}, S. and {Rivi{\`e}re-Marichalar}, P. and {Mart{\'\i}n-Dom{\'e}nech}, R. and {Mu{\~n}oz Caro}, G.~M.},
        title = "{Gas phase Elemental abundances in Molecular cloudS (GEMS). VI. A sulphur journey across star-forming regions: study of thioformaldehyde emission}",
      journal = {\aap},
         year = 2022,
        month = jun,
       volume = {662},
          eid = {A52},
        pages = {A52},
          doi = {10.1051/0004-6361/202142936},
archivePrefix = {arXiv},
       eprint = {2204.02645},
 primaryClass = {astro-ph.SR},
       adsurl = {https://ui.adsabs.harvard.edu/abs/2022A&A...662A..52E}
}

@ARTICLE{2022Ha_SFR_turbulence,
       author = {{Ha}, Trung and {Li}, Yuan and {Kounkel}, Marina and {Xu}, Siyao and {Li}, Hui and {Zheng}, Yong},
        title = "{Turbulence in Milky Way Star-forming Regions Traced by Young Stars and Gas}",
      journal = {\apj},
         year = 2022,
        month = jul,
       volume = {934},
       number = {1},
          eid = {7},
        pages = {7},
          doi = {10.3847/1538-4357/ac76bf},
archivePrefix = {arXiv},
       eprint = {2205.00012},
 primaryClass = {astro-ph.GA},
       adsurl = {https://ui.adsabs.harvard.edu/abs/2022ApJ...934....7H}
}

@article{2022Garrod_non-diffusive,
   author = {Garrod, Robin T. and Jin, Mihwa and Matis, Kayla A. and Jones, Dylan and Willis, Eric R. and Herbst, Eric},
   title = {Formation of Complex Organic Molecules in Hot Molecular Cores through Nondiffusive Grain-surface and Ice-mantle Chemistry},
   journal = {The Astrophysical Journal Supplement Series},
   volume = {259},
   pages = {1},
   ISSN = {0067-0049},
   DOI = {10.3847/1538-4365/ac3131},
   url = {https://ui.adsabs.harvard.edu/abs/2022ApJS..259....1G
https://iopscience.iop.org/article/10.3847/1538-4365/ac3131/pdf},
   year = {2022},
   type = {Journal Article}
}

@article{2022Kalvans_desorption,
   author = {Kalvāns, Juris and Silsbee, Kedron},
   title = {Icy molecule desorption in interstellar grain collisions},
   journal = {Monthly Notices of the Royal Astronomical Society},
   volume = {515},
   pages = {785-794},
   ISSN = {0035-8711},
   DOI = {10.1093/mnras/stac1792},
   url = {https://ui.adsabs.harvard.edu/abs/2022MNRAS.515..785K},
   year = {2022},
   type = {Journal Article}
}

@article{2022Lin_L1544_CH3OH_HNCO,
   author = {Lin, Y. and Spezzano, S. and Sipilä, O. and Vasyunin, A. and Caselli, P.},
   title = {Multiline observations of CH<SUB>3</SUB>OH, c-C<SUB>3</SUB>H<SUB>2</SUB>, and HNCO toward L1544. Dissecting the core structure with chemical differentiation},
   journal = {Astronomy and Astrophysics},
   volume = {665},
   pages = {A131},
   ISSN = {0004-6361},
   DOI = {10.1051/0004-6361/202243657},
   url = {https://ui.adsabs.harvard.edu/abs/2022A&A...665A.131L},
   year = {2022},
   type = {Journal Article}
}

@ARTICLE{2022Tatematsu_ALMASOP_inward,
       author = {{Tatematsu}, Ken'ichi and {Yeh}, You-Ting and {Hirano}, Naomi and {Liu}, Sheng-Yuan and {Liu}, Tie and {Dutta}, Somnath and {Sahu}, Dipen and {Evans}, II, Neal J. and {Juvela}, Mika and {Yi}, Hee-Weon and {Lee}, Jeong-Eun and {Sanhueza}, Patricio and {Li}, Shanghuo and {Eden}, David and {Kim}, Gwanjeong and {Lee}, Chin-Fei and {Wu}, Yuefang and {Kim}, Kee-Tae and {T{\'o}th}, L. Viktor and {Choi}, Minho and {Kang}, Miju and {Thompson}, Mark A. and {Fuller}, Gary A. and {Li}, Di and {Wang}, Ke and {Sakai}, Takeshi and {Kandori}, Ryo and {Hsu}, Shih-Ying and {Chiong}, Chau-Ching and {''Almasop'' Collaboration}},
        title = "{Nobeyama Survey of Inward Motions toward Cores in Orion Identified by SCUBA-2}",
      journal = {\apj},
         year = 2022,
        month = may,
       volume = {931},
       number = {1},
          eid = {33},
        pages = {33},
          doi = {10.3847/1538-4357/ac6100},
archivePrefix = {arXiv},
       eprint = {2203.12885},
 primaryClass = {astro-ph.GA},
       adsurl = {https://ui.adsabs.harvard.edu/abs/2022ApJ...931...33T}
}

@ARTICLE{2022Xia_SFR_UV,
       author = {{Xia}, Jifeng and {Tang}, Ningyu and {Zhi}, Qijun and {Jiao}, Sihan and {Xie}, Jinjin and {Fuller}, Gary A. and {Goldsmith}, Paul F. and {Li}, Di},
        title = "{The Distribution of UV Radiation Field in the Molecular Clouds of Gould Belt}",
      journal = {Research in Astronomy and Astrophysics},
         year = 2022,
        month = aug,
       volume = {22},
       number = {8},
          eid = {085017},
        pages = {085017},
          doi = {10.1088/1674-4527/ac784e},
archivePrefix = {arXiv},
       eprint = {2208.05250},
 primaryClass = {astro-ph.GA},
       adsurl = {https://ui.adsabs.harvard.edu/abs/2022RAA....22h5017X}
}

@article{2023Hsu_ALMASOP,
   author = {Hsu, Shih-Ying and Liu, Sheng-Yuan and Johnstone, Doug and Liu, Tie and Bronfman, Leonardo and Chen, Huei-Ru Vivien and Dutta, Somnath and Eden, David J. and Evans, Neal J., II and Hirano, Naomi and Juvela, Mika and Kuan, Yi-Jehng and Kwon, Woojin and Lee, Chin-Fei and Lee, Chang Won and Lee, Jeong-Eun and Li, Shanghuo and Liu, Chun-Fan and Liu, Xunchuan and Luo, Qiuyi and Qin, Sheng-Li and Rawlings, Mark G. and Sahu, Dipen and Sanhueza, Patricio and Shang, Hsien and Tatematsu, Ken'ichi and Yang, Yao-Lun},
   title = {ALMA Survey of Orion Planck Galactic Cold Clumps (ALMASOP): The Warm-envelope Origin of Hot Corinos},
   journal = {The Astrophysical Journal},
   volume = {956},
   pages = {120},
   ISSN = {0004-637X},
   DOI = {10.3847/1538-4357/acefcf},
   url = {https://ui.adsabs.harvard.edu/abs/2023ApJ...956..120H},
   year = {2023},
   type = {Journal Article}
}

@ARTICLE{2023Navarro-Almaida_chem,
       author = {{Navarro-Almaida}, D. and {Bop}, C.~T. and {Lique}, F. and {Esplugues}, G. and {Rodr{\'\i}guez-Baras}, M. and {Kramer}, C. and {Romero}, C.~E. and {Fuente}, A. and {Caselli}, P. and {Rivi{\`e}re-Marichalar}, P. and {Kirk}, J.~M. and {Chac{\'o}n-Tanarro}, A. and {Roueff}, E. and {Mroczkowski}, T. and {Bhandarkar}, T. and {Devlin}, M. and {Dicker}, S. and {Lowe}, I. and {Mason}, B. and {Sarazin}, C.~L. and {Sievers}, J.},
        title = "{Linking the dust and chemical evolution: Taurus and Perseus. New collisional rates for HCN, HNC, and their C, N, and H isotopologues}",
      journal = {\aap},
         year = 2023,
        month = feb,
       volume = {670},
          eid = {A110},
        pages = {A110},
          doi = {10.1051/0004-6361/202245000},
archivePrefix = {arXiv},
       eprint = {2212.07675},
 primaryClass = {astro-ph.GA},
       adsurl = {https://ui.adsabs.harvard.edu/abs/2023A&A...670A.110N}
}

@article{2023Sahu_ALMASOP_density,
   author = {Sahu, Dipen and Liu, Sheng-Yuan and Johnstone, Doug and Liu, Tie and Evans, I. I. Neal J. and Hirano, Naomi and Tatematsu, Ken'ichi and Di Francesco, James and Lee, Chin-Fei and Kim, Kee-Tae and Dutta, Somnath and Hsu, Shih-Ying and Li, Shanghuo and Luo, Qiu-Yi and Sanhueza, Patricio and Shang, Hsien and Traficante, Alessio and Juvela, Mika and Lee, Chang Won and Eden, David J. and Goldsmith, Paul F. and Bronfman, Leonardo and Kwon, Woojin and Lee, Jeong-Eun and Kuan, Yi-Jehng and Ristorcelli, Isabelle},
   title = {ALMA Survey of Orion Planck Galactic Cold Clumps (ALMASOP): Density Structure of Centrally Concentrated Prestellar Cores from Multiscale Observations},
   journal = {The Astrophysical Journal},
   volume = {945},
   pages = {156},
   ISSN = {0004-637X},
   DOI = {10.3847/1538-4357/acbc26},
   url = {https://ui.adsabs.harvard.edu/abs/2023ApJ...945..156S
https://iopscience.iop.org/article/10.3847/1538-4357/acbc26/pdf},
   year = {2023},
   type = {Journal Article}
}

@article{2024Hsu_HOPS87,
   author = {Hsu, Shih-Ying and Lee, Chin-Fei and Liu, Sheng-Yuan and Johnstone, Doug and Liu, Tie and Takahashi, Satoko and Bronfman, Leonardo and Chen, Huei-Ru Vivien and Dutta, Somnath and Eden, David J. and Evans, Neal J., II and Hirano, Naomi and Juvela, Mika and Kuan, Yi-Jehng and Kwon, Woojin and Lee, Chang Won and Lee, Jeong-Eun and Li, Shanghuo and Liu, Chun-Fan and Liu, Xunchuan and Luo, Qiuyi and Qin, Sheng-Li and Sahu, Dipen and Sanhueza, Patricio and Shang, Hsien and Tatematsu, Kenichi and Yang, Yao-Lun},
   title = {ALMASOP. The Localized and Chemically Rich Features near the Bases of the Protostellar Jet in HOPS 87},
   journal = {The Astrophysical Journal},
   volume = {976},
   pages = {29},
   ISSN = {0004-637X},
   DOI = {10.3847/1538-4357/ad7e25},
   url = {https://ui.adsabs.harvard.edu/abs/2024ApJ...976...29H
https://iopscience.iop.org/article/10.3847/1538-4357/ad7e25},
   year = {2024},
   type = {Journal Article}
}

@article{2024DeSimone_IRAS4A2_COMdiskwind,
   author = {De Simone, M. and Podio, L. and Chahine, L. and Codella, C. and Chandler, C. J. and Ceccarelli, C. and Lopez-Sepulcre, A. and Loinard, L. and Svoboda, B. and Sakai, N. and Johnstone, D. and Menard, F. and Aikawa, Y. and Bouvier, M. and Sabatini, G. and Miotello, A. and Vastel, C. and Cuello, N. and Bianchi, E. and Caselli, P. and Caux, E. and Hanawa, T. and Herbst, E. and Segura-Cox, D. and Zhang, Z. and Yamamoto, S.},
   title = {FAUST XV. A disk wind mapped by CH$_3$OH and SiO in the inner 300 au of the NGC 1333 IRAS 4A2 protostar},
   journal = {arXiv e-prints},
   pages = {arXiv:2404.19690},
   DOI = {10.48550/arXiv.2404.19690},
   url = {https://ui.adsabs.harvard.edu/abs/2024arXiv240419690D},
   year = {2024},
   type = {Journal Article}
}

@ARTICLE{2024Ichimura_SFR_12C13C,
       author = {{Ichimura}, Ryota and {Nomura}, Hideko and {Furuya}, Kenji},
        title = "{Carbon Isotope Fractionation of Complex Organic Molecules in Star-forming Cores}",
      journal = {\apj},
         year = 2024,
        month = jul,
       volume = {970},
       number = {1},
          eid = {55},
        pages = {55},
          doi = {10.3847/1538-4357/ad47ba},
archivePrefix = {arXiv},
       eprint = {2406.02961},
 primaryClass = {astro-ph.SR},
       adsurl = {https://ui.adsabs.harvard.edu/abs/2024ApJ...970...55I}
}

@ARTICLE{2024Jensen_HCN_isotope,
       author = {{Jensen}, S.~S. and {Spezzano}, S. and {Caselli}, P. and {Sipil{\"a}}, O. and {Redaelli}, E. and {Giers}, K. and {Ferrer Asensio}, J.},
        title = "{Fractionation in young cores: Direct determinations of nitrogen and carbon fractionation in HCN}",
      journal = {\aap},
         year = 2024,
        month = may,
       volume = {685},
          eid = {A149},
        pages = {A149},
          doi = {10.1051/0004-6361/202449344},
archivePrefix = {arXiv},
       eprint = {2403.04408},
 primaryClass = {astro-ph.GA},
       adsurl = {https://ui.adsabs.harvard.edu/abs/2024A&A...685A.149J}
}

@article{2024Scibelli_COM_Perseus,
   author = {Scibelli, Samantha and Shirley, Yancy and Megías, Andrés and Jiménez-Serra, Izaskun},
   title = {Survey of Complex Organic Molecules in Starless and Prestellar Cores in the Perseus Molecular Cloud},
   journal = {arXiv e-prints},
   pages = {arXiv:2408.11613},
   DOI = {10.48550/arXiv.2408.11613},
   url = {https://ui.adsabs.harvard.edu/abs/2024arXiv240811613S
http://arxiv.org/pdf/2408.11613},
   year = {2024},
   type = {Journal Article}
}

@article{2025Hsu_ALMASOP_starless,
       author = {{Hsu}, Shih-Ying and {Liu}, Sheng-Yuan and {Liu}, Xunchuan and {Li}, Pak Shing and {Liu}, Tie and {Sahu}, Dipen and {Tatematsu}, Ken'ichi and {Li}, Shanghuo and {Hirano}, Naomi and {Lee}, Chin-Fei and {Lin}, Sheng-Jun},
        title = "{ALMASOP: Detection of Turbulence-induced Mass Assembly Shocks in Starless Cores}",
      journal = {\apjl},
         year = 2025,
        month = may,
       volume = {984},
       number = {2},
          eid = {L58},
        pages = {L58},
          doi = {10.3847/2041-8213/adcd6a},
archivePrefix = {arXiv},
       eprint = {2504.11776},
 primaryClass = {astro-ph.SR},
       adsurl = {https://ui.adsabs.harvard.edu/abs/2025ApJ...984L..58H}
}

@article{2025Hsu_G192,
       author = {{Hsu}, Shih-Ying and {Lee}, Chin-Fei and {Johnstone}, Doug and {Liu}, Sheng-Yuan and {Liu}, Tie and {Bronfman}, Leonardo and {Chen}, Huei-Ru Vivien and {Dutta}, Somnath and {Eden}, David J. and {Hirano}, Naomi and {Juvela}, Mika and {Kim}, Kee-Tae and {Kuan}, Yi-Jehng and {Kwon}, Woojin and {Lee}, Chang Won and {Lee}, Jeong-Eun and {Li}, Shanghuo and {Lin}, Sheng-Jun and {Liu}, Chun-Fan and {Liu}, Xunchuan and {L{\'o}pez-V{\'a}zquez}, J.~A. and {Luo}, Qiuyi and {Rawlings}, Mark G. and {Sahu}, Dipen and {Sanhueza}, Patricio and {Shang}, Hsien and {Tatematsu}, Ken'ichi and {Yang}, Yao-Lun},
        title = "{ALMASOP. A Rotating Feature Rich in Complex Organic Molecules in a Protostellar Core}",
      journal = {\apj},
         year = 2025,
        month = aug,
       volume = {989},
       number = {1},
          eid = {56},
        pages = {56},
          doi = {10.3847/1538-4357/ade7fc},
       adsurl = {https://ui.adsabs.harvard.edu/abs/2025ApJ...989...56H}
}

@ARTICLE{2025Borshcheva_COM_reactive_desorption,
       author = {{Borshcheva}, Katerina and {Fedoseev}, Gleb and {Punanova}, Anna F. and {Caselli}, Paola and {Jim{\'e}nez-Serra}, Izaskun and {Vasyunin}, Anton I.},
        title = "{Formation of Complex Organic Molecules in Prestellar Cores: The Role of Nondiffusive Grain Chemistry}",
      journal = {\apj},
         year = 2025,
        month = sep,
       volume = {990},
       number = {2},
          eid = {163},
        pages = {163},
          doi = {10.3847/1538-4357/adea73},
archivePrefix = {arXiv},
       eprint = {2507.00338},
 primaryClass = {astro-ph.GA},
       adsurl = {https://ui.adsabs.harvard.edu/abs/2025ApJ...990..163B}
}

@article{2025Ichimura_COM_12C13C,
   title={Isotopomer-Specific Carbon Isotope Ratio of Complex Organic Molecules in Star-Forming Cores},
   volume={10},
   ISSN={2472-3452},
   url={http://dx.doi.org/10.1021/acsearthspacechem.5c00180},
   DOI={10.1021/acsearthspacechem.5c00180},
   number={1},
   journal={ACS Earth and Space Chemistry},
   publisher={American Chemical Society (ACS)},
   author={Ichimura, Ryota and Nomura, Hideko and Furuya, Kenji and Hama, Tetsuya and Millar, T. J.},
   year={2025},
   month=dec, pages={43–56} }

@ARTICLE{2025Lin_G205.46-14.56M3_H2Dp,
       author = {{Lin}, Sheng-Jun and {Liu}, Sheng-Yaun and {Sahu}, Dipen and {Pagani}, Laurent and {Hsieh}, Tien-Hao and {Hirano}, Naomi and {Lai}, Shih-Ping and {Liu}, Tie and {Hsu}, Shih-Ying and {Li}, Shanghuo and {Kim}, Kee-Tae},
        title = "{Unveiling Central Ortho-H$_{2}$D$^{+}$ Depletion at Sub-kau Scales in Prestellar Core G205.46-14.56 M3: The First Interferometric Evidence and Implications for Deuterium Chemistry}",
      journal = {\apj},
         year = 2025,
        month = dec,
       volume = {995},
       number = {1},
          eid = {36},
        pages = {36},
          doi = {10.3847/1538-4357/ae0cb5},
archivePrefix = {arXiv},
       eprint = {2509.21158},
 primaryClass = {astro-ph.GA},
       adsurl = {https://ui.adsabs.harvard.edu/abs/2025ApJ...995...36L}
}

@ARTICLE{2025Tasa-Chaveli_GEMS,
       author = {{Tasa-Chaveli}, A. and {Fuente}, A. and {Esplugues}, G. and {Navarro-Almaida}, D. and {Majumdar}, L. and {Rayalacheruvu}, P. and {Rivi{\`e}re-Marichalar}, P. and {Rodr{\'\i}guez-Baras}, M.},
        title = "{Gas phase Elemental abundances in Molecular cloudS (GEMS): XI. The evolution of HCN, HNC, and N$_{2}$H$^{+}$ isotopic ratios in starless cores}",
      journal = {\aap},
         year = 2025,
        month = aug,
       volume = {700},
          eid = {A226},
        pages = {A226},
          doi = {10.1051/0004-6361/202554121},
archivePrefix = {arXiv},
       eprint = {2507.10380},
 primaryClass = {astro-ph.GA},
       adsurl = {https://ui.adsabs.harvard.edu/abs/2025A&A...700A.226T}
}

@ARTICLE{2026Hsu_starless_core,
       author = {{Hsu}, Shih-Ying and {Liu}, Sheng-Yuan and {Liu}, Xunchuan and {Li}, Pak Shing and {Tatematsu}, Ken'ichi and {Hirano}, Naomi and {Lin}, Sheng-Jun and {Kim}, Kee-Tae and {Li}, Shanghuo and {Liu}, Tie and {Sahu}, Dipen},
        title = "{Ubiquity of Methanol and Its Related Chemical Segregation in Orion Starless Cores: The ALMASOP Sample}",
      journal = {\apj},
         year = 2026,
        month = jan,
       volume = {997},
       number = {1},
          eid = {16},
        pages = {16},
          doi = {10.3847/1538-4357/ae25fd},
archivePrefix = {arXiv},
       eprint = {2512.00498},
 primaryClass = {astro-ph.SR},
       adsurl = {https://ui.adsabs.harvard.edu/abs/2026ApJ...997...16H}
}

@article{2026Hsu_HOPS-288,
   author = {Hsu, Shih-Ying and Murillo, Nadia M. and Lee, Chin-Fei and Johnstone, Doug and Hsieh, Tien-Hao and Hirano, Naomi and Bronfman, Leonardo and Chuang, Yo-Ling and Eden, David J. and Li, Shanghuo and Lin, Sheng-Jun and Rawlings, Mark G. and Tatematsu, Ken'ichi and Liu, Sheng-Yuan and Chen, Huei-Ru Vivien and Kim, Kee-Tae and Kuan, Yi-Jehng and Kwon, Woojin and Lee, Chang Won and Lee, Jeong-Eun and Liu, Tie and Luo, Qiuyi and Sanhueza, Patricio and Shang, Hsien},
   title = {HOPS-288: A Laboratory for Complex Organics in Proto-binary/Proto-multiple Systems},
   journal = {The Astrophysical Journal},
   volume = {999},
   pages = {62},
   ISSN = {0004-637X},
   DOI = {10.3847/1538-4357/ae38e1},
   url = {https://ui.adsabs.harvard.edu/abs/2026ApJ...999...62H},
   year = {2026},
   type = {Journal Article}
}
\bibliographystyle{aasjournal}




\end{CJK*}
\end{document}